\documentclass[aip,jcp,reprint,amsmath,amssymb,floatfix]{revtex4-2}
\usepackage[T1]{fontenc}
\usepackage{graphicx}
\usepackage{mathrsfs}
\usepackage{hyperref} 
\hypersetup{hidelinks,pdftitle={Extended Onsager's variational principle for nonlinear chemical reaction kinetics in active matter}}
\begin{document}
\title{Extended Onsager's variational principle for nonlinear chemical reaction kinetics in active matter}
% \title{Extending Onsager's variational principle to nonlinear chemical reaction kinetics in active matter}

\author{Xinpeng Xu}\thanks{Corresponding author}
\email{xu.xinpeng@gtiit.edu.cn}
\affiliation{Department of Physics and Guangdong Key Laboratory of Materials and Technologies for Energy Conversion, Guangdong Technion --- Israel Institute of Technology, 241 Daxue Road, Shantou, Guangdong 515063, China}
\affiliation{Technion --- Israel Institute of Technology, Haifa 3200003, Israel} 

\author{Changhao Li} 
\affiliation{College of Energy Engineering and State Key Laboratory of Clean Energy Utilization, Zhejiang University, Hangzhou, Zhejiang, 310003, China}
\affiliation{Joint Research Centre on Medicine, Xiangshan Hospital of Wenzhou Medical University, Ningbo, Zhejiang 315700, China}

\author{Zhongqiang Xiong}
\affiliation{Zhejiang Key Laboratory of Soft Matter Biomedical Materials, Wenzhou Institute, University of Chinese Academy of Sciences, Wenzhou, Zhejiang 325001, China}

\author{Bin Zheng} 
\affiliation{Zhejiang Key Laboratory of Soft Matter Biomedical Materials, Wenzhou Institute, University of Chinese Academy of Sciences, Wenzhou, Zhejiang 325001, China}

\author{Kento Yasuda}
\affiliation{Laboratory of Physics, College of Science and Technology, Nihon University, Funabashi, Chiba 274-8501, Japan}

\author{Shigeyuki Komura} 
\affiliation{Zhejiang Key Laboratory of Soft Matter Biomedical Materials, Wenzhou Institute, University of Chinese Academy of Sciences, Wenzhou, Zhejiang 325001, China}

\date{\today}
\begin{abstract}
Onsager's variational principle (OVP) constructs dissipative dynamics from a free energy, dissipation potential, and input power, but a deterministic rate law alone does not identify its fluctuations or chemical energy supply. We examine these distinctions for nonlinear chemical reaction kinetics in active matter using established variational and fluctuation structures. Comparing Gaussian diffusion with resolved Poisson reaction events separates statistical assumptions from thermodynamic matching. For the reaction model, matching requires a dissipation potential compatible with the counting statistics and local
detailed balance linking the event bias to chemical affinity; kinetic scales and their state dependence remain constitutive inputs. The resulting nonquadratic Rayleighian selects deterministic rates, and its normalized excess gives their leading fluctuation cost in the frozen-state, many-event regime. Three models sharing passive and fueled conversion channels illustrate the consequences. Spatially uniform reactions exhibit protocol-dependent nonreciprocal response compatible with fixed-state OVP. Distinct Rayleighians reproduce the same selected rate and dissipation power without fixing the fluctuation cost. A nonideal reacting mixture couples phase separation to chemical
input; retaining a finite fuel pool accounts for the free-energy
transfer and dense-domain dissolution after replenishment stops in
the example studied. A membrane--cytosol model admits distinct channel realizations with identical protein-density trajectory statistics but different chemical input power. 
Under a separate kinetic assumption, an exact equation for the
total protein density still requires its initial rate and does
not determine chemical input power or fluctuations. 
Together, these results clarify the physical and statistical inputs required for constructing coupled reaction--transport models and the conditions under which the Rayleighian excess represents their fluctuations.  
\end{abstract}

\maketitle

\section{Introduction}
\label{sec:intro}

Chemical reactions maintain many active systems away from equilibrium
and couple internal chemistry to motion, transport, and structure.
Examples include the cytoskeleton and chemically maintained biomolecular
condensates~\cite{Kruse2005,Julicher2018,Weber2019}.
Reactions change composition and material properties, while mechanical
forces and intermolecular interactions can change reaction rates.
The modeling challenge is to combine these mechanisms under shared
conservation laws and energy exchanges, and to identify which physical
inputs determine the dynamics, its fluctuations, and the chemical power
that sustains it.

Onsager's variational principle (OVP) organizes dissipative dynamics
through a scalar Rayleighian~\cite{Onsager1931a,Onsager1931b,Doi2011,Doi2021}.
For an isothermal system, its physical decomposition is
\begin{equation}
\mathcal R=\Phi+\dot{\mathcal F}-\mathcal P_{\rm in}.
\label{eq:R}
\end{equation}
Here, $\Phi$ is the dissipation potential, $\mathcal F$ the retained free
energy, and $\mathcal P_{\rm in}$ the input power; a dot denotes a time
derivative. All three terms have power units, and positive
$\mathcal P_{\rm in}$ denotes supply to the retained system.
Their physical meaning comes from specified resistance, storage, and
work mechanisms. Minimization over admissible rates at fixed state and
controls selects the potential-generated dissipative dynamics.
Interpreting the Rayleighian excess $\mathcal R-\mathcal R_{\min}$, the difference from its fixed-state minimum, as a fluctuation cost requires matching to a specified stochastic mechanism~\cite{Wang2021}.

\begin{figure*}[t]
\centering
\includegraphics[width=\textwidth]{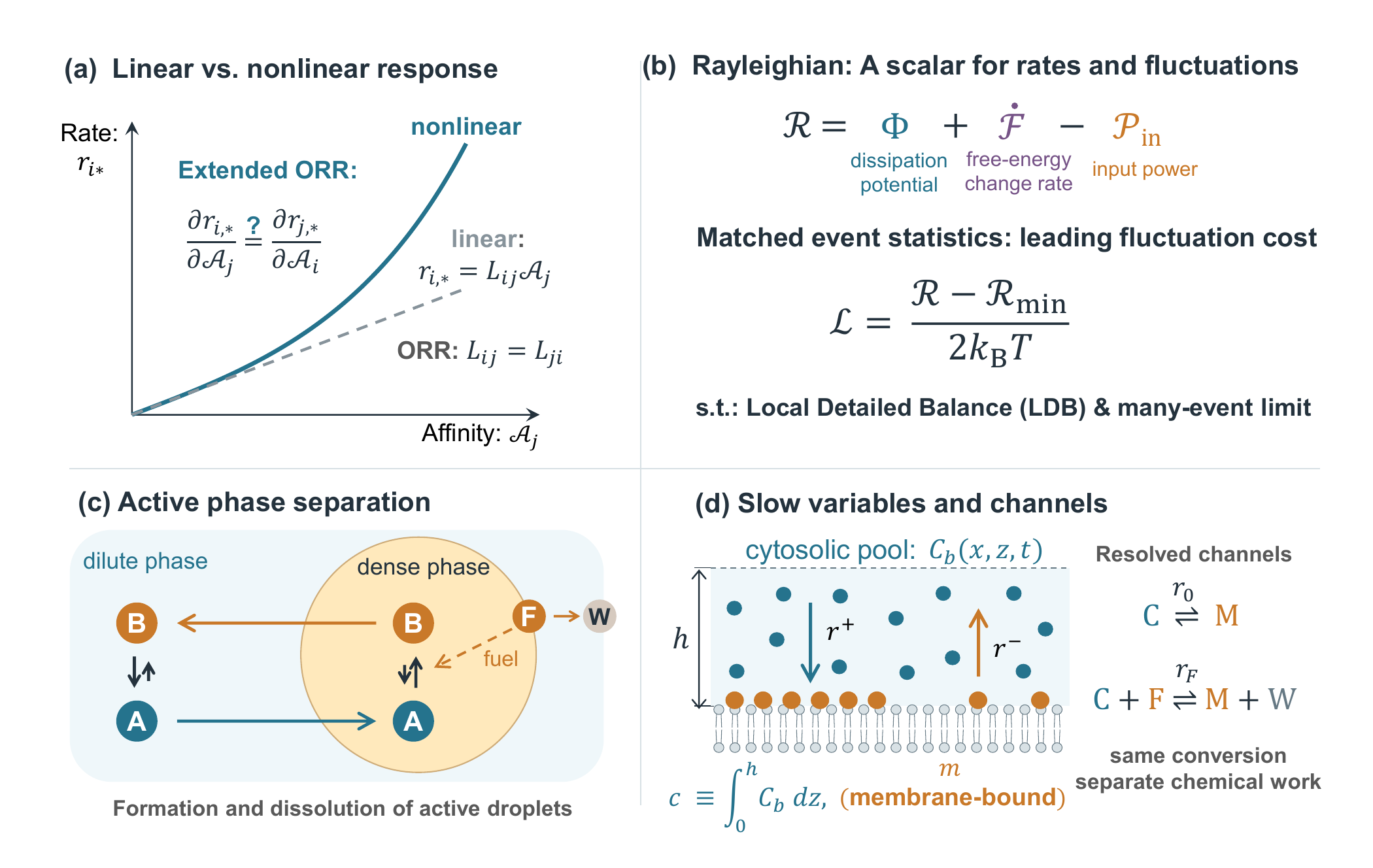}
\caption{\textbf{Nonlinear response, fluctuation costs, and chemically driven organization.}
(a) Near equilibrium, $L_{ij}$ relates selected molar reaction-rate densities $r_{i,*}$ to molar affinities $\mathcal A_j$ and satisfies
the Onsager reciprocal relations (ORR), $L_{ij}=L_{ji}$.  
The solid curve schematically represents the response of a selected reaction-rate density $r_{i,*}$ to a molar affinity $\mathcal A_j$ under a chosen control protocol; the dashed line shows its near-equilibrium linear approximation.
% Here, $\Gamma$ is a fixed reference value of the kinetic scale introduced in Sec.~\ref{sub:Poisson}, used only to normalize
% the plotted rate; $R$ is the gas constant and $T$ the bath temperature. 
The questioned equality concerns a proposed extension to finite-driving differential response across  compositions or controls; it need not hold and is distinct from the symmetry
of a fixed-state trial-rate Hessian (Sec.~\ref{sec:catalyst}).
(b) The physical Rayleighian combines the dissipation potential $\Phi$, retained free-energy rate $\dot{\mathcal F}$, and chemical
input power $\mathcal P_{\rm in}$.
For physically matched event statistics, the normalized Rayleighian excess $\mathscr L=(\mathcal R-\mathcal R_{\min})/(2k_{\mathrm B}T)$
gives the leading rate-fluctuation cost per unit time. Here, the statistics are resolved Poisson counting with local detailed balance (LDB) in the frozen-state, many-event regime~\cite{Shiraishi2023,MPPR2017}. $k_{\mathrm B}$ is Boltzmann's constant and $T$ the bath temperature.
(c) A qualitative reaction--transport cycle couples $A\rightleftharpoons B$ conversion and transport between dilute and dense phases to fuel consumption $F\to W$. Here, $F,W$ denote fuel and waste. Adapted from the mechanism in Fig.~1(a) of Haugerud \emph{et al.}~\cite{Haugerud2026} under CC BY 4.0.
(d) Membrane binding beneath a solution layer of fixed height $h$. The bulk molar concentration $C_b(\boldsymbol x,z,t)$ gives the
cytosolic amount per unit area $c=\int_0^h C_b\,dz$, which exchanges with the membrane-bound molar surface density $m$. The coordinates $\boldsymbol x,z$ are lateral and perpendicular to the membrane; rapid perpendicular mixing is assumed. The arrows $r^\pm$ denote total one-way attachment and detachment rate densities. The passive and fueled channels, with net rates $r_0,r_F$, produce the same conversion between cytosolic and bound protein states $C,M$, but carry different chemical work. The membrane schematic is informed by Weyer, Roth, and Frey~\cite{Weyer2026}. }
\label{fig:overview}
\end{figure*}
 % Copyright 2026 the authors; published by the American Physical Society; adapted under the \href{https://creativecommons.org/licenses/by/4.0/}{Creative Commons Attribution 4.0 license}. Geometry, colors, and labels are redrawn. 

Using established nonlinear variational and reaction-fluctuation structures~\cite{Gyarmati1970,Shiraishi2023,MPPR2017,LMPR2017},
we construct and compare coupled active-matter models sharing passive and fueled conversion channels.
The comparisons make three distinctions explicit.
First, a two-channel model shows how nonreciprocal net-affinity response under specified controls is compatible with partial-affinity response symmetries and a fixed-state dissipation potential.
Second, reproducing the selected rate $r_*$ and dissipation power $\mathcal D$ does not determine the Rayleighian excess or its fluctuation interpretation. 
Third, the membrane--cytosol model shows that distinct passive and fueled channel realizations can have identical total attachment and detachment rates, and hence identical protein-density trajectory statistics, but different chemical input power, showing that composition dynamics alone cannot identify fuel consumption. 
The coupled applications further track free-energy exchange with a finite fuel pool and distinguish the loss of reaction-channel information from the elimination of dynamical state variables.

The practical value of OVP lies in constructing and coupling models.
Free energy specifies interactions and storage, dissipation potentials specify resistance, and input power identifies external supply.
These ingredients also underlie active-gel constitutive
models~\cite{Kruse2005,Julicher2018}. Related continuum descriptions couple drift, diffusion, and recombination in semiconductors~\cite{vanRoosbroeck1950}, and ionic transport, electrostatics, and interfacial kinetics
in electrochemistry~\cite{BazantThorntonAjdari2004}.
Varying the Rayleighian under shared constraints incorporates reaction and transport consistently into their coupled equations and power balance. An equivalent direct construction gives the same equations; the advantage is their systematic assembly rather than a different prediction for already fixed inputs. Restricting trial fields before variation also provides a variational approximation method~\cite{Doi2015Approximation,Doi2019}.

Chemical reaction kinetics makes the distinction between deterministic dynamics and fluctuations particularly clear.
The selected net rate $r_{\rho,*}=r_\rho^+-r_\rho^-$ determines only the difference between the mean forward and reverse
rates $r_\rho^\pm$, not their sum or the associated counting statistics.
Different Rayleighians can therefore select the same deterministic rate while implying different fluctuation costs under their respective stochastic models. 
Onsager--Machlup theory connects near-equilibrium linear relaxation and Gaussian fluctuations~\cite{OnsagerMachlup1953,MachlupOnsager1953}. For nonlinear chemical kinetics, the question is when the same Rayleighian retains both physical rate selection and a statistical meaning. A nonideal free energy changes chemical driving forces; a nonquadratic dissipation potential changes the force--rate relation
at fixed kinetic scales. These are distinct constitutive choices, and both must be specified when reactions couple to phase separation.

Nonlinear variational chemistry has substantial precedents, including thermokinetic and dissipation-potential approaches~\cite{Li1962,Gyarmati1970} and formulations using separate forward and reverse rates~\cite{Lengyel1988}.
The Gyarmati--Li generalized reciprocal relations propose symmetry of
nonlinear differential response for specified conjugate forces and rates~\cite{Li1962,Gyarmati1970}.
Bataille, Edelen, and Kestin found chemical-kinetic counterexamples when net rates are differentiated with respect to net affinities at fixed complementary thermodynamic variables~\cite{Bataille1978}.
Lengyel obtained separate symmetric blocks for partial-affinity
responses~\cite{Lengyel1988}; these statements concern different
derivatives and are not contradictory.
Reciprocal linear response can also persist in selected microscopically irreversible or driven models, whereas other driven states are nonreciprocal~\cite{Gabrielli1996,WangDobnikarFrenkel2022,DeCorato2022}.
Neither outcome establishes a universal finite-affinity symmetry.
For a common convex dissipation potential and its conjugate, the
Gyarmati-type functional and Rayleighian excess coincide up to sign.
Their variational equivalence must therefore be distinguished from
identifying the physical inputs and fluctuation law of a reaction model.

The statistical connection also builds on established results.
Grabert and Green extended short-time transition probabilities and
path integrals to nonlinear irreversible processes~\cite{GrabertGreen1979}.
Large-deviation formulations related trajectory probabilities to generalized dissipative dynamics~\cite{Shiraishi2023,MPR2014,MPR2016}; for reversible chemical jump models, the event statistics yield nonquadratic reaction
potentials~\cite{Shiraishi2023,MPPR2017,LMPR2017}.
For driven Markov systems, local detailed balance and time-symmetric kinetic activity enter nonlinear Onsager--Machlup structures~\cite{MaesNetocny2008,KaiserJackZimmer2018,MaesNetocny2026}; flux-based variational formulations extend beyond detailed balance~\cite{PRS2024}.
Kobayashi \emph{et al.}~\cite{Kobayashi2022Hessian} developed the Hessian geometry of the dual nonquadratic reaction potentials and associated entropy-production decompositions, and discussed quadratic alternatives representing the same kinetics. 
Energetic variational approaches provide related constructions coupling nonlinear reactions and diffusion~\cite{WangLiu2020,WangLiu2022Review,Hao2025GrayScott}.
The extension considered here uses these established structures, rather than proposing a distinct general extremum principle.

The retained variables and reaction channels carry different information.
A description is dynamically sufficient at a stated resolution when its present variables and imposed controls determine the subsequent transition law without unresolved slow memory.
An observed membrane density can fail this test when attachment depends on an evolving cytosolic pool~\cite{Brauns2020,Weyer2026}.
Conditional local equilibrium of eliminated thermal degrees of freedom permits retained free energies without requiring chemical equilibrium.
Resolving reaction channels is a separate requirement: channels with the same retained-state change can exchange different chemical work, so the retained-state dynamics can close without channel labels while concealing fuel consumption~\cite{Polettini2014,RaoEsposito2016}.
Neither dynamical closure nor timescale separation alone establishes local detailed balance~\cite{HartichGodec2023}.
The slow variables, thermodynamically distinct channels, and nonlinear dissipation with its event statistics are therefore complementary modeling choices.

% Our contribution is an explicit comparison of these choices in coupled active-matter models sharing a passive/fueled reaction topology. We connect the physical Rayleighian to both the Gaussian and resolved Poisson descriptions, distinguishing constitutive kinetic scales from thermodynamic matching. The models then show how nonreciprocal chemical response can coexist with fixed-state OVP, how finite fuel changes the free-energy balance of a nonideal mixture, and how aggregate reaction statistics or a density-only equation can leave chemical power and fluctuations undetermined. These comparisons specify what is retained in a physical variational model and what requires additional information; they do not infer a unique mechanism from a deterministic rate law.

These three modeling choices guide the analysis summarized in Fig.~\ref{fig:overview}. Section~\ref{sec:theory} establishes the statistical and thermodynamic foundations by comparing Gaussian and Poisson fluctuations and identifying the conditions under which a physical Rayleighian also determines their fluctuation costs.
We then examine three models sharing passive and fueled conversion channels.
Section~\ref{sec:catalyst} distinguishes nonlinear response across controlled states from fixed-state variational symmetry, and deterministic rates from their fluctuation costs.
Section~\ref{sec:mixture} couples the reactions to nonideal phase separation and a finite fuel pool. 
Section~\ref{sec:duality} examines how channel aggregation and state-variable reduction affect membrane--cytosol dynamics and chemical-power accounting. 
Section~\ref{sec:discussion} discusses the broader modeling implications and limitations.
Appendices~\ref{app:reciprocity}--\ref{app:activecost} provide supporting response relations and stochastic matching conditions.

\section{Extending Onsager--Machlup theory to nonlinear chemical reaction kinetics}\label{sec:theory}

In this section, we first revisit Onsager and Machlup's linear Gaussian theory~\cite{OnsagerMachlup1953,MachlupOnsager1953} 
% and review extensions of Onsager reciprocity and fluctuation theory to nonlinear response, together with their limitations. 
to establish the deterministic and statistical roles of the physical Rayleighian. Our construction of nonlinear chemical kinetics rests on three complementary modeling choices: the retained slow variables, thermodynamically distinct reaction channels, and their nonlinear dissipation and event statistics. For a resolved Markovian reaction model, we derive the nonquadratic reaction-rate fluctuation cost from frozen-state Poisson counting statistics in the many-event regime and impose local detailed balance (LDB) to identify the statistical bias with the chemical affinity. The resulting physical Rayleighian retains the classical form of Eq.~\eqref{eq:R}, $\mathcal R=\Phi+\dot{\mathcal F}-\mathcal P_{\rm in}$, now with a nonquadratic dissipation potential $\Phi$. It also preserves both roles: its minimization over admissible trial rates at fixed state and controls selects the deterministic kinetics, while its normalized excess $(\mathcal R-\mathcal R_{\min})/(2k_{\mathrm B}T)$ gives the leading reaction-rate fluctuation cost per unit time, where $\mathcal R_{\min}$ is the instantaneous minimum.

\subsection{Onsager--Machlup theory and the physical Rayleighian}
\label{sub:OM}

\textbf{Gaussian fluctuations and irreversible dynamics.} 
Onsager and Machlup connected near-equilibrium Gaussian fluctuations with linear irreversible relaxation~\cite{OnsagerMachlup1953}. Following Wang, Qian, and Xu~\cite{Wang2021}, we revisit this connection using an additive It\^o stochastic model for $N_\alpha$ independent retained slow coordinates $\boldsymbol\alpha=\{\alpha_1,\dots,\alpha_{N_\alpha}\}$:
\begin{equation}
\mathrm d\boldsymbol\alpha
=\boldsymbol L\boldsymbol X(\boldsymbol\alpha,t)\,\mathrm dt
+\sqrt2\boldsymbol B\,\mathrm d\boldsymbol W.
\label{eq:SDE}
\end{equation}
Here, $\boldsymbol L$ is a constant response matrix and $\boldsymbol X(\boldsymbol\alpha,t)$ is a prescribed generalized force. The drift is linear in $\boldsymbol X$ but may be nonlinear in $\boldsymbol\alpha$. The components of $\boldsymbol W$ are independent standard Wiener processes, and the constant noise-amplitude matrix $\boldsymbol B$ defines the symmetric diffusion matrix $\boldsymbol D=\boldsymbol B\boldsymbol B^{\mathsf T}$, which is assumed positive definite. We first derive the transition density without imposing Onsager reciprocity on $\boldsymbol L$ or the Einstein relation between $\boldsymbol D$ and $\boldsymbol L$.

For the Markov model in Eq.~\eqref{eq:SDE}, approximate the force over a sufficiently short interval $[t_n,t_{n+1}]$ by its initial value $\boldsymbol X_n=\boldsymbol X(\boldsymbol\alpha_n,t_n)$. This gives the Euler--Maruyama update
\begin{equation}
\Delta\boldsymbol\alpha_n
=\boldsymbol L\boldsymbol X_n\Delta t
+\sqrt{2\Delta t}\boldsymbol B\boldsymbol\xi_n,
\label{eq:EM}
\end{equation}
where
$\Delta\boldsymbol\alpha_n=\boldsymbol\alpha_{n+1}-\boldsymbol\alpha_n$, $\Delta t=t_{n+1}-t_n$, and
$\boldsymbol\xi_n=
[\boldsymbol W(t_{n+1})-\boldsymbol W(t_n)]/\sqrt{\Delta t}$.
These Gaussian vectors have zero mean and unit covariance,
with each $\boldsymbol\xi_n$ independent of the history up to $t_n$. For this update, the conditional increment has mean
$\boldsymbol L\boldsymbol X_n\Delta t$ and covariance
$2\boldsymbol D\Delta t$. Writing $\dot{\boldsymbol\alpha}_n
=\Delta\boldsymbol\alpha_n/\Delta t$ for the measured step rate, the conditional transition density is
\begin{equation}
p_{\rm EM}(\boldsymbol\alpha_{n+1},t_{n+1}|
\boldsymbol\alpha_n,t_n)
=\frac{\exp[-\Delta t\,\mathscr L_D(\dot{\boldsymbol\alpha}_n)]}
{(4\pi\Delta t)^{N_\alpha/2}\sqrt{\det\boldsymbol D}},
\label{eq:EMkernel}
\end{equation}
with the local fluctuation cost
\begin{equation}
\mathscr L_D(\dot{\boldsymbol\alpha})
=\frac14
(\dot{\boldsymbol\alpha}-\boldsymbol L\boldsymbol X_n)^{\mathsf T}
\boldsymbol D^{-1}
(\dot{\boldsymbol\alpha}-\boldsymbol L\boldsymbol X_n).
\label{eq:GaussianCost}
\end{equation}
Here, $\dot{\boldsymbol\alpha}$ denotes a trial rate and
$\mathscr L_D$ has units of inverse-time; the measured step rate $\dot{\boldsymbol\alpha}_n$
does not imply differentiability of a Brownian trajectory. Equation~\eqref{eq:EMkernel} is exact for the frozen update but generally only a short-time approximation to the transition density of Eq.~\eqref{eq:SDE}. The Gaussian cost in Eq.~\eqref{eq:GaussianCost} selects
$\boldsymbol L\boldsymbol X_n$ as the most probable step rate,
but does not by itself identify the physical dissipation
and energy exchanges. 

\textbf{Physical Rayleighian and deterministic rate.} We now describe the deterministic energetics of the retained variables $\boldsymbol{\alpha}$, with the free energy, resistance, and applied forces specified independently of the noise covariance. 
% The resulting Rayleighian generates their overdamped force balance; its interpretation as a fluctuation cost requires a separate noise--response matching condition.
% We now formulate the deterministic force balance independently of the noise law. 
For an isothermal system, let $\boldsymbol X=-\nabla_{\boldsymbol\alpha}\mathcal F+\boldsymbol F_a$, where $\mathcal F(\boldsymbol\alpha)$ is the free energy of the retained system, with no explicit time dependence, and $\boldsymbol F_a(\boldsymbol\alpha,t)$ is a prescribed nonconservative force. For a trial rate $\dot{\boldsymbol\alpha}$, the rate of free energy change and the input power are $\dot{\mathcal F}=\nabla_{\boldsymbol\alpha}\mathcal F
\cdot\dot{\boldsymbol\alpha}$ and $\mathcal P_{\rm in}=\boldsymbol F_a\cdot\dot{\boldsymbol\alpha}$, respectively, so that $\dot{\mathcal F}-\mathcal P_{\rm in}
=-\boldsymbol X\cdot\dot{\boldsymbol\alpha}$. 

For time-reversal-even coordinates, we impose Onsager reciprocity, $\boldsymbol L=\boldsymbol L^{\mathsf T}$. This symmetry follows from microscopic reversibility in equilibrium linear response without time-reversal-odd
controls~\cite{Onsager1931a,Onsager1931b,Casimir1945};
beyond that regime, it is a constitutive assumption. The second law requires non-negative dissipation; we assume $\boldsymbol L$ to be positive definite so that
$\boldsymbol L^{-1}$ defines the resistance matrix. Integrating the linear resistance force $\boldsymbol L^{-1}\dot{\boldsymbol\alpha}$ over the trial rates gives the dissipation potential, normalized by $\Phi(\boldsymbol0)=0$, 
\[
\Phi(\dot{\boldsymbol\alpha})
=\frac12\dot{\boldsymbol\alpha}^{\mathsf T}
\boldsymbol L^{-1}\dot{\boldsymbol\alpha}.
\]
Since $\partial\Phi/\partial\dot{\boldsymbol\alpha}
=\boldsymbol L^{-1}\dot{\boldsymbol\alpha}$, the overdamped force balance between resistance and driving force is the stationarity condition of the physical Rayleighian
\begin{equation}
\mathcal R(\dot{\boldsymbol\alpha};\boldsymbol\alpha,t)
=\Phi-\boldsymbol X\cdot\dot{\boldsymbol\alpha}
=\Phi+\dot{\mathcal F}-\mathcal P_{\rm in},
\label{eq:physicalGaussianR}
\end{equation}
which has the classical form of Eq.~\eqref{eq:R}~\cite{OnsagerMachlup1953,Doi2019,Wang2021}. Its strict convexity gives the unique minimizing rate at fixed state and controls:
\begin{align}
\dot{\boldsymbol\alpha}_*
&=\boldsymbol L\boldsymbol X.
\label{eq:linearOVP}
\end{align}
This recovers the deterministic drift in Eq.~\eqref{eq:SDE}.
Substitution into Eq.~\eqref{eq:physicalGaussianR} gives
$\mathcal R_{\min}(\boldsymbol\alpha,t)
=-\frac12\boldsymbol X^{\mathsf T}\boldsymbol L\boldsymbol X$.
Along the selected deterministic evolution, the free-energy balance is
\begin{equation}
\dot{\mathcal F}=\mathcal P_{\rm in}-\mathcal D,
\label{eq:Frate}
\end{equation}
where $\mathcal D$ is the nonnegative dissipation power, 
\begin{equation*}
\mathcal D =\boldsymbol X\cdot\dot{\boldsymbol\alpha}_*
=\dot{\boldsymbol\alpha}_*^{\mathsf T}
\boldsymbol L^{-1}\dot{\boldsymbol\alpha}_*
=2\Phi(\dot{\boldsymbol\alpha}_*)\ge0. 
\end{equation*}
Positive $\mathcal P_{\rm in}$ denotes work supplied to the retained system, while positive $\dot{\mathcal F}$ denotes free-energy accumulation. The dissipation power $\mathcal D$ measures irreversible free-energy loss along the selected deterministic evolution, not necessarily the heat current or the total dissipation of the stochastic model.
The identity  $\mathcal D=2\Phi(\dot{\boldsymbol\alpha}_*)$ follows from quadratic homogeneity and does not generally hold for nonquadratic dissipation potentials.

\textbf{Einstein relation and fluctuation cost.} At the same frozen state and controls as in Eq.~\eqref{eq:GaussianCost}, completing the square in Eq.~\eqref{eq:physicalGaussianR} using the minimum $\mathcal R_{\min}$ gives
\[
\mathcal R-\mathcal R_{\min}
=\frac12
(\dot{\boldsymbol\alpha}-\dot{\boldsymbol\alpha}_*)^{\mathsf T}
\boldsymbol L^{-1}
(\dot{\boldsymbol\alpha}-\dot{\boldsymbol\alpha}_*).
\]
Here, $\dot{\boldsymbol\alpha}_*=\boldsymbol L\boldsymbol X_n$.
For the present constant-coefficient model at bath temperature $T$, impose the Einstein relation $\boldsymbol D=k_{\mathrm B}T\boldsymbol L$,
where $k_{\mathrm B}$ is Boltzmann's constant. Comparison with Eq.~\eqref{eq:GaussianCost} then gives
\begin{equation}
\begin{gathered}
\mathscr L_D(\dot{\boldsymbol\alpha})
=\frac{\mathcal R(\dot{\boldsymbol\alpha})-\mathcal R_{\min}}
{2k_{\mathrm B}T},
\end{gathered}
\label{eq:Einstein}
\end{equation}
for all independent trial rates $\dot{\boldsymbol\alpha}$. Under this Einstein relation, the same physical Rayleighian both selects the deterministic rate and determines the Gaussian fluctuation cost through its excess. With $\boldsymbol F_a=0$ and a quadratic free energy about a stable equilibrium, the model reduces to the equilibrium linear Gaussian setting. Appendix~\ref{app:reciprocity} derives the equilibrium counterpart for reaction extents from resolved reaction events.

\textbf{Onsager--Machlup path integral and variational principle.} For $N$ Euler--Maruyama steps with $\Delta t=(t_f-t_0)/N$, multiplying the one-step transition densities in Eq.~\eqref{eq:EMkernel} gives the joint probability density of a discretized trajectory, conditioned on its initial state:
\begin{equation} 
p_N(\boldsymbol\alpha_1,\ldots,\boldsymbol\alpha_N
\mid\boldsymbol\alpha_0)
=\frac{\exp\left[{-\mathcal I_N}\right]}
{\left[(4\pi\Delta t)^{N_\alpha/2}
\sqrt{\det\boldsymbol D}\right]^N}, 
\label{eq:EMhistory}
\end{equation}
with $\mathcal I_N=\Delta t\sum_{n=0}^{N-1}
\mathscr L_D(\dot{\boldsymbol\alpha}_n)$. Each cost is evaluated at the current state and time $(\boldsymbol\alpha_n,t_n)$. Integrating over the $N-1$ intermediate states, with $\boldsymbol\alpha_N=\boldsymbol\alpha_f$, and taking $N\to\infty$ at fixed $t_f-t_0$ gives the formal path integral
\begin{equation}
p(\boldsymbol\alpha_f,t_f|\boldsymbol\alpha_0,t_0)
=\int_{\boldsymbol\alpha_0}^{\boldsymbol\alpha_f}
\mathcal D\boldsymbol\alpha\,
e^{-\mathcal I[\boldsymbol\alpha]}.
\label{eq:OMpath}
\end{equation}
The measure carries all $N$ one-step normalization factors in Eq.~\eqref{eq:EMhistory}. Under Einstein matching, the dimensionless Onsager--Machlup action in this left-endpoint representation is 
\begin{equation}
\mathcal I[\boldsymbol\alpha]
=\frac{1}{2k_{\mathrm B}T}\int_{t_0}^{t_f}
\left[
\mathcal R(\dot{\boldsymbol\alpha};\boldsymbol\alpha,t)
-\mathcal R_{\min}(\boldsymbol\alpha,t)
\right]\mathrm dt.
\label{eq:OMaction}
\end{equation}
The action is evaluated as a sum over successive time steps, with $\mathcal R$ and $\mathcal R_{\min}$ evaluated at the current state and time. For a smooth trial trajectory, the discrete action $\mathcal I_N$ converges to the time integral in
Eq.~\eqref{eq:OMaction}. At fixed time resolution, minimizing $\mathcal I_N$ maximizes the corresponding discrete trajectory density, since its normalization factor is independent of the trajectory. The Onsager--Machlup variational principle (OMVP) minimizes $\mathcal I$ over admissible smooth trajectories with prescribed
initial and, when required, final states~\cite{Wang2021,Doi2019}. With only the initial state fixed, the deterministic trajectory attains $\mathcal I=0$. With both endpoints fixed, minimization gives the classical
linear Gaussian fluctuation path and, for nonlinear drift, the leading weak-noise transition path. 
% The normalized left-endpoint path integral requires no
% additional drift-divergence term. At finite noise strength,
% however, comparing probabilities of narrow tubes around
% smooth trajectories involves an additional contribution
% $\frac12\int_{t_0}^{t_f}
% \nabla_{\boldsymbol\alpha}\cdot(\boldsymbol L\boldsymbol X)
% \,\mathrm dt$, up to trajectory-independent terms
% ~\cite{DurrBach1978,CugliandoloLecomte2017,Cates2022}.
% This contribution can affect the most probable trajectory
% when the drift divergence depends on the state, but does
% not alter the physical Rayleighian or the one-step
% Einstein matching in Eq.~\eqref{eq:Einstein}.
 
\textbf{Scope of the Onsager--Machlup description.} 
Constant coefficients $\boldsymbol D$ and $\boldsymbol L$ simplify the stochastic description; Onsager reciprocity $\boldsymbol L=\boldsymbol L^{\mathsf T}$ allows the linear resistance to be generated by a scalar dissipation potential $\Phi$; and Einstein matching relation $\boldsymbol D=k_{\mathrm B}T\boldsymbol L$ connects the physical Rayleighian to the Gaussian fluctuation cost. We now examine how relaxing these assumptions affects deterministic OVP and its statistical interpretation.
(i) Allowing state dependence while retaining symmetric positive-definite mobility preserves deterministic OVP: $\boldsymbol L(\boldsymbol\alpha)$ is held fixed during trial-rate variation, not during state evolution~\cite{Doi2019}.
For thermal diffusion in flat independent coordinates at uniform temperature, with $\boldsymbol D=k_{\mathrm B}T\boldsymbol L$ and no additional reversible currents, the conditional It\^o drift is $\dot{\boldsymbol\alpha}_{\rm I}
=\dot{\boldsymbol\alpha}_*+\nabla_{\boldsymbol\alpha}\cdot\boldsymbol D$, where $\dot{\boldsymbol\alpha}_*=\boldsymbol L\boldsymbol X$ is the deterministic rate in Eq.~\eqref{eq:linearOVP} selected by OVP. 
For this multiplicative-noise model, the Einstein matching relation gives the same quadratic coefficients, but the frozen Gaussian cost is minimized at $\dot{\boldsymbol\alpha}_{\rm I}$, whereas the physical Rayleighian excess is minimized at
$\dot{\boldsymbol\alpha}_*$. A nonzero noise-induced $\nabla_{\boldsymbol\alpha}\cdot\boldsymbol D$ therefore prevents their equality for the same trial rate.
Appendix~\ref{app:ensemble} gives the stochastic consistency conditions~\cite{LauLubensky2007}.
(ii) Onsager reciprocity must be assessed for the specified conjugate rates and forces, response protocol, and time-reversal parities~\cite{Onsager1931a,Casimir1945}. A genuinely nonsymmetric linear resistance cannot be generated by a scalar dissipation potential alone, whose rate Hessian matrix is symmetric. 
The present minimization cannot then reproduce the complete response, although a physical Rayleighian may still describe a potential-generated dissipative part, with nonpotential contributions treated separately~\cite{Goddard2015}.
Nonsymmetric finite-affinity response under changing composition or
reservoir controls is a different issue and need not preclude a
fixed-state reaction Rayleighian (Sec.~\ref{sec:catalyst} and
Appendix~\ref{app:reciprocity}).
(iii) Even with constant coefficients and an unchanged reciprocal response, additional active Gaussian white noise can violate Einstein matching. If matching fails, the Gaussian cost remains computable and deterministic OVP remains valid, but the Rayleighian excess no longer gives the total fluctuation cost at the bath temperature; see Appendix~\ref{app:activecost}.

\textbf{Extension to nonlinear chemical kinetics.}
We retain the physical Rayleighian in Eq.~\eqref{eq:R},
but allow nonquadratic reaction dissipation potentials.
We first choose dynamically sufficient slow variables,
resolve reaction channels by their state changes and
chemical work, and specify their one-way reaction rates.
We then derive the net-rate fluctuation cost from Poisson
counts and impose local detailed balance to identify the
statistical bias with the physical chemical affinity.
The resulting Rayleighian selects the deterministic rates
by minimization, while its excess determines their leading
fluctuation cost in the frozen-state, many-event regime.
This fixed-state variational construction does not require
universal nonlinear differential reciprocity under changes
in composition or reservoir controls, as illustrated in
Sec.~\ref{sec:catalyst}.

\subsection{Slow variables, reaction channels, and chemical driving}
\label{sub:reaction_inputs}

\textbf{Slow variables and Markov closure.}
For the time-local stochastic description sought here, dynamical
sufficiency is a property of the retained variables and observation resolution, within a specified range of states and controls. With the imposed control protocol fixed, it means that the conditional transition law is independent, to the stated accuracy, of earlier retained-state history:
\[
\Pr(\boldsymbol\alpha_{n+1}\mid\boldsymbol\alpha_n,
\boldsymbol\alpha_{n-1},\ldots)
\simeq
\Pr(\boldsymbol\alpha_{n+1}\mid\boldsymbol\alpha_n).
\]
Here, $\boldsymbol\alpha_n$ is the retained state at
$t_n=t_0+n\Delta t$. To obtain such a description by eliminating rapidly
relaxing variables, we seek an intermediate observation interval
\begin{equation}
\tau_{\rm fast}\ll\Delta t\ll\tau_{\rm slow}.
\label{eq:slowwindow}
\end{equation}
Here, $\tau_{\rm fast}$ is the longest relevant conditional relaxation or
correlation time among eliminated variables, and $\tau_{\rm slow}$ is
the shortest relevant evolution time of retained variables or imposed
conditions. At fixed retained state, the eliminated variables must lose
relevant dependence on their preparation before that state changes
appreciably. A large number of microscopic degrees of freedom does not
by itself ensure this conditional relaxation or the scale separation.

A proposed closure should reproduce transition probabilities at several
lags from the same fitted evolution law and show no significant residual
dependence on earlier history. Chapman--Kolmogorov consistency and
relaxation predictions provide practical tests, but finite data do not
prove exact Markovianity.\cite{Prinz2011}
Agreement for mean relaxation alone is insufficient when the model is
also used to predict rate fluctuations or rare histories.
If relevant memory persists, additional variables or a history-dependent
description are required.\cite{Mori1965}
The window in Eq.~\eqref{eq:slowwindow} specifies the regime of our local
approximation, not a prerequisite for every valid coarse-grained theory:
exact Markov projections and descriptions retaining memory provide
other possibilities.\cite{RogersPitman1981,Mori1965}
For example, the membrane--cytosol model in Sec.~\ref{sec:duality}
retains the evolving cytosolic pool because the membrane density alone
does not determine attachment.

For the thermodynamic description, we additionally assume conditional
local equilibrium (LE): at fixed slow variables, the eliminated thermal
degrees of freedom are approximately equilibrated. This permits a retained
free energy and chemical potentials at the current composition without
requiring reaction equilibrium or small affinities.
\cite{RaoEsposito2016}
Kinetic closure, conditional LE, and the local detailed-balance condition
introduced in Sec.~\ref{sub:Poisson} are distinct requirements.
\cite{HartichGodec2023}

\textbf{Reaction channels and material balances.}
Consider a well-mixed isothermal reactor of fixed volume $V$ containing
$N_s$ retained species and $N_r$ resolved reversible reaction channels.
Let $c_i$ be the retained molar concentrations, $r_\rho$ the trial net
reaction-rate densities, and $\nu_{i\rho}$ the products-minus-reactants
stoichiometric coefficients. The material balances are
\begin{equation}
\dot c_i=\sum_{\rho=1}^{N_r}\nu_{i\rho}r_\rho,
\qquad i=1,\ldots,N_s.
\label{eq:stoich}
\end{equation}
A channel denotes one specified reaction event and its reverse; a multistep pathway can contain several channels. Conserved amounts restrict the independent concentrations. Channel rates must nevertheless
be distinguished from independent state rates: different channel allocations, including cycles, can produce the same concentration change.
These stoichiometric distinctions matter both for reciprocal response and for chemical power
\cite{Schnakenberg1976,LengyelGyarmati1981,Polettini2014}. Keeping forward and reverse event counts distinct does not by itself add dynamically necessary slow-state variables.

\textbf{Free energy and chemical work.}
For a retained free-energy density $f(\boldsymbol c)$ with no explicit
time dependence, the free energy and its rate are
\begin{equation}
\mathcal F=Vf(\boldsymbol c),\qquad
\dot{\mathcal F}
=V\sum_\rho\left(\sum_i\nu_{i\rho}\mu_i\right)r_\rho.
\label{eq:reactionFdot}
\end{equation}
Here, $\boldsymbol c$ is the concentration vector and
$\mu_i=\partial f/\partial c_i$ is a molar chemical potential. The second
relation follows from the chain rule and Eq.~\eqref{eq:stoich}.

Reservoir exchanges supply chemical work. Let $a$ label reservoir species
omitted from the retained state, $\nu^{\rm res}_{a\rho}$ their
products-minus-reactants stoichiometry, and $\mu_a^{\rm res}$ their
maintained chemical potentials. The work per mole of forward events and
the input power are
\begin{equation}
\mathcal W_\rho^{\rm res}=-\sum_a\nu^{\rm res}_{a\rho}\mu_a^{\rm res},
\qquad
\mathcal P_{\rm in}=V\sum_\rho\mathcal W_\rho^{\rm res} r_\rho.
\label{eq:reservoirwork}
\end{equation}
The resolved rates and reservoir chemical potentials determine
the chemical free-energy input in this system--reservoir partition.
Converting that input into a heat balance additionally requires
the associated energy and entropy transfers; chemical work is
not itself a heat current~\cite{Seifert2011,RaoEsposito2016}.
Combining the two powers identifies the complete channel affinity:
\begin{equation}\label{eq:affinity0}
\begin{aligned}
\dot{\mathcal F}-\mathcal P_{\rm in}=-V\sum_\rho\mathcal A_\rho r_\rho, 
\end{aligned} 
\end{equation}
with the affinity
\begin{equation}\label{eq:affinity} 
\mathcal A_\rho =\mathcal W_\rho^{\rm res}-\sum_i\nu_{i\rho}\mu_i. 
\end{equation}
Retained and reservoir species contribute once each; no extra driving
is added to an affinity that already includes the reservoirs. A chemostat
fixes its own chemical potential, not generally the complete affinity.
\cite{Polettini2014,RaoEsposito2016}
For example, maintained fuel and waste in
$A+F\rightleftharpoons B+W$ supply
$\mu_F^{\rm res}-\mu_W^{\rm res}$ per mole of forward events. If
replenishment stops, their finite amounts must instead be retained and
that transfer enters the enlarged system's free-energy change. Explicit
changes of energy parameters would require separate parameter-work terms
and are excluded here.

Two channels with the same retained stoichiometry can have different
$\mathcal W_\rho^{\rm res}$. Their rates enter the material balance through their
sum, but their separate values are needed in Eq.~\eqref{eq:reservoirwork}.
They must therefore remain resolved for chemical power accounting even
when the concentration dynamics already closes. Choosing more state
variables and resolving thermodynamically distinct channels are not the
same operation.\cite{Polettini2014,RaoEsposito2016}

For comparison with Sec.~\ref{sub:OM}, independent molar reaction extents
can be used as $\alpha_\rho=\xi_\rho$, with
$\dot\alpha_\rho=Vr_\rho$ and $X_\rho=\mathcal A_\rho$, when this choice
spans the resolved rates. For one independent concentration with
$\dot c=r$, the corresponding choice is $\alpha=c$ and $X=V\mathcal A$.
Such a coordinate identification must respect the conservation constraints
and must not discard work-carrying channel cycles.

These inputs specify the retained thermodynamics and channel work,
but not the mean one-way reaction rates. We next specify the stochastic
reaction mechanism, derive its reaction-rate fluctuation cost, and only then identify its statistical bias with the affinity through LDB. 
 
\subsection{Poisson reaction-event statistics and nonquadratic dissipation} \label{sub:Poisson}

\textbf{Poisson event counting statistics.}
Let $r_\rho^+(\boldsymbol c)$ and $r_\rho^-(\boldsymbol c)$ be specified
positive mean one-way molar rate densities, with any imposed-control
dependence left implicit. For independent Markov event clocks, freezing
the current state and controls gives the counting model
\begin{equation}
\Delta N_\rho^\pm\sim\operatorname{Poisson}(N_{\mathrm A}Vr_\rho^\pm\Delta t),
\qquad q_\rho^\pm=\frac{\Delta N_\rho^\pm}{N_{\mathrm A}V\Delta t}.
\label{eq:counts}
\end{equation}
Here, $N_{\mathrm A}$ is Avogadro's constant, $\Delta N_\rho^\pm$ are event counts,
and $q_\rho^\pm$ are measured one-way rate densities. A forward event
changes $c_i$ by $\nu_{i\rho}/(N_{\mathrm A}V)$. Independent Markov event clocks
are specified at the resolved level; fixed-rate independent Poisson
counts over a finite interval additionally use the frozen-state
approximation.\cite{Gillespie2000,PattersonRenger2019}
The rate spacing is $1/(N_{\mathrm A}V\Delta t)$. Relevant counts must be large
for the continuous-rate approximation, while the state changes little.
Large volume can provide this window; LE alone cannot. We work where
both directions have positive rates. A forbidden direction has zero
count and must instead be treated on its restricted event support.

For one reaction direction (forward or reverse), let $n$ be the number of reaction events observed over an interval $\Delta t$ at a fixed state. Assuming Poisson statistics with mean one-way rate density $a>0$, the exact count probability is $\Pr(n)=e^{-\lambda}\lambda^n/n!$, where $\lambda=N_{\mathrm A}Va\Delta t$ is the mean event count and $q=n/(N_{\mathrm A}V\Delta t)$ is the measured one-way rate density. Applying Stirling's approximation for large counts gives
\begin{equation}\label{eq:onewaycost}
-\ln\Pr(n)\simeq N_{\mathrm A}V\Delta t\,h(q|a),
\end{equation}
where $h(q|a)=q\ln(q/a)-q+a$ for $q\geq0$, with $h(0|a)=a$ by continuity. Here, $h(q|a)$ is the fluctuation cost per unit volume and time in molar-rate units, minimized at the mean rate $q=a$. The approximation retains the nonquadratic dependence on $q$; a Gaussian approximation would require a further expansion
about $q=a$.

A prescribed net rate sums over compatible forward/reverse counts. The leading minimization follows from
\begin{equation}
-\Omega^{-1}\ln\sum_j e^{-\Omega g_j}
=g_{\min}-\Omega^{-1}\ln\sum_j e^{-\Omega(g_j-g_{\min})}.
\label{eq:Laplace}
\end{equation}
Here, $\Omega$ is a dimensionless large scale, $g_j$ are dimensionless
costs, and $g_{\min}=\min_jg_j$. If the remaining weights and number of
alternatives supply no additional exponential factor, the final term is subleading.  
Contracting the forward and reverse counting costs at fixed net rate gives the standard net-current construction~\cite{Gingrich2016,PattersonRenger2019,Shiraishi2023}:
\begin{equation}
\ell(r)=\min_{q^\pm\geq0,\ q^+-q^-=r}
\{h(q^+|r^+)+h(q^-|r^-)\}.
\label{eq:netmin}
\end{equation}
The channel index is suppressed. Differentiation at fixed $r$ gives
$\ln(q^+/r^+)+\ln(q^-/r^-)=0$, hence $q^+q^-=r^+r^-$.
Writing $s=\sqrt{r^2+4r^+r^-}$, the minimizing one-way rates are
$q^\pm=(s\pm r)/2$, and substitution gives
\begin{equation}\label{eq:netcostRaw}
\ell(r)=r\ln\frac{r+s}{2r^+}-s+r^++r^-.
\end{equation}
Here, $s$ and $\ell$ have rate-density units. The cost has zero minimum
at $r_*=r^+-r^-$.

To display its even and odd parts under rate reversal, define
\begin{subequations}\label{eq:rates}
\begin{equation}\label{eq:ratesa}
\vartheta_\rho=\ln\frac{r_\rho^+}{r_\rho^-}, \qquad
\Gamma_\rho=2\sqrt{r_\rho^+r_\rho^-}, 
\end{equation}
\begin{equation}\label{eq:ratesb}
\sigma_\rho =r_\rho^++r_\rho^-, \qquad
r_\rho^\pm =\frac{\Gamma_\rho}{2} e^{\pm\vartheta_\rho/2}.
\end{equation}
\end{subequations} 
Here, $\vartheta_\rho$ is the dimensionless forward--reverse rate bias, $\Gamma_\rho$ is the geometric-mean kinetic scale, invariant
under exchange of the forward and reverse rates, and $\sigma_\rho$ is the sum of the one-way rates. 
These quantities simply reparametrize the prescribed positive one-way rates; their definitions neither impose local detailed
balance nor assume equal activation barriers. 
The fluctuation cost in Eq.~\eqref{eq:netcostRaw} can now be rewritten as
\begin{equation}\label{eq:netcost}
\ell(r)=r\operatorname{arsinh}(r/\Gamma)
-\sqrt{r^2+\Gamma^2}+\sigma-\frac{\vartheta r}{2},
\end{equation}
where $\operatorname{arsinh}x=\ln(x+\sqrt{1+x^2})$ and the channel index is again suppressed. No thermodynamic identification has entered.
The kinetic scales may change with composition or controls, but their current values are fixed during trial-rate variation.

\textbf{Physical Rayleighian and thermodynamic matching.}
The affinity $\mathcal A_\rho$ is fixed by the retained free
energy and reservoir work in Eq.~\eqref{eq:affinity},
independently of the statistical bias $\vartheta_\rho$.
A physical reaction Rayleighian additionally requires a
constitutive dissipation potential $\Phi$.
For the even potentials considered here, with $\Phi(\boldsymbol0)=0$,
it has the form
$\mathcal R=\Phi-V\sum_\rho\mathcal A_\rho r_\rho$.
Whether its excess reproduces the counting cost is a separate
matching question.
Comparison of their rate-reversal-odd parts requires
% \textbf{Thermodynamic identification and rate selection.}
% To identify the statistical bias with the free energy and reservoir work already specified, now impose
\begin{equation}
RT\vartheta_\rho
=RT\ln\frac{r_\rho^+}{r_\rho^-}
=\mathcal A_\rho
=\mathcal W_\rho^{\rm res}-\sum_i\nu_{i\rho}\mu_i.
\label{eq:reactionLDB}
\end{equation}
Here, $R=N_{\mathrm A}k_{\mathrm B}$ is the gas constant and $T$ is the same isothermal bath temperature used to define the thermodynamics. 
This macroscopic LDB is the small-concentration-change form of the finite-number relation comparing a forward transition with its reverse from the arrival state. It does not require equal forward and reverse rates.  
It may hold far from chemical equilibrium and does not determine $\Gamma_\rho$~\cite{SchmiedlSeifert2007,RaoEsposito2016,Maes2021}.
Temperature is a bath property, not a consequence of Poisson or Gaussian statistics. Unresolved active forcing need not satisfy this identification.

The resolved jumps may represent elementary reactions or effective
transitions whose Markov description is independently justified.
A balanced overall reaction equation and timescale separation alone
do not establish either the required counting law or physical
LDB~\cite{AndersonKurtz2011,HartichGodec2023}. 

% With the LDB relation in Eq.~\eqref{eq:reactionLDB}imposed, the even and odd parts of the reaction-rate fluctuation cost lead to the established nonquadratic dissipation potential~\cite{Shiraishi2023,MPPR2017,LMPR2017}, written here in molar-rate and power-density units:
The LDB relation matches the statistical bias to the physical affinity, but
does not determine the rate-reversal-even contribution.
For one channel, matching this contribution with
$\varphi(0;\Gamma)=0$ requires
\begin{equation*}
\varphi(r;\Gamma)
=RT\bigl[\ell(r)+\ell(-r)-2\ell(0)\bigr].
\end{equation*} 
Substituting the Poisson cost gives the established nonquadratic reaction dissipation potential~\cite{Shiraishi2023,MPPR2017,LMPR2017}. It is equivalent to the flux-side potential in Eq.~(12) of
Kobayashi \emph{et al.}~\cite{Kobayashi2022Hessian}, written here in molar-rate and power-density units:
\begin{equation}\label{eq:potential}
\varphi(r;\Gamma)=2RT\left[r\operatorname{arsinh}(r/\Gamma)-\sqrt{r^2+\Gamma^2}+\Gamma
\right].
\end{equation}
Here, $\varphi$ has units of power per volume, vanishes at $r=0$, and is strictly convex, with
$\partial_r\varphi=2RT\operatorname{arsinh}(r/\Gamma)$ and $\partial_r^2\varphi=2RT/\sqrt{r^2+\Gamma^2}>0$. 
For the spatially uniform reaction system, the total dissipation potential is
\begin{equation}
\Phi=V\sum_\rho\varphi(r_\rho;\Gamma_\rho).
\end{equation}
The corresponding physical Rayleighian is $\mathcal R=\Phi+\dot{\mathcal F}-\mathcal P_{\rm in}$,
or equivalently,
\begin{equation}\label{eq:chemicalR}
\mathcal R =V\sum_\rho\left[\varphi(r_\rho;\Gamma_\rho)-\mathcal A_\rho r_\rho\right].
\end{equation}
Minimizing $\mathcal R$ over the trial reaction rates $r_\rho$ at fixed composition, controls, and kinetic scales $\Gamma_\rho$ gives the unique selected rates
\begin{equation}\label{eq:selection}
r_{\rho,*} =\Gamma_\rho\sinh\frac{\mathcal A_\rho}{2RT}
=r_\rho^+-r_\rho^-.
\end{equation}
The star denotes the selected deterministic rate. 
The constitutive choice in Eq.~\eqref{eq:potential} matches the even part of the event cost, while Eq.~\eqref{eq:reactionLDB} matches its odd part;
together they yield the excess relation in
Eq.~\eqref{eq:reactionkernel}.
Substituting Eq.~\eqref{eq:selection} into Eq.~\eqref{eq:stoich} determines the concentration dynamics. 
Along the selected evolution, the free-energy balance is $\dot{\mathcal F}=\mathcal P_{\rm in}-\mathcal D$, where $\mathcal D=V\sum_\rho\mathcal A_\rho r_{\rho,*}\geq0$ is the dissipation power.

The dual dissipation potential $\Psi$ is the convex conjugate of $\Phi$ at fixed state and controls~\cite{OnsagerMachlup1953,Gyarmati1970,Kobayashi2022Hessian}:
\begin{equation}
\Psi(\boldsymbol X)
=\sup_{\boldsymbol s}
\left[\boldsymbol X\cdot\boldsymbol s-\Phi(\boldsymbol s)\right].
\label{eq:dualDissipation}
\end{equation}
Here, $\boldsymbol s=(s_\rho)$ denotes trial rate densities and $\boldsymbol X$ their conjugate force argument.
% Since the selected rates satisfy $\partial_{r_\rho}\Phi(\boldsymbol r_*)=V\mathcal A_\rho$, convex duality gives 
At fixed state and kinetic scales, the nonlinear dual relations
$\boldsymbol X=\partial_{\boldsymbol r}\Phi(\boldsymbol r_*)$
and $\boldsymbol r_*=\partial_{\boldsymbol X}\Psi(\boldsymbol X)$,
with $\boldsymbol X=V\boldsymbol{\mathcal A}$, imply
\begin{equation}
\mathcal D
=\Phi(\boldsymbol r_*)+\Psi(V\boldsymbol{\mathcal A}),
\end{equation}
where $\boldsymbol{\mathcal A}=(\mathcal A_\rho)$.
For a quadratic homogeneous dissipation potential,
$\Psi(V\boldsymbol{\mathcal A})=\Phi(\boldsymbol r_*)$,
so $\mathcal D=2\Phi(\boldsymbol r_*)$.
For the present nonquadratic reaction potential, this factor-of-two relation holds to leading quadratic order near equilibrium, but not in general.
% The power balance is $\dot{\mathcal F}=\mathcal P_{\rm in}-\mathcal D$, with $\mathcal D=V\sum_\rho\mathcal A_\rho r_{\rho,*}
% =V\sum_\rho r_{\rho,*}\partial_{r_\rho}\varphi\geq0$.
% Nonlinear dissipation remains a conjugate force--rate product, but
% $\mathcal D$ is generally not $2\Phi$. 
For the thermodynamically resolved isothermal reaction model, channelwise LDB identifies $\mathcal D=T\dot S_{\rm prod}^{\rm mac}$, where $\dot S_{\rm prod}^{\rm mac}$ is the entropy production rate of the macroscopic rate equations~\cite{RaoEsposito2016}. This identification neither makes each reaction contribution a heat current nor equates $\mathcal D$ with the complete finite-system stochastic entropy production. 
% It is free-energy dissipation, or temperature times the macroscopic entropy production in this resolved isothermal model, not an identification of every reaction term with heat. 
Equation~\eqref{eq:netmin} is minimized over unobserved counts at fixed $r$; Eq.~\eqref{eq:selection} instead selects the typical $r$. 
This also differs from minimizing the dissipation potential at a prescribed concentration-change rate, as in Kobayashi \emph{et al.}~\cite[Eq.~(D12)]{Kobayashi2022Hessian}. That constrained minimization determines a reference allocation of reaction-channel rates, not generally the actual driven rates.

\textbf{Quadratic limit and mobility normalization.}
Expanding Eq.~\eqref{eq:potential} about zero trial rate at fixed
state gives
\begin{equation*}
\varphi(r_\rho;\Gamma_\rho)
=\frac{RT}{\Gamma_\rho}r_\rho^2
+O\!\left(\frac{RT\,r_\rho^4}{\Gamma_\rho^3}\right)
\simeq\frac{r_\rho^2}{2L_\rho},
\qquad
L_\rho=\frac{\Gamma_\rho}{2RT},
\end{equation*}
for $|r_\rho|/\Gamma_\rho\ll1$.
Here, $L_\rho$ is the reaction-rate-density mobility of this
quadratic approximation, and $L_\rho^{-1}$ is its resistance.
For $|\mathcal A_\rho|\ll RT$, variation gives
$r_{\rho,*}\simeq L_\rho\mathcal A_\rho$.
A strict linearization about chemical equilibrium evaluates
$L_\rho$ there; it does not determine the full composition
dependence of $\Gamma_\rho$.
This small-affinity limit is distinct from the local Gaussian
expansion of fluctuations about a finite-affinity mean in
Sec.~\ref{sub:ratefluct}.

The mobility also depends on the chosen rate coordinate and its
conjugate force.
For a single independent conversion with $\dot c=r_\rho$ and
molar extent rate $\dot\xi=Vr_\rho$, the corresponding scalar
entry of $\boldsymbol L$ in Sec.~\ref{sub:OM} is $VL_\rho$
for $\alpha=\xi$, $X=\mathcal A_\rho$, and $L_\rho/V$
for $\alpha=c$, $X=V\mathcal A_\rho$.
Appendix~\ref{app:reciprocity} checks the extent normalization
through its equilibrium response and counting variance.

\textbf{Independent event drivers and coupled dynamics.}
The channelwise sum in Eq.~\eqref{eq:chemicalR} follows from the
specified independent Poisson event drivers.
At fixed composition and controls,
$\partial^2\Phi/(\partial r_\rho\partial r_{\rho'})=0$
for $\rho\ne\rho'$, since each term depends on one trial channel rate.
This separability belongs to the resolved event-rate representation;
it does not imply an uncoupled chemical network.
\cite{AndersonKurtz2011,PattersonRenger2019}
The diagonal kinetic matrix in the equilibrium linear limit is
Onsager-symmetric: reciprocity permits cross coefficients but does
not require them to be nonzero.
Changing to other independent rate coordinates, with consistently
transformed conjugate forces, can introduce reciprocal cross
coefficients.~\cite{LengyelGyarmati1981}

Channels remain coupled through their composition-dependent
affinities and kinetic scales and through the material balances.
An event in one channel can change subsequent rates in other
channels, so independent local event drivers do not imply
independent accumulated counts along an evolving trajectory.
Moreover, one event can change several species.
Since a forward or reverse event in channel $\rho$ changes $c_i$
by $\pm\nu_{i\rho}/(N_{\mathrm A}V)$, the Poisson counting variances give
\begin{equation}
\operatorname{Cov}(\mathrm dc_i,\mathrm dc_j|\boldsymbol c)
=\frac{\mathrm dt}{N_{\mathrm A}V}
\sum_\rho\nu_{i\rho}\nu_{j\rho}\sigma_\rho
+o(\mathrm dt),
\label{eq:chemicalcov}
\end{equation}
where $\sigma_\rho=r_\rho^++r_\rho^-$ is the sum of the mean
one-way rate densities.
The concentration covariance can therefore have off-diagonal
entries even when the local channel cost is additive.
Changing rate coordinates, or contracting the channel cost subject
to prescribed concentration rates, can likewise produce a
nonseparable reduced cost.
The two-channel model in Sec.~\ref{sec:catalyst} illustrates
correlated material and fuel conversion and distinguishes
fixed-state rate variation from response under changing
composition or reservoir controls.

Thus, resolving channels does not remove all coupling, and observed
cross correlations alone do not establish that channels are missing.
If chosen rate variables count common events, their joint statistics
must be derived from the underlying event model.
A simultaneous physical event is represented as one Markov jump
with its complete stoichiometric change.
If eliminating intermediates instead leaves slow memory, a
description retaining that memory or the relevant intermediates
is required.
Cross terms in a fluctuation-derived dissipation potential must
follow from this stochastic construction; inserting them
independently requires rechecking the fluctuation interpretation.

\textbf{Kinetic scales and mass-action specialization.}
For the specified Poisson event model, matching the counting
statistics fixes the trial-rate dependence of
$\varphi(r_\rho;\Gamma_\rho)$, while LDB relates the
forward--reverse rate ratio to the physical channel affinity.
The kinetic scale $\Gamma_\rho$ and its dependence on composition and controls remain constitutive inputs, analogous to specifying mobility or resistance in quadratic OVP~\cite{Doi2011,Doi2021,Lyu2026Glassy}.
These inputs must be supplied from the reaction mechanism,
kinetic measurements, or an explicit phenomenological assumption; they are not determined by the free energy, LDB, or variational minimization.
Once $\Gamma_\rho$ is specified, Eq.~\eqref{eq:selection}
determines the selected net rate without a separate constitutive law for that rate.

To express this kinetic input in chemical terms, let
$\nu_{s\rho}^{\rm R},\nu_{s\rho}^{\rm P}\geq0$ be the dimensionless stoichiometric coefficients of species $s$ on the reactant and product sides of channel $\rho$. 
They specify the reaction, not the amounts or concentrations present. For retained species,
$\nu_{i\rho}=\nu_{i\rho}^{\rm P}-\nu_{i\rho}^{\rm R}$
is the coefficient in Eq.~\eqref{eq:stoich}; reservoir coefficients use the same products-minus-reactants convention.
The index $s$ below includes both retained and reservoir species, with $\mu_s$ denoting the corresponding molar chemical potential.
Define the stoichiometrically weighted chemical-potential sums
\begin{equation*}
\mathcal G_\rho^{\rm R}
=\sum_s\nu_{s\rho}^{\rm R}\mu_s,
\qquad
\mathcal G_\rho^{\rm P}
=\sum_s\nu_{s\rho}^{\rm P}\mu_s.
\end{equation*}
Both have molar-energy units, and their difference is the complete
affinity of Eq.~\eqref{eq:affinity}:
$\mathcal A_\rho=\mathcal G_\rho^{\rm R}-\mathcal G_\rho^{\rm P}$.
For example, $A+F\rightleftharpoons B+W$ has
$\mathcal G_F^{\rm R}=\mu_A+\mu_F$ and
$\mathcal G_F^{\rm P}=\mu_B+\mu_W$.

A useful parametrization is
\begin{equation}
\Gamma_\rho=2\kappa_\rho
\exp\!\left[
\frac{\mathcal G_\rho^{\rm R}+\mathcal G_\rho^{\rm P}}{2RT}
\right],
\label{eq:complexscale}
\end{equation}
where $\kappa_\rho>0$ is a rate-density prefactor that may depend
on composition and controls.
Using $\mathcal A_\rho=\mathcal G_\rho^{\rm R}
-\mathcal G_\rho^{\rm P}$ in Eq.~\eqref{eq:selection} gives
\begin{equation}
r_{\rho,*}=\kappa_\rho
\left[
e^{\mathcal G_\rho^{\rm R}/(RT)}
-e^{\mathcal G_\rho^{\rm P}/(RT)}
\right].
\label{eq:massactionselected}
\end{equation}
With unrestricted $\kappa_\rho$, this parametrizes a positive
kinetic scale rather than predicting it. Chemical-potential exponential rate laws with a common transition-state prefactor
are also used in nonequilibrium reaction
kinetics~\cite{Bazant2009,Bazant2013}.
Specifying the prefactor and its state dependence supplies kinetic information,
such as activation-barrier and attempt-frequency effects,
that neither LE nor LDB determines. 

Write $\mu_s=\mu_s^\circ+RT\ln a_s$, where $\mu_s^\circ$ is the
standard molar chemical potential and $a_s$ the dimensionless
activity.
Define the activity-based one-way prefactors by
\begin{equation*}
\begin{aligned}
k_\rho^{+,a}
&=\kappa_\rho
\exp\!\left[
\frac{\sum_s\nu_{s\rho}^{\rm R}\mu_s^\circ}{RT}
\right],\\
k_\rho^{-,a}
&=\kappa_\rho
\exp\!\left[
\frac{\sum_s\nu_{s\rho}^{\rm P}\mu_s^\circ}{RT}
\right].
\end{aligned}
\end{equation*}
The superscript $a$ distinguishes them from concentration-based
coefficients.
Both have rate-density units and inherit any composition
dependence of $\kappa_\rho$.
Equation~\eqref{eq:complexscale} then becomes
\begin{equation}
\Gamma_\rho
=2\sqrt{k_\rho^{+,a}k_\rho^{-,a}}\,
\prod_s a_s^{(\nu_{s\rho}^{\rm R}+\nu_{s\rho}^{\rm P})/2}.
\label{eq:activityGamma}
\end{equation}
Combining this scale with
$r_\rho^\pm=(\Gamma_\rho/2)e^{\pm\mathcal A_\rho/(2RT)}$
gives
\begin{equation}
r_\rho^+=k_\rho^{+,a}\prod_s a_s^{\nu_{s\rho}^{\rm R}},
\qquad
r_\rho^-=k_\rho^{-,a}\prod_s a_s^{\nu_{s\rho}^{\rm P}}.
\label{eq:activityRateLaw}
\end{equation}
For an elementary channel, choosing $\kappa_\rho$ to be
composition independent at fixed temperature, controls, and
standard states makes $k_\rho^{\pm,a}$ constant and yields the constant-coefficient activity-based law of mass action.
This is an additional kinetic specialization, not a consequence of stoichiometry, LE, or LDB alone. Effective multistep reaction kinetics need not have this form; composition-dependent prefactors can preserve LDB without obeying this constant-coefficient law of mass action.
% If eliminating slow intermediate variables leaves relevant memory, additional state variables or a memory-dependent description are required, rather than merely a modified instantaneous prefactor.
The Poisson assumption specifies the counting statistics and the associated cost of fluctuations once the mean one-way rates are prescribed; however, it does not constrain how the mean one-way rates depend on composition. 

For a dilute ideal mixture, $a_s=c_s/c^\circ$, where $c^\circ>0$
is a fixed reference concentration.
Absorbing its powers into the coefficients gives
\begin{equation*}
k_\rho^+
=k_\rho^{+,a}(c^\circ)^{-\sum_s\nu_{s\rho}^{\rm R}},
\qquad
k_\rho^-
=k_\rho^{-,a}(c^\circ)^{-\sum_s\nu_{s\rho}^{\rm P}},
\end{equation*}
and hence the usual concentration form of mass action.
The units of these concentration-based constants depend on the
total stoichiometric power on the corresponding reaction side.
\cite{RaoEsposito2016,Goddard2023}

\textbf{Constitutive choices in the applications.}
For the passive and fueled conversions of
Sec.~\ref{sec:catalyst}, constant-coefficient concentration
mass action gives
$\Gamma_0=2\sqrt{k_0^+k_0^-c_Ac_B}$ and
$\Gamma_F=2\sqrt{k_F^+k_F^-c_Ac_Bc_Fc_W}$,
as used in Eq.~\eqref{eq:catrates}.
These factors express reactant and product availability through
$\Gamma_\rho=2\sqrt{r_\rho^+r_\rho^-}$.
Their square roots, and the half powers in
Eq.~\eqref{eq:activityGamma}, do not represent fractional
reaction orders.

The nonideal mixture in Sec.~\ref{sec:mixture} instead retains
chemical-potential exponentials in Eq.~\eqref{eq:fuelrates},
including nonideal and gradient contributions.
Its exchange-potential representation absorbs solvent factors into
$k_0,\kappa_F$ without determining their composition dependence.
The membrane model in Sec.~\ref{sec:duality},
Eq.~\eqref{eq:memGamma}, retains ideal component activities but
uses state-dependent channel prefactors to describe saturating
cooperative recruitment and its allocation to the fueled channel.
It is therefore not a constant-prefactor mass-action special case:
ideal component thermodynamics does not fix the kinetic prefactors.
The application models in Secs.~\ref{sub:coupledphysical},
\ref{sec:mixture}, and \ref{sub:membrane_model} use the same
Poisson-matched potential in Eq.~\eqref{eq:potential}, with different constitutive functions $\Gamma_\rho$.
Section~\ref{sub:ratefluct} instead compares different Rayleighian
representations of the same selected rate.  

Inhibition, saturation, and interactions involving activated states
can invalidate the constant-prefactor approximation.
If eliminated intermediates relax on the retained timescale,
additional state variables or memory may be required rather than
merely a modified instantaneous prefactor.
\cite{Bauermann2022,Bazant2013}
Throughout, holding $\Gamma_\rho$ fixed during trial-rate variation
does not make it constant along the evolving state.
The further approximation
$\Gamma_\rho\simeq\Gamma_{\rho,\rm eq}$, where
$\Gamma_{\rho,\rm eq}$ is its equilibrium value, is justified only
over a neighborhood in which its relative variation is negligible.
It supplies a constant coefficient for the local Onsager limit,
not a general kinetic law over finite changes in
composition.~\cite{Doi2011}

\subsection{The extended Rayleighian: deterministic and statistical roles} \label{sub:comparison}

\textbf{One Rayleighian for rates and fluctuations.}
The resolved LDB-compatible channels realize the established nonlinear Onsager--Machlup structure~\cite{KaiserJackZimmer2018,PRS2024}. Their frozen-state, many-event rate probability is
\begin{equation}
\begin{aligned}
\Pr(\{r_\rho\}|\boldsymbol c)
&\asymp\exp[-N_{\mathrm A}V\Delta t\sum_\rho\ell_\rho(r_\rho)]\\
&=\exp[-\Delta t(\mathcal R-\mathcal R_{\min})/(2k_{\mathrm B}T)].
\end{aligned}
\label{eq:reactionkernel}
\end{equation}
Here, $\asymp$ denotes leading exponential equality, not a normalized exact finite-count density. The local fluctuation cost per unit time is
$\mathscr L=N_{\mathrm A}V\sum_\rho\ell_\rho(r_\rho)$.
The rates must satisfy the concentration constraints. 
An endpoint probability further sums over unresolved rate allocations.
% The instantaneous minimum is
% $\mathcal R_{\min}=-2RTV\sum_\rho(\sigma_\rho-\Gamma_\rho)$.
% The corresponding joint history cost is
The instantaneous minimum is
$\mathcal R_{\min}=-\Psi(V\boldsymbol{\mathcal A})
=-2RTV\sum_\rho(\sigma_\rho-\Gamma_\rho)$.
The corresponding Gyarmati-type local functional,
expressed in power units~\cite{Gyarmati1970}, satisfies
\begin{subequations}\label{eq:GyarmatiReactionCost}
\begin{equation}\label{eq:GyarmatiReactionCosta} 
\mathcal G=V\sum_\rho\mathcal A_\rho r_\rho
-\Phi(\boldsymbol r)-\Psi(V\boldsymbol{\mathcal A}), 
\end{equation}
\begin{equation}\label{eq:GyarmatiReactionCostb} 
-\mathcal G=\mathcal R-\mathcal R_{\min}
=2RTV\sum_\rho\ell_\rho(r_\rho)\ge0. 
\end{equation}
\end{subequations}
At fixed state and controls, maximizing $\mathcal G$ is equivalent to minimizing $\mathcal R$.
The nonnegative excess is the standard convex-dual gap, also written as a mixed Bregman divergence by Kobayashi \emph{et al.}~\cite[Eq.~(30)]{Kobayashi2022Hessian}.
Convex duality establishes the variational equivalence; the specified Poisson statistics and LDB identify $-\mathcal G/(2k_{\mathrm B}T)$ with the local fluctuation cost in the stated many-event limit. 
Equation~\eqref{eq:Einstein} is the quadratic, Einstein-matched counterpart.
The corresponding dimensionless trajectory action $\mathcal I$ is 
\begin{equation}
\mathcal I=N_{\mathrm A}V\int_{t_0}^{t_f}\sum_\rho\ell_\rho(r_\rho)\,dt
=\frac1{2k_{\mathrm B}T}\int_{t_0}^{t_f}(\mathcal R-\mathcal R_{\min})\,dt.
\label{eq:gap}
\end{equation}
Here, $t_0$ and $t_f$ are the initial and final times. Conditional on the initial state, the trajectory weight scales as
$e^{-\mathcal I}$ at the stated resolution and leading exponential order.
The state-dependent minimum $\mathcal R_{\min}$ must be retained when comparing trajectories. 
% The same physical Rayleighian selects
% the deterministic rates and weights fluctuations: this combined property,
% not merely the existence of an arbitrary scalar minimum, is the statistical
% extension established here.

The role of LDB is also evident from the rate-reversal asymmetry:
\begin{equation}\label{eq:costreversal}
\ell(r)-\ell(-r)=-\vartheta r
\mathrel{\mathop{=}^{\rm LDB}}-\frac{\mathcal A r}{RT}.
\end{equation}
The first equality follows from Poisson counting statistics, whereas LDB identifies the statistical bias with the physical channel affinity $\mathcal A$ through
$RT\vartheta=\mathcal A$.
Thus, LDB determines the physical rate-reversal asymmetry, but neither the kinetic scale $\Gamma$ nor the full fluctuation cost $\ell(r)$.
Along the selected evolution, the rate--affinity products give the dissipation power $\mathcal D$ and the corresponding macroscopic entropy production by
$\mathcal D=V\sum_\rho\mathcal A_\rho r_{\rho,*} =T\dot S_{\rm prod}^{\rm mac}$,
connecting the fluctuation asymmetry to the deterministic free-energy balance. 

This frozen-state relation is consistent with fluctuation-theorem descriptions~\cite{Maes2021,Yasuda2026Covariant}, but does not by itself give the entropy production of a finite stochastic trajectory. The latter requires the forward and reversed path probabilities, including their initial distributions and control protocols~\cite{SchmiedlSeifert2007,Seifert2012,Maes2021}. Identifying heat and matter-related entropy flows further requires the corresponding energetic information.
LDB permits driven chemical cycles and does not imply equilibrium detailed balance or universal finite-affinity differential reciprocity~\cite{RaoEsposito2016}.

\textbf{Physical interpretation and scope.}
Near equilibrium the expansion in Sec.~\ref{sub:Poisson} recovers linear chemical response. A different expansion at finite affinity gives $\ell(r)\simeq(r-r_*)^2/(2\sigma)$ and local measured-rate variance $\sigma/(N_{\mathrm A}V\Delta t)$. This Gaussian central approximation neither linearizes the constitutive law nor retains the non-Gaussian tails.
\cite{Horowitz2015} Conversely, chemical event counts give a local equilibrium Einstein relation for reaction-extent fluctuations without a diffusion approximation (Appendix~\ref{app:reciprocity}). Gaussianity, whiteness, temperature, and thermodynamic matching are different properties.

A physical Rayleighian can retain its deterministic role when additional active fluctuations are not matched to the specified mobility and bath temperature, provided its constitutive inputs still describe the deterministic response. Its minimum then selects the deterministic rates, but its excess is not generally the total fluctuation cost. Appendix~\ref{app:activecost}, Eq.~\eqref{eq:activeCurvatureComparison}, compares the quadratic forms set by physical resistance and inverse diffusion covariance in the same retained coordinates. 
Their common minimizer does not establish the Einstein matching required for identical fluctuation costs. The total fluctuations must instead be calculated from the specified stochastic model; Eq.~\eqref{eq:gap} is not asserted for arbitrary active noise.
Appendix~\ref{app:ensemble} derives the drift consistency condition for state-dependent thermal diffusion and contrasts it with the chemical Langevin approximation obtained from reaction counts. 
% Appendix~\ref{app:ensemble} starts from composition-dependent reaction counts, distinguishes their chemical Langevin approximation from a thermodynamically matched diffusion, and states when a single physical Rayleighian also describes the empirical concentration distribution and its probability flux. This is not an exact single-reactor propagator or a restoration of Poisson tails lost through a diffusion approximation.

The stochastic and thermodynamic descriptions must be reassessed when
slow memory is discarded, channels with different work are merged, or
unresolved forcing is treated as an equilibrium bath. Timescale separation
alone does not guarantee physical LDB.\cite{HartichGodec2023}
For general colored or heavy-tailed processes, even a local exponential
rate cost may require a different state space or asymptotic description.
% A formula for a scalar cannot replace these checks.

% \textbf{Physical meaning and nonuniqueness.}
% \label{sub:physicalmeaning}\label{app:physicalmeaning}
A physical Rayleighian requires independently justified storage, actual
work mechanisms, and dissipative forces generated by the chosen
potential. For independent admissible rates, write
$\mathcal R=\Phi-\boldsymbol X\cdot\dot{\boldsymbol\alpha}$.
If $\Phi$ is differentiable and convex with $\Phi(0)=\min\Phi=0$,
variation selects $\partial_{\dot{\boldsymbol\alpha}}\Phi=\boldsymbol X$ and
$\mathcal D=\dot{\boldsymbol\alpha}_*\cdot\partial_{\dot{\boldsymbol\alpha}}\Phi
\geq\Phi(\dot{\boldsymbol\alpha}_*)\geq0$.
Homogeneous rate constraints that do no work preserve $\mathcal D=\boldsymbol X\cdot\dot{\boldsymbol\alpha}_*$; driven constraints require additional input-power terms. These checks do not validate an arbitrary proposed material law. Nonpotential dissipative responses and reversible couplings need their appropriate formulation.\cite{Doi2011,Wang2021}

A local chemical relaxation illustrates the inverse ambiguity.
Let $\xi$ be the molar reaction-extent displacement from equilibrium.
For any resistance $\zeta>0$, the inputs
$\Phi_\zeta=\zeta\dot\xi^2/2$ and $\mathcal F_\zeta=\zeta\lambda\xi^2/2$
select the same equation $\dot\xi=-\lambda\xi$, where $\lambda>0$ is a
relaxation frequency. Different $\zeta$ describe different physical storage
and resistance, not a freedom to alter measured parameters in one model; the corresponding thermally matched stochastic model has reaction-extent diffusion coefficient $k_{\mathrm B}T/\zeta$. 
A different ambiguity is
\begin{equation}
\mathcal F\mapsto\mathcal F+G,\qquad
\mathcal P_{\rm in}\mapsto\mathcal P_{\rm in}+\dot G
\quad\Longrightarrow\quad\widetilde{\mathcal R}=\mathcal R.
\label{eq:physicalGauge}
\end{equation}
Here, $G$ is a differentiable state function. This changes the decomposition,
not the scalar; arbitrary $G$ need not describe the same experimental
storage and reservoirs. Finally, only after physical matching is established
can a local cost with zero minimum fix
$\mathcal R(\dot{\boldsymbol\alpha})=2k_{\mathrm B}T[\mathscr L(\dot{\boldsymbol\alpha})-\mathscr L(0)]$
by the convention $\mathcal R(0)=0$, when the zero-rate cost is finite.
Such normalization does not itself determine hidden chemistry or physical
resistance. 
These distinctions explain why a deterministic model can be
constructively useful without automatically specifying its fluctuations.

\textbf{A recipe for constructive modeling with OVP.}
OVP constructs coupled dynamics from specified physical inputs, without requiring the evolution equations in advance.
First, choose dynamically sufficient slow variables and resolve reaction channels by their stoichiometric changes, including reservoir exchanges, to establish the material balances. 
Second, specify the retained free energy and reservoir chemical potentials to determine the free-energy rate, chemical input power, and channel affinities.
Third, model the dissipation mechanisms and kinetic scales, and minimize the physical Rayleighian
$\mathcal R=\Phi+\dot{\mathcal F}-\mathcal P_{\rm in}$
over admissible trial rates at fixed state and controls.
The selected rates and material balances then determine the deterministic evolution and its power balance.
Predicting fluctuations requires a specified stochastic mechanism; obtaining them from the same Rayleighian additionally requires matching its excess to the corresponding fluctuation cost at the same resolution and in the stated probability limit. 

For the resolved Poisson reaction model, the specified kinetic scales $\Gamma_\rho$ and LDB give the compatible one-way rates
\begin{equation*}
r_\rho^\pm=\frac{\Gamma_\rho}{2}
\exp\!\left(\pm\frac{\mathcal A_\rho}{2RT}\right).
\end{equation*}
The Poisson counting statistics determine the fluctuation cost, while Eq.~\eqref{eq:potential} matches its rate-reversal-even part and LDB matches its bias to the physical affinity.
Consequently, the Rayleighian excess gives the leading
rate-fluctuation cost in the frozen-state, many-event limit. 
A finite-number realization additionally requires allowed
reaction jumps and reverse-edge LDB. 
Without such statistical and thermodynamic matching, the
physical Rayleighian still selects the deterministic dynamics but does not determine their fluctuations.

\section{Spatially uniform reactions: reciprocity and fluctuations}
\label{sec:catalyst}

\subsection{Physical channels and variational dynamics}
\label{sub:coupledphysical}

\textbf{Reaction channels and thermodynamics.} Fuel-driven state conversion is a basic mechanism in dissipative self-assembly. For example, carbodiimide fuel activates dicarboxylate precursors into transient anhydrides, which can assemble before hydrolyzing back~\cite{TenaSolsona2017}.
Motivated by this system, we consider a spatially uniform model with two reversible reaction channels, passive conversion and fuel-coupled conversion:
\begin{equation}
A\rightleftharpoons B,
\qquad
A+F\rightleftharpoons B+W,
\label{eq:coupledMotif}
\end{equation}
where $A,B$ denote precursor and activated states, and $F,W$ denote fuel and waste.
The subscripts $0$ and $F$ label the passive and fuel-coupled channels, respectively. The model omits independent fuel consumption and assembly feedback rather than reproducing the full experimental reaction network.
We distinguish partial- and net-affinity responses under different chemical-potential control protocols from fixed-state Rayleighian rate variation, and compare fluctuation models that yield the same selected rate.

Consider a well-mixed isothermal reactor of volume $V$ at temperature $T$, containing the four species with positive molar concentrations $c_s$, $s=A,B,F,W$. The trial net reaction-rate densities $r_0,r_F$ are positive for forward $A\to B$ conversion and have units of moles per volume per time.
For the closed reactor, the concentration balances are
\begin{equation}
\begin{aligned}
\dot c_A&=-r_0-r_F,&\qquad \dot c_B&=r_0+r_F,\\
\dot c_F&=-r_F,&\dot c_W&=r_F.
\end{aligned}
\label{eq:coupledBalances}
\end{equation}
Thus, $c_A+c_B$ and $c_F+c_W$ are conserved.

For this ideal four-species mixture, we choose the total free energy
\begin{equation}\label{eq:coupledFreeEnergy}
\mathcal F_{\rm T} =V\sum_{s=A,B,F,W}c_s \left[\mu_s^\circ+RT\left(\ln\frac{c_s}{c^\circ}-1\right)\right],
\end{equation}
where $c^\circ>0$ is a standard concentration and $\mu_s^\circ$ is the standard molar chemical potential of species $s$. The corresponding chemical potential is 
\begin{equation}\label{eq:ideal-mus}
\mu_s=\mu_s^\circ+RT\ln(c_s/c^\circ).    
\end{equation} 
We incorporate any fixed solvent or buffer contributions to the effective reaction free energies into the standard chemical potentials $\mu_s^\circ$. Using Eq.~\eqref{eq:coupledBalances}, we obtain the rate of free-energy change
\begin{equation}\label{eq:catalyst-dotF}
\dot{\mathcal F}_{\rm T}=-V(\mathcal A_0r_0+\mathcal A_Fr_F),
\end{equation}
where the channel affinities are identified as 
\begin{equation}\label{eq:catalyst-A0AF}
\mathcal A_0=\mu_A-\mu_B, \quad \mathcal A_F=\mu_A+\mu_F-\mu_B-\mu_W.    
\end{equation} 

\textbf{Dissipation and rate selection.} For the closed four-species system, use the nonquadratic dissipation potential $\varphi$ in Eq.~\eqref{eq:potential} to construct the total dissipation potential
\begin{equation}\label{eq:catalyst-Phi}
\Phi=V\sum_{\rho=0,F}\varphi(r_\rho;\Gamma_\rho), 
\end{equation}
where the positive kinetic scales $\Gamma_\rho(\boldsymbol c)$ may depend on the retained concentrations $\boldsymbol c=(c_A,c_B,c_F,c_W)$. Minimizing the Rayleighian $\mathcal R=\Phi+\dot{\mathcal F}_{\rm T}$
over the trial rates $r_0,r_F$ at fixed state and controls, including $\Gamma_\rho$, gives the unique selected rates 
\begin{equation}\label{eq:coupledAffinityRates}
r_{\rho,*}=\Gamma_\rho\sinh\frac{\mathcal A_\rho}{2RT}, 
\end{equation}
for $\rho\in\{0,F\}$. This nonlinear rate--affinity relation does not yet specify the concentration dependence of the kinetic scales, which remains a constitutive input. To recover the law of mass action, we choose
\begin{equation}\label{eq:catrates}
\Gamma_0=2\sqrt{k_0^+k_0^-c_Ac_B},\qquad
\Gamma_F=2\sqrt{k_F^+k_F^-c_Ac_Bc_Fc_W}.
\end{equation}
Here, the positive rate constants $k_0^\pm$ have units of inverse time, while $k_F^\pm$ have units of inverse concentration per unit time. They satisfy $RT\ln(k_\rho^+/k_\rho^-)=\mathcal A_\rho^\circ$,
where $\mathcal A_\rho^\circ$ is the channel affinity defined by the standard chemical potentials. Substitution into Eq.~\eqref{eq:coupledAffinityRates} yields
\begin{subequations}\label{eq:coupledRates} 
\begin{equation}\label{eq:coupledRatesa}
r_{0,*}=k_0^+c_A-k_0^-c_B\equiv r_0^+-r_0^-,
\end{equation}
\begin{equation}\label{eq:coupledRatesb}
r_{F,*}=k_F^+c_Ac_F-k_F^-c_Bc_W\equiv r_F^+-r_F^-.
\end{equation}
\end{subequations}     
Here, $r_\rho^+$ and $r_\rho^-$ denote the mean forward and reverse one-way reaction-rate densities, respectively, and satisfy $RT\ln(r_\rho^+/r_\rho^-)=\mathcal A_\rho$.
Their concentration dependence follows from the kinetic choice in Eq.~\eqref{eq:catrates}, not from thermodynamics or minimization alone. 
The selected rates determine the concentration dynamics through Eq.~\eqref{eq:coupledBalances}, with
$\dot{\mathcal F}_{\rm T}=-\mathcal D$ and
\begin{equation}\label{eq:ideal-D}
\mathcal D=V(\mathcal A_0r_{0,*}+\mathcal A_Fr_{F,*})\ge0.    
\end{equation} 

\textbf{Chemical input from maintained fuel.}
We now consider the same reaction channels with $F,W$ maintained by external reservoirs, retaining only $A,B$ as dynamical components.
Writing $\Delta\mu_F=\mu_F-\mu_W$, the retained free energy $\mathcal F$ is the $A,B$ part of the total free energy $\mathcal F_{\rm T}$ in Eq.~\eqref{eq:coupledFreeEnergy}, and
\begin{equation}
\dot{\mathcal F}=-V\mathcal A_0(r_0+r_F),
\qquad
\mathcal P_{\rm in}=V\Delta\mu_F r_F.
\label{eq:coupledPower}
\end{equation}
Since $\mathcal A_F=\mathcal A_0+\Delta\mu_F$, the same Rayleighian is $\mathcal R=\Phi+\dot{\mathcal F}-\mathcal P_{\rm in}$.
Along the selected evolution,
$\dot{\mathcal F}=\mathcal P_{\rm in}-\mathcal D$.
A single fueled channel can supply chemical power, but with that channel
alone, constant $\Delta\mu_F$, and no other driving, $\dot c_B=r_{F,*}$ implies
\begin{equation}
\frac{d}{dt}(\mathcal F -\Delta\mu_F\,c_BV)=-\mathcal D.
\label{eq:singlechannel}
\end{equation}
Such a model can relax under a tilted free energy but cannot sustain positive dissipation at a deterministic steady state.
With both channels, $r_{0,*}+r_{F,*}=0$ can coexist with $r_{F,*}\ne0$:
fueled activation followed by passive return consumes fuel without net
$A\to B$ conversion, with
$\mathcal P_{\rm in}=\mathcal D=V\Delta\mu_F\,r_{F,*}$. 

\subsection{Nonlinear response and fixed-state dissipation potentials}
\label{sub:coupledresponse}
 
The mass-action kinetics described above admit a fixed-state Rayleighian formulation but can exhibit \emph{nonreciprocal} net-affinity responses~\cite{Bataille1978}. We first explain why these results are compatible, then examine the partial-affinity response symmetries identified by Lengyel~\cite{Lengyel1988}.

\textbf{Conversion rates and conjugate net affinities.} The two physical observables are total net $A\to B$ conversion and net $F\to W$ fuel consumption. Their rates and conjugate affinities are 
\begin{subequations}\label{eq:conversionCoordinates}
\begin{equation}\label{eq:conversionRates}
r_1=r_0+r_F,\qquad r_2=r_F,
\end{equation}
\begin{equation}\label{eq:conversionAffinities}
\mathcal A_1=\mathcal A_0,\qquad
\mathcal A_2=\mathcal A_F-\mathcal A_0.
\end{equation}
\end{subequations}
Thus, $\mathcal A_1=\mu_A-\mu_B$ is the net conversion affinity, while $\mathcal A_2=\mu_F-\mu_W=\Delta\mu_F$ is the fuel affinity (the fuel--waste chemical-potential difference). The full fueled-channel affinity is $\mathcal A_F=\mathcal A_1+\mathcal A_2$.  This invertible transformation preserves the power-density pairing, $\mathcal A_0r_0+\mathcal A_Fr_F =\mathcal A_1r_1+\mathcal A_2r_2$, without introducing additional reaction channels or driving forces. Along the selected evolution, the dissipation power is
$\mathcal D=V(\mathcal A_0r_{0,*}+\mathcal A_Fr_{F,*})=V(\mathcal A_1r_{1,*}+\mathcal A_2r_{2,*})$.

\textbf{Net-affinity response.} A direct nonlinear extension of Onsager reciprocity asks whether
the Jacobian of selected net rates with respect to their conjugate net affinities is  symmetric~\cite{Bataille1978} [see Fig.~\ref{fig:overview}(a)]. Such symmetry is necessary for a scalar force potential to generate these rates under a specified control protocol. Since the two net affinities $\mathcal A_1$ and $\mathcal A_2$ do not uniquely determine the four chemical potentials, two complementary variables must also be held fixed. Following the type of response test considered by Bataille, Edelen, and Kestin~\cite{Bataille1978}, we fix the product and fuel chemical potentials $\mu_B,\mu_F$ and vary $\mu_A,\mu_W$ at fixed $T$, $V$, and kinetic constants. This independently varies the net affinities, $d\mathcal A_1=d\mu_A$ and $d\mathcal A_2=-d\mu_W$. For the ideal mixture, fixing $\mu_B,\mu_F$ is equivalent to fixing $c_B,c_F$. The chosen protocol illustrates the type of nonlinear response test considered by Bataille, Edelen, and Kestin using the present reaction network, rather than reproducing their original reaction example~\cite{Bataille1978}.

Under this control protocol, varying $\mathcal A_1$ changes $c_A$ at fixed $c_B$, whereas varying $\mathcal A_2$ changes $c_W$ at fixed $c_F$. Using the ideal chemical potentials and the mass-action rates in Eq.~\eqref{eq:coupledRates}, we obtain, for $i,j=1,2$, 
\begin{equation}
\left[\left.
\frac{\partial r_{i,*}}{\partial\mathcal A_j}
\right|_{c_B,c_F}\right]
=\frac1{RT}
\begin{pmatrix}
r_0^++r_F^+&r_F^-\\
r_F^+&r_F^-
\end{pmatrix}.
\label{eq:catresponse}
\end{equation}
In particular, the two cross responses are $\partial_{\mathcal A_2}r_{1,*}=r_F^-/(RT)$ and $\partial_{\mathcal A_1}r_{2,*}=r_F^+/(RT)$ under the specified controls. Their difference is $-r_{F,*}/(RT)$, so reciprocity generally fails away from equilibrium when $r_{F,*}\ne0$, even though both channels satisfy LDB. At equilibrium, $r_F^+=r_F^-$ and reciprocal linear response is recovered. 
% The result depends on the control protocol: fixing $\mu_B,\mu_W$ instead gives both cross derivatives equal to $r_F^+/(RT)$. Thus, nonlinear differential reciprocity depends on the specified controls, not solely on the reaction network.  

\textbf{Compatibility with fixed-state OVP.} The response in Eq.~\eqref{eq:catresponse} changes concentrations and hence kinetic scales $\Gamma_\rho$. OVP instead holds the state and controls fixed while varying trial rates. The driving term is then linear in those rates, so
\begin{equation}
\frac{\partial^2\mathcal R}{\partial r_\rho\partial r_\eta}
=\frac{\partial^2\Phi}{\partial r_\rho\partial r_\eta}
=\frac{2RTV}{\sqrt{r_\rho^2+\Gamma_\rho^2}}\,
\delta_{\rho\eta},
\label{eq:cathessian}
\end{equation}
where $\rho,\eta\in\{0,F\}$ and $\delta_{\rho\eta}$ is the Kronecker symbol. In conversion coordinates, the same potential is $\Phi=V[\varphi(r_1-r_2;\Gamma_0)+\varphi(r_2;\Gamma_F)]$;
its rate Hessian remains symmetric and positive definite. 

In the conversion-rate coordinates of
Eq.~\eqref{eq:conversionCoordinates}, $\Psi$ denotes the convex conjugate of the transformed $\Phi$, with conjugate forces $V(\mathcal A_1,\mathcal A_2)$.
% The dual potential $\Psi(V\boldsymbol{\mathcal A};\boldsymbol c)$
% also exists at each fixed state, as defined in
% Eq.~\eqref{eq:dualDissipation}.
Since the physical affinities
$\boldsymbol{\mathcal A}(\boldsymbol c)=(\mathcal A_1(\boldsymbol c),\mathcal A_2(\boldsymbol c))$ are determined by the
concentrations, the corresponding Rayleighian minimum satisfies
$\mathcal R_{\min}(\boldsymbol c)
=-\Psi(V\boldsymbol{\mathcal A}(\boldsymbol c);\boldsymbol c)$.
This pointwise convex-duality relation does not require varying affinities independently of the state. In contrast, the response in Eq.~\eqref{eq:catresponse} compares different compositions, where both the affinities and kinetic scales $\Gamma_\rho(\boldsymbol c)$ change. Its nonsymmetric Jacobian therefore does not contradict the existence of the fixed-state dissipation potential $\Phi$ or its convex conjugate $\Psi$.

\textbf{Partial-affinity response symmetries.} Nonreciprocity of the net-affinity response does not exclude other differential symmetries of mass-action
kinetics. Lengyel obtained two separate symmetric response matrices by differentiating the selected rates with respect to the reactant- and product-side contributions to the affinities~\cite{Lengyel1988}. For the conversion coordinates in Eq.~\eqref{eq:conversionCoordinates}, we define 
\begin{subequations}\label{eq:coupledPartialForces}
\begin{equation}\label{eq:coupledPartialForcesa}
X_1'=\mu_A/T,\qquad X_1''=-\mu_B/T,
\end{equation}
\begin{equation}\label{eq:coupledPartialForcesb}
X_2'=\mu_F/T,\qquad X_2''=-\mu_W/T.
\end{equation}
\end{subequations}
Then, $\mathcal A_i=T(X_i'+X_i'')$ for $i=1,2$. The fueled channel contains the reactant-side combination $X_1'+X_2'$ and the product-side combination $X_1''+X_2''$. These partial quantities separate chemical-potential contributions; only their sums give the net affinities. 

Each response derivative below measures the change of a selected rate when one chemical potential is varied, with the other three chemical potentials, $T$, $V$, and the kinetic constants fixed.
These derivatives compare selected rates at different controlled compositions, whereas Rayleighian minimization varies trial rates at a fixed composition. The constants $k_\rho^\pm$ remain fixed, but the concentration-dependent $\Gamma_\rho$ can change. Using $\partial c_s/\partial(\mu_s/T)=c_s/R$ in Eq.~\eqref{eq:coupledRates} gives, for $i,j=1,2$, 
\begin{subequations}\label{eq:coupledPartialResponses}
\begin{equation}\label{eq:coupledPartialResponsesa}
\left[\frac{\partial r_{i,*}}{\partial X_j'}\right]=\frac1R
\begin{pmatrix}
r_0^++r_F^+&r_F^+\\
r_F^+&r_F^+
\end{pmatrix},
\end{equation}
\begin{equation}\label{eq:coupledPartialResponsesb}
\left[\frac{\partial r_{i,*}}{\partial X_j''}\right]=\frac1R
\begin{pmatrix}
r_0^-+r_F^-&r_F^-\\
r_F^-&r_F^-
\end{pmatrix}.
\end{equation}
\end{subequations}
For example, increasing $X_2'=\mu_F/T$ changes only $r_F^+$, giving $\partial_{X_2'}r_{1,*}=r_F^+/R=\partial_{X_1'}r_{2,*}$. Both blocks are symmetric because the same fueled event contributes to both conversions. They have Lengyel's composite-rate matrix structure; here the rate transformation is invertible and retains both channels.

The two symmetric response matrices in Eq.~\eqref{eq:coupledPartialResponses} are generally different because they involve different one-way rates.
Under the control protocol of Eq.~\eqref{eq:catresponse}, varying $\mathcal A_1$
changes $\mu_A$, whereas varying $\mathcal A_2$ changes $-\mu_W$. 
The resulting net-affinity Jacobian therefore combines the first column of the forward partial-affinity response matrix with the second column of the reverse matrix, each divided by $T$, and need not be symmetric. By contrast, fixing $\mu_B,\mu_W$ and varying $\mu_A,\mu_F$ selects both columns from the forward
matrix, giving equal cross responses $r_F^+/(RT)$. Thus, both types of response compare selected rates at different controlled compositions, but use different
force variables and control conditions. The partial-affinity symmetries therefore do not imply net-affinity reciprocity under arbitrary controls. Appendix~\ref{app:reciprocity} gives the corresponding higher-order response and fluctuation constraints. We next examine whether the selected deterministic
rates also determine their fluctuation statistics. 

\begin{figure}[t]
\centering
\includegraphics[width=\columnwidth]{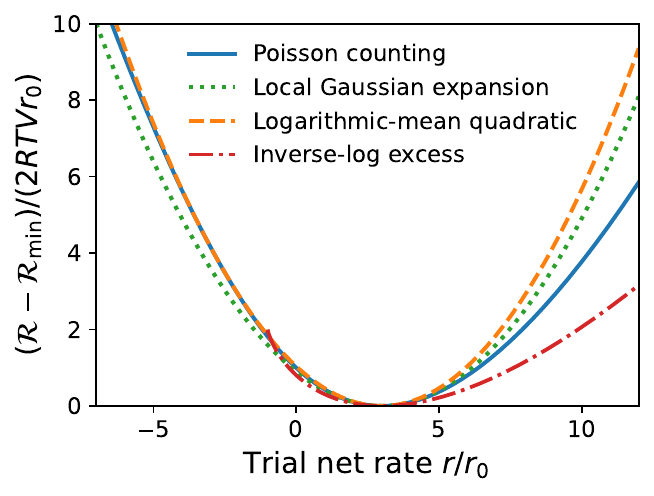}
\caption{\textbf{One selected rate and four Rayleighian representations.}
The axes are $r/r_0$ and $\Delta \mathcal R /(2RTVr_0)$, where $r_0>0$ is a reference molar rate density, distinct from the passive-channel rate; $\Delta \mathcal R=\mathcal R-\mathcal R_{\min}$.
For $r^+/r_0=4$ and $r^-/r_0=1$, the common minimizing rate is $r_*/r_0=3$, with $\Gamma/r_0=4$, $\sigma/r_0=5$, and $\mathcal A/(RT)=\ln4$.
The solid, dotted, and dashed curves equal $\ell_P/r_0$, $\ell_{\rm loc}/r_0$, and $\ell_G/r_0$ in Eqs.~\eqref{eq:PoissonCostComparison}, \eqref{eq:localPoissonCost},
and \eqref{eq:logmeanCost}, respectively. They represent the Poisson model, its local Gaussian approximation, and a separately specified logarithmic-mean Gaussian model.
The dash-dotted curve shows $\Delta\mathcal R_{\mathrm I}/(2RTVr_0)$ from Eq.~\eqref{eq:inverseLogExcess}; no fluctuation interpretation is assigned to it here.
It is shown only for $r>-r^-=-r_0$. The local Gaussian approximation is valid only near $r_*$; its continuation is shown for comparison.}
\label{fig:cost}
\end{figure}

\subsection{The selected rate does not determine its fluctuations}
\label{sub:ratefluct}

The preceding response tests compare deterministic rates at different chemical potentials. We now fix the state and controls and compare variational and statistical descriptions with the same selected net rate. For either resolved channel, suppress the index $\rho$ and use the prescribed mean forward and reverse rate densities $r^\pm>0$ as reference parameters, with
$r_*=r^+-r^-$, $\Gamma=2\sqrt{r^+r^-}$, $\sigma=r^++r^-$, and
$\mathcal A=RT\ln(r^+/r^-)$. These quantities remain fixed when varying the trial net rate $r$. 

For each representation, write the dissipation potential as
$\Phi=V\varphi$ and the Rayleighian as 
$\mathcal R=V[\varphi(r)-\mathcal A r]$. If its minimizing rate is $r_*$, the Rayleighian excess, defined as the difference from its fixed-state minimum, is
\begin{equation*}
\Delta\mathcal R\equiv\mathcal R-\mathcal R_{\min}
=V[\varphi(r)-\varphi(r_*)-\mathcal A(r-r_*)].
\end{equation*}
A fluctuation cost $\ell(r)$ instead specifies the leading probability weight $\exp[-N_{\mathrm A}V\Delta t\,\ell(r)]$ for observing a net rate $r$ over an interval $\Delta t$.
The equality $\ell=\Delta\mathcal R/(2RTV)$ must be established for the specified stochastic model; it does not follow from a shared minimizing rate. We compare a Poisson model, its local Gaussian approximation, a separately specified Gaussian model, and an inverse-log deterministic representation.

\textbf{Poisson event statistics.}
Independent forward and reverse Poisson counts $\Delta N^\pm$ have
means $N_{\mathrm A}Vr^\pm\Delta t$, which should be
distinguished from the measured net rate,  
$\widehat r=(\Delta N^+-\Delta N^-)/(N_{\mathrm A}V\Delta t)$.
The matched dissipation potential density in Eq.~\eqref{eq:potential} is
\begin{equation*}
\varphi_P(r)=2RT\left[
r\operatorname{arsinh}\left(\frac r\Gamma\right)
-\sqrt{r^2+\Gamma^2}+\Gamma\right].
\end{equation*}
In the frozen-state, many-event regime, the net-rate fluctuation cost satisfies $\ell_P(r)=\Delta\mathcal R_P/(2RTV)$, with 
\begin{equation}\label{eq:PoissonCostComparison} 
\ell_P(r)=r\operatorname{arsinh}\left(\frac r\Gamma\right)-\frac{\mathcal A r}{2RT}-\sqrt{r^2+\Gamma^2}+\sigma.
\end{equation} 
This cost is obtained by minimizing the joint forward and reverse
counting cost at fixed net rate $r$; subleading probability prefactors are omitted.

\textbf{Local Gaussian approximation.}
Expanding Eq.~\eqref{eq:PoissonCostComparison} about $r_*$ gives
$\ell_P=\ell_{\rm loc}+O((r-r_*)^3/\sigma^2)$.
The corresponding second-order Taylor approximation to the
dissipation potential density is
\begin{equation*}
\varphi_{\rm loc}(r)=\varphi_P(r_*)+\mathcal A(r-r_*)
+\frac{RT}{\sigma}(r-r_*)^2.
\end{equation*}
Its Rayleighian excess satisfies
\begin{equation}\label{eq:localPoissonCost}
\frac{\Delta\mathcal R_{\rm loc}}{2RTV}
=\ell_{\rm loc}(r)=\frac{(r-r_*)^2}{2\sigma}.
\end{equation}
The linear term ensures $\partial_r\varphi_{\rm loc}(r_*)=\mathcal A$, so minimization still selects $r_*$.  
This is a local Taylor approximation near $r_*$, not an independently specified physical dissipation potential or reaction-event model.

\textbf{Logarithmic-mean quadratic representation.}
The alternative dissipation potential density
$\varphi_{\log}(r)=r^2/(2L_{\log})$ reproduces the same deterministic rate~\cite{Mielke2011,Kobayashi2022Hessian}, with the logarithmic-mean mobility~\cite{Mielke2011,Arroyo2018}
\begin{subequations}\label{eq:logmean}
\begin{equation}\label{eq:logmeanMobility}
L_{\log}=\frac{r^+-r^-}{RT\ln(r^+/r^-)}
=\frac{\Gamma}{2RT}
\frac{\sinh[\mathcal A/(2RT)]}{\mathcal A/(2RT)}.
\end{equation}
The mean rates $r^\pm=(\Gamma/2)e^{\pm\mathcal A/(2RT)}$ determine
$L_{\log}$, which is held fixed during variation of $r$.
Its continuous value at $\mathcal A=0$ is $\Gamma/(2RT)$.
Minimizing $\mathcal R_{\log}$ gives $r=L_{\log}\mathcal A=r_*$.
Assigning the normalized Rayleighian excess as a fluctuation cost
specifies a separate Gaussian model:
\begin{equation}\label{eq:logmeanCost}
\frac{\Delta\mathcal R_{\log}}{2RTV}
=\ell_G(r)=\frac{(r-r_*)^2}{4RTL_{\log}}.
\end{equation}
\end{subequations}
This model differs from the local Gaussian approximation
$\ell_{\rm loc}$ to the Poisson cost.
The same quadratic function also provides the upper bound
$\ell_P(r)\leq\ell_G(r)$~\cite{Gingrich2016}.

\textbf{Inverse-log deterministic representation.}
In our notation, the logarithmic dissipative force used in energetic variational chemical kinetics~\cite{WangLiu2020,WangLiu2022Review} is $RT\ln(1+r/r^-)$.
Integrating this force with respect to $r$ at fixed $r^-$ gives the dissipation potential density
\begin{subequations}\label{eq:inverseLogComparison}
\begin{equation}\label{eq:inverseLogPotential}
\varphi_{\mathrm I}(r)
=RT\left[(r+r^-)\ln\left(1+\frac r{r^-}\right)-r\right],
\end{equation}
defined for $r>-r^-$. The product $r\,\partial_r\varphi_{\mathrm I}=RT\,r\ln(1+r/r^-)$ is the dissipation power density, not
$\varphi_{\mathrm I}$ itself.
The condition $\partial_r\varphi_{\mathrm I}=\mathcal A$ again
selects $r_*=r^+-r^-$, while the Rayleighian excess is
\begin{equation}\label{eq:inverseLogExcess}
\frac{\Delta\mathcal R_{\mathrm I}}{2RTV}
=\frac12\left[(r+r^-)\ln\frac{r+r^-}{r^+}-(r-r_*)\right].
\end{equation}
\end{subequations}
This normalized Rayleighian excess differs from $\ell_P(r)$ for
the prescribed two-way Poisson events. 
The deterministic construction does not specify a stochastic law,
so identifying this function as a fluctuation cost would require
an additional stochastic model.

All four Rayleighians have the same minimizing rate $r_*$, where
their excesses $\Delta\mathcal R$ vanish and the selected dissipation power is $\mathcal D=V\mathcal A r_*$.
The curvatures of $\Delta\mathcal R/(2RTV)$ at $r_*$ are
$1/\sigma$, $1/\sigma$, $1/(2RTL_{\log})$, and $1/(2r^+)$ for the
Poisson, local Gaussian, logarithmic-mean, and inverse-log
representations, respectively. 
% Figure~\ref{fig:cost} therefore compares a statistically matched Rayleighian excess, its local approximation, a separately specified Gaussian model, and a deterministic representation without an assigned fluctuation law.
Figure~\ref{fig:cost} compares these differences in $\Delta\mathcal R(r)$. A common minimizing rate does not make the
four representations statistically equivalent.
We next couple the passive and fueled reaction channels to spatial
transport and nonideal interactions in Sec.~\ref{sec:mixture}.
Section~\ref{sec:duality} then examines how membrane--cytosol dynamics and chemical input power depend on state and channel resolution.

\section{Active emulsions: phase separation and fuel conversion}
\label{sec:mixture}

We now consider the passive and fueled conversions of Sec.~\ref{sec:catalyst} in a spatially varying nonideal fluid mixture. Unlike the ideal mixture in that section, the nonideal free energy can be nonconvex, although equilibrium phase coexistence also depends on which component amounts are conserved.
We examine how the same free energy determines the driving forces for diffusion and reaction, and how the free-energy balance changes when externally maintained fuel and waste are replaced by a finite fuel pool~\cite{Weber2019,Bauermann2022,Avanzini2021,Avanzini2024}.

\subsection{Nonideal thermodynamics and material balances}
\label{sub:mixture_thermodynamics}

\textbf{Reaction channels, fields, and constraints.} Consider a nonadvecting incompressible mixture with two reversible reaction channels, 
\begin{equation}\label{eq:fuelnetwork}
A\rightleftharpoons B,\qquad A+F\rightleftharpoons B+W.
\end{equation}
The chemical system contains five species: interconverting components $A,B$, a nonreacting solvent $C$, and fuel and waste $F,W$. The subscripts $0,F$ distinguish passive and fueled conversion, with trial net molar rate densities $r_0,r_F$ positive for $A\to B$. 
We retain the composition fields of $A,B,C$, while the activities of $F,W$ are initially prescribed by external reservoirs. Neglecting the volume contributions of fuel and waste, we describe the retained fluid as an incompressible ternary mixture. For equal, constant molecular volumes of $A,B,C$, their molar concentrations satisfy $c_i=c_0\phi_i$, where $\phi_i>0$ are volume fractions and $c_0=c_A+c_B+c_C$ is constant. The volume-filling constraint $\phi_C=1-\phi_A-\phi_B$ leaves two independent composition fields. On the fixed domain $\Omega$, the material balances are
\begin{subequations}\label{eq:mixturekin}
\begin{equation}\label{eq:mixturekina}
c_0\partial_t\phi_A=-\nabla\cdot\boldsymbol J_A-r_0-r_F,
\end{equation}
\begin{equation}\label{eq:mixturekinb}
c_0\partial_t\phi_B=-\nabla\cdot\boldsymbol J_B+r_0+r_F.
\end{equation}
\end{subequations}
The independent molar diffusive fluxes are $\boldsymbol J_A,\boldsymbol J_B$, with $\boldsymbol J_C=-\boldsymbol J_A-\boldsymbol J_B$. Periodic boundaries, or impermeable walls with $\hat{\boldsymbol n}\cdot\boldsymbol J_i=\hat{\boldsymbol n}\cdot\nabla\phi_i=0$ for $i=A,B$ and outward unit normal $\hat{\boldsymbol n}$, remove boundary power terms and conserve the total
amount of $A+B$.

\textbf{Free energy and its rate.} Choose the nonideal mixture free energy
\begin{equation}\label{eq:mixF}
\mathcal F[\phi_A,\phi_B]=c_0\int_\Omega\left[f+\frac{K_A}{2}|\nabla\phi_A|^2
+\frac{K_B}{2}|\nabla\phi_B|^2\right]dV.
\end{equation}
Here, the homogeneous molar free energy $f(\phi_A,\phi_B,\phi_C)=RT\sum_{s=A,B,C}\phi_s\ln\phi_s+
\chi\phi_B(\phi_A+\phi_C)+h_B\phi_B$ has the Flory--Huggins form for a ternary mixture with $\phi_C=1-\phi_A-\phi_B$. The interaction term corresponds to $\chi_{AB}=\chi_{BC}=\chi>0$ and $\chi_{AC}=0$: mixing $B$ with either $A$ or $C$ is unfavorable, whereas no nonideal $A$--$C$ mixing term is included. The coefficient $h_B$ specifies the standard molar free-energy difference
between $B$ and $A$; terms proportional to the conserved amounts of $A+B$ and $C$ are omitted. Both $\chi$ and $h_B$ have molar energy units, while $K_A,K_B>0$ have units of molar energy times length squared. For $\chi=0$, the bulk free energy has the ideal-mixing structure of Eq.~\eqref{eq:coupledFreeEnergy}. With $c_i=c_0\phi_i$ and consistent standard states, the volume-fraction and concentration forms differ only by composition-independent terms for the same retained species at fixed $c_0,V,T$. The standard-state bias $h_B$ remains; the gradient terms account separately for spatial inhomogeneity.

For the homogeneous mixture, use pure-component standard states at the same temperature and pressure. In the chosen free-energy convention, $\mu_A^\circ=\mu_C^\circ=0$ and $\mu_B^\circ=h_B$.
The component molar chemical potentials are
\begin{equation}\label{eq:mu-nonideal}
\mu_s=\mu_s^\circ+RT\ln a_s,
\end{equation}
where $a_s=\gamma_s\phi_s$ are dimensionless activities for $s=A,B,C$, with
\begin{equation}
\gamma_A=\gamma_C=\exp\!\left(\frac{\chi\phi_B^2}{RT}\right),\quad
\gamma_B=\exp\!\left(\frac{\chi(1-\phi_B)^2}{RT}\right).
\label{eq:mixActivities}
\end{equation}
The activity coefficients follow from the partial molar derivatives of $f$; equal molecular volumes identify volume fractions with mole fractions. The common pressure contribution is included in the standard-state convention.

Including gradient contributions gives component potentials $\hat\mu_s$, defined up to a common incompressibility term. For admissible variations, $\delta\mathcal F=c_0\int_\Omega\sum_{s=A,B,C}\hat\mu_s\delta\phi_s\,dV$. Using $\delta\phi_C=-\delta\phi_A-\delta\phi_B$ gives $\delta\mathcal F=c_0\int_\Omega\sum_{i=A,B}\bar\mu_i\delta\phi_i\,dV$, where
\begin{subequations}\label{eq:barmuAB}
\begin{equation}\label{eq:barmuA}
\bar\mu_A\equiv\hat\mu_A-\hat\mu_C
=RT\ln(a_A/a_C)-K_A\nabla^2\phi_A,
\end{equation}
\begin{equation}\label{eq:barmuB}
\bar\mu_B\equiv\hat\mu_B-\hat\mu_C
=h_B+RT\ln(a_B/a_C)-K_B\nabla^2\phi_B.
\end{equation}
\end{subequations}
These exchange potentials enter both diffusion and the reaction affinities. Substituting Eq.~\eqref{eq:mixturekin} and integrating by parts gives
\begin{equation}
\dot{\mathcal F}=\int_\Omega\left[
\sum_{i=A,B}\nabla\bar\mu_i\cdot\boldsymbol J_i
-(\bar\mu_A-\bar\mu_B)(r_0+r_F)\right]dV.
\label{eq:mixFdot}
\end{equation}
The ideal-mixing contribution is convex, whereas $\chi>2RT$
permits interaction-driven demixing over part of the composition space.
If both channels of chemical reactions are disabled, the amount of $B$ is separately
conserved, so the linear term $c_0 h_B\int_\Omega\phi_B\,dV$ is constant and does not affect diffusive equilibration. With interconversion allowed, $h_B$ instead contributes to the chemical driving.

\subsection{Variational dynamics and chemical maintenance}\label{sub:mixture_variational}

\textbf{Chemical driving and rate selection.} The chemical potentials $\mu_s^{\rm res}$ of fuel and waste also obey Eq.~\eqref{eq:mu-nonideal}, but their activities $a_s$ are reservoir inputs rather than consequences of $\mathcal F[\phi_A,\phi_B]$ with $s=F,W$. For the ideal fuel-pool model used here, take $a_s=c_s^R/c_R^\circ$, $s=F,W$, with a common concentration scale $c_R^\circ$ and equal standard chemical potentials $\mu_s^\circ$. Thus, $\Delta\mu_F\equiv\mu_F^{\rm res}-\mu_W^{\rm res}=RT\ln(a_F/a_W)$; initially $a_F,a_W$ are uniform and maintained constant. Each mole of forward fueled conversion consumes one mole of $F$ and produces one mole of $W$. The spatial form of Eqs.~\eqref{eq:reservoirwork} and \eqref{eq:coupledPower} is therefore
\begin{equation}\label{eq:fuelpower}
\mathcal P_{\rm in}=\int_\Omega\Delta\mu_F r_F\,dV,
\end{equation} 
\begin{equation}\label{eq:fuelpower2}
\dot{\mathcal F}-\mathcal P_{\rm in}
=\int_\Omega\left[\sum_{i=A,B}\nabla\bar\mu_i\cdot\boldsymbol J_i
-\mathcal A_0r_0-\mathcal A_Fr_F\right]dV,
\end{equation}
where $\mathcal A_0=\bar\mu_A-\bar\mu_B$ and $\mathcal A_F=\bar\mu_A-\bar\mu_B+\Delta\mu_F$.
Combining transport resistances with the reaction dissipation potential $\varphi$ of Eq.~\eqref{eq:potential} gives the total
dissipation potential $\Phi$:  
\begin{equation}
\Phi=\int_\Omega\left[
\sum_{i=A,B}\frac{|\boldsymbol J_i|^2}{2M_i}
+\sum_{\rho=0,F}\varphi(r_\rho;\Gamma_\rho)\right]dV.
\label{eq:mixtureDissipation}
\end{equation}
The $M_i>0$ are prescribed constant transport mobilities on the independent composition directions. The positive $\Gamma_\rho$ may depend on the instantaneous composition fields and fuel controls; their values are fixed during trial-rate variation. Minimizing the Rayleighian $\mathcal R=\Phi+\dot{\mathcal F}-\mathcal P_{\rm in}$ gives \begin{subequations}\label{eq:mixtureAffinityRates}
\begin{equation}\label{eq:mixtureAffinityRatesa}
\boldsymbol J_{i,*}=-M_i\nabla\bar\mu_i,
\end{equation}
\begin{equation}\label{eq:mixtureAffinityRatesb}
r_{\rho,*}=\Gamma_\rho\sinh\frac{\mathcal A_\rho}{2RT},
\end{equation}
\end{subequations}
for $i=A,B$ and $\rho=0,F$. Equation~\eqref{eq:mixtureAffinityRatesb} gives the nonlinear rate--affinity relation at fixed $\Gamma_\rho$; their state dependence remains a constitutive input.

\textbf{Kinetic specialization and dissipation.} To obtain the activity-based reaction laws used here, we choose 
\begin{subequations}\label{eq:fuelrates}
\begin{equation}\label{eq:fuelratesa}
\Gamma_0=2k_0e^{(\bar\mu_A+\bar\mu_B)/(2RT)},
\end{equation}
\begin{equation}\label{eq:fuelratesb}
\Gamma_F=2\kappa_F\sqrt{a_Fa_W}\,e^{(\bar\mu_A+\bar\mu_B)/(2RT)}.
\end{equation}
\end{subequations}
Here, $k_0,\kappa_F>0$ are rate-density prefactors that may depend on composition. Since $\sqrt{a_Fa_W}e^{\Delta\mu_F/(2RT)}=a_F$ and $\sqrt{a_Fa_W}e^{-\Delta\mu_F/(2RT)}=a_W$, substitution into Eq.~\eqref{eq:mixtureAffinityRatesb} gives
\begin{subequations}\label{eq:selectedmixture}
\begin{equation}\label{eq:selectedmixturea}
r_{0,*}=k_0(e^{\bar\mu_A/(RT)}-e^{\bar\mu_B/(RT)}),
\end{equation}
\begin{equation}\label{eq:selectedmixtureb}
r_{F,*}=\kappa_F(a_Fe^{\bar\mu_A/(RT)}-a_We^{\bar\mu_B/(RT)}).
\end{equation}
\end{subequations}
These laws follow from the chosen $\Gamma_{0,F}$, not from thermodynamics alone. The identity $e^{\bar\mu_i/(RT)}=e^{-\hat\mu_C/(RT)}e^{\hat\mu_i/(RT)}$ shows that changing from component to exchange chemical potentials rescales a common kinetic prefactor. This change of representation does not fix $k_0$ or $\kappa_F$; their composition dependence models kinetic effects such as catalysis or activation barriers~\cite{Bauermann2022,Bazant2013}. In a homogeneous ideal mixture at fixed solvent concentration, consistent standard states and prefactors recover the concentration dependence of Eq.~\eqref{eq:coupledRates}. Nonideal activities and gradient contributions must otherwise remain in the chemical potentials; idealization alone does not fix the same kinetic coefficients.

Equations~\eqref{eq:mixtureAffinityRatesa} and \eqref{eq:selectedmixture}, inserted into Eq.~\eqref{eq:mixturekin}, determine the coupled dynamics. Along the selected evolution, the free-energy balance $\dot{\mathcal F}=\mathcal P_{\rm in}-\mathcal D$ of Eq.~\eqref{eq:Frate} holds with 
\begin{equation}\label{eq:mixD}
\mathcal D=\int_\Omega\left[\sum_{i=A,B}M_i|\nabla\bar\mu_i|^2
+\mathcal A_0r_{0,*}+\mathcal A_Fr_{F,*}\right]dV\ge0.
\end{equation}
For this resolved macroscopic model, $\mathcal D=T\dot S_{\rm prod}^{\rm mac}$. It is not generally a heat current: the entropy balance also includes system entropy changes and entropy carried by exchanged matter. At a single-bath steady state with no net entropy transport by matter, the two powers coincide. Heat exchange in the present chemically open mixture requires this additional thermodynamic information~\cite{Seifert2011,RaoEsposito2016}.
The deterministic balance alone does not specify fluctuations. 
At a fixed spatial resolution, independent thermal transport noise with covariance amplitude $2k_{\mathrm B}T M_i$ and resolved Poisson reaction events give additive local fluctuation costs, extending the
excess relation in Eq.~\eqref{eq:GyarmatiReactionCostb} to coupled reaction--transport dynamics in the stated frozen-state, many-event limit~\cite{Cates2022}. 

\textbf{Chemical maintenance and spatial structure.}
The selected kinetics distinguish stationary composition from zero fuel consumption. First consider a homogeneous stationary state at
maintained $a_F>a_W$. With no spatial gradients,
Eq.~\eqref{eq:mixturekinb} requires $r_{0,*}+r_{F,*}=0$, while
Eq.~\eqref{eq:selectedmixture} gives $r_{F,*}=e^{\bar\mu_A/(RT)}
{k_0\kappa_F(a_F-a_W)}/({k_0+\kappa_Fa_W})>0$ and $r_{0,*}=-r_{F,*}<0$. 
Thus, fueled conversion and passive return consume fuel without
changing composition, as in Sec.~\ref{sec:catalyst}.
A nonuniform stationary profile instead obeys
$r_{0,*}+r_{F,*}=\nabla\cdot\boldsymbol J_{B,*}$:
local net conversion need not vanish.
With the stated boundaries,
$\int_\Omega(r_{0,*}+r_{F,*})\,dV=0$, while
$\mathcal P_{\rm in}=\Delta\mu_F\int_\Omega r_{F,*}\,dV
=\mathcal D>0$~\cite{Polettini2014,RaoEsposito2016}.
The input power follows from the selected rates rather than being prescribed independently. 

\subsection{Finite-fuel relaxation and domain dissolution}\label{sub:mixture_finitefuel}

\begin{figure*}[t]
\centering
% \includegraphics[width=.49\textwidth]{figures/fig4a_fuel.pdf}\hfill
%%{figures/figV_length.pdf}\hfill
% \includegraphics[width=.49\textwidth]{figures/fig4b_energy.pdf}
%%{figures/figV_energy.pdf}\\[1ex]
% \includegraphics[width=.64\textwidth]{figures/figV_profiles_log.pdf}
\includegraphics[width=\textwidth]{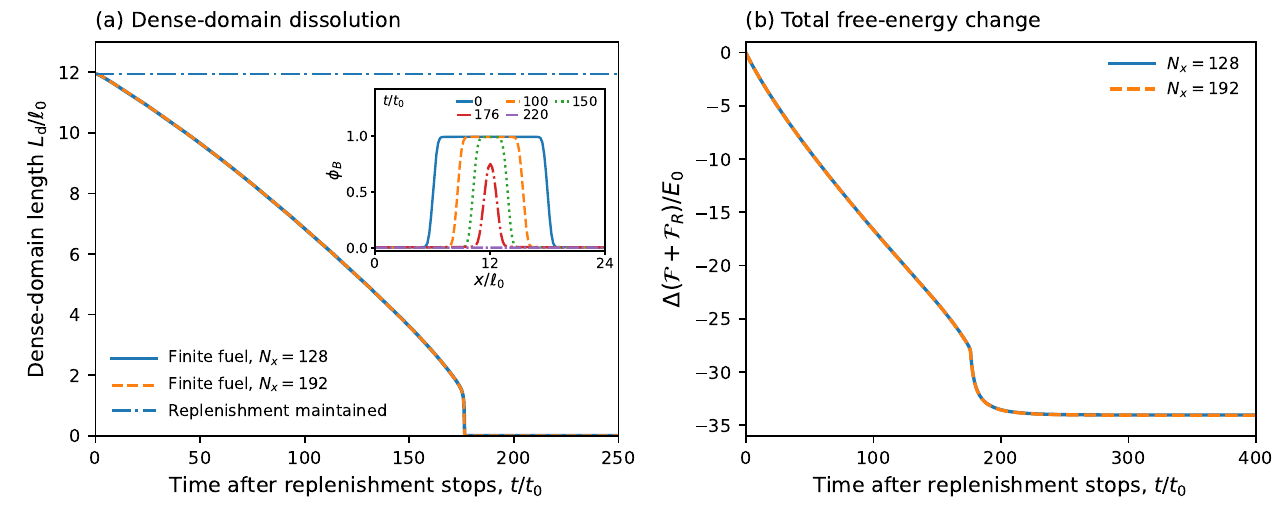}
\caption{\textbf{Dissolution of a chemically maintained dense domain
after external replenishment stops.}
(a) Threshold length
$L_{\rm d}=\int_0^L\mathbf1_{\{\phi_B>0.5\}}\,dx$,
compared with a maintained-activity control.
Inset: $B$ profiles on 128 cells at
$t/t_0=0,100,150,176,220$.
Loss of the threshold region does not imply extinction of $B$.
(b) Change in the combined mixture--pool free energy,
$\Delta(\mathcal F+\mathcal F_R)/E_0$, relative to $t=0$.
The deterministic calculations use $N_x=128,192$ spatial grid cells.
Parameters are $L/\ell_0=24$,
$K_A/(RT\ell_0^2)=K_B/(RT\ell_0^2)=1$,
$M_ARTt_0/(c_0\ell_0^2)=M_BRTt_0/(c_0\ell_0^2)=0.15$,
$\chi/(RT)=5$, $h_B/(RT)=1$, $Kt_0/c_0=0.002$, and
$V_Rc_R^\circ/(c_0S_\perp\ell_0)=0.02$.
The kinetic prefactors are
$k_0=K(e^6-z)/(e^6-1)$ and
$\kappa_F=K(z-1)/(e^6-1)$, with
$z=\exp\{3[1+\tanh((\psi-0.25)/0.03)]\}$ and
$\psi=(\phi_A+\phi_B)/2$. The switching form of $\ln z$ is adapted from Haugerud \emph{et al.}~\cite{Haugerud2026}.
The factor $e^6$ in these functions remains fixed as $a_F/a_W$ evolves.
Initial activities are $a_F=e^6$, $a_W=1$, and the conserved
spatial mean is $\langle\phi_A+\phi_B\rangle=0.51$.}
\label{fig:finitefuel}
\end{figure*}

\textbf{Preparation of a chemically maintained domain.}
Figure~\ref{fig:finitefuel} considers a periodic one-dimensional slab
of length $L$ and cross-sectional area $S_\perp$.
For the prefactors specified in the caption,
$K=k_0+\kappa_F>0$ is constant, while $\kappa_F/K$ increases with
local solute content $\phi_A+\phi_B$. 
At the maintained preparation activities, this allocation realizes the smooth composition-dependent reaction bias of
Haugerud \emph{et al.}~\cite{Haugerud2026}
through resolved passive and fueled channels. The phase selectivity is a phenomenological constitutive choice, not a prediction of OVP or LDB.
We use $\ell_0=(K_A/RT)^{1/2}$, $Kt_0/c_0=0.002$, and
$E_0=c_0RTS_\perp\ell_0$.
A smooth seed with $\phi_B\simeq0.94$ in the central region and
$\phi_B\simeq0.025$ outside is evolved for $8000t_0$ at maintained
$a_F=e^6$, $a_W=1$, with $\langle\phi_A+\phi_B\rangle=0.51$;
brackets denote spatial averages.
The resulting driven stationary profile defines $t=0$ and has
$\langle\phi_B\rangle\simeq0.498$.
This preparation selects a nonuniform steady state; it does not
test nucleation or convergence from arbitrary initially mixed states.

In this prepared state, fueled $A\to B$ conversion in the dense
region and passive $B\to A$ return in the dilute region are coupled
by inward $A$ and outward $B$ transport
[Fig.~\ref{fig:overview}(c)].
Here, interactions favor demixing, while fuel maintains the selected
$B$-rich composition and reaction--transport cycle.

\textbf{Release to a finite fuel pool.}
At $t=0$, external fuel replenishment and waste removal cease,
while both reactions remain active and a fuel/waste pool of fixed
volume $V_R$ remains coupled to the mixture.
The fields and activities are continuous across this switch;
interaction parameters and kinetic functions remain unchanged.
A control branch retains the original maintained activities.
Assume sufficiently fast mixing and fuel/waste exchange that uniform activities $a_F(t),a_W(t)$ describe the pool and its coupling to the mixture. For the ideal-pool convention already specified, the molar amounts are $n_s^R=V_Rc_R^\circ a_s$, $s=F,W$.
Fuel conversion gives $\dot a_W=-\dot a_F$ and
\begin{equation}
V_Rc_R^\circ\dot a_F=-\int_\Omega r_{F,*}\,dV.
\label{eq:finitefuel}
\end{equation}
Up to a conserved standard-state term, the pool free energy is
$\mathcal F_R=RTV_Rc_R^\circ
[a_F(\ln a_F-1)+a_W(\ln a_W-1)]$.
Using Eq.~\eqref{eq:finitefuel} gives
$\dot{\mathcal F}_R=-\mathcal P_{\rm in}$ and
$\dot{\mathcal F}+\dot{\mathcal F}_R=-\mathcal D$.
Thus, $\mathcal P_{\rm in}$ is now the rate of free-energy transfer
from the pool to the mixture, not external power supplied to their
combined system.  
Neither $a_F$ nor $\Delta\mu_F$ is set to zero at release;
the constant-bias argument of Eq.~\eqref{eq:singlechannel}
does not replace this balance when $\Delta\mu_F(t)$ evolves.

\textbf{Numerical implementation.}
Spatial derivatives were approximated by second-order central
finite differences on uniform periodic grids of 128 and 192 cells.
The resulting dimensionless ordinary differential equations were
integrated using the adaptive-step, variable-order backward
differentiation formula (BDF) solver in SciPy's \texttt{solve\_ivp}, with relative and absolute error tolerances of $2\times10^{-7}$ and $2\times10^{-9}$, respectively.
The discrete free energy was evaluated with a gradient operator
consistent with the discrete Laplacian in the evolution equations.
Threshold crossings defining $L_{\rm d}$ were located by linear
interpolation between adjacent grid points.
The initial profiles, code, and data are available as specified in the Data Availability statement. ChatGPT (GPT-6 Pro; OpenAI) assisted analytical consistency checks and numerical-code development; its uses, purposes, and the authors' verification procedures are detailed in the Declaration of Generative AI Use.

\textbf{Dissolution and transient fuel regeneration.}
Initially, net fuel consumption lowers $a_F/a_W$, moving the mixture
away from its maintained reaction--transport balance.
The $B$-rich domain gradually narrows before its remaining core
decays rapidly near $176t_0$ [Fig.~\ref{fig:finitefuel}(a)].
The threshold length $L_{\rm d}$ of the region with $\phi_B>0.5$
vanishes at $t/t_0\simeq176.4$; neither extinction of $B$ nor complete
homogenization has occurred.
During this terminal decay, composition and gradient contributions
raise $\bar\mu_B$ in the core.
At $t/t_0=176.5$, the computed affinity at the $B$-concentration
maximum is $\mathcal A_F/(RT)\simeq-0.57$ on the 128-cell grid,
despite $\Delta\mu_F>0$.
Reverse fueled conversion and transport contribute to the peak decay.

The steeper decrease of the combined mixture--pool free energy
[Fig.~\ref{fig:finitefuel}(b)] reflects increased total dissipation
through $\dot{\mathcal F}+\dot{\mathcal F}_R=-\mathcal D$.
The integrated fueled rate briefly reverses, giving
$\mathcal P_{\rm in}<0$ and $\dot{\mathcal F}_R>0$ while the combined
free energy continues to decrease: this is transient fuel
regeneration, not fuel exhaustion.
For these parameters, net $B\to A$ conversion and transport lead
to a homogeneous $B$-poor state with
$\phi_B\simeq1.28\times10^{-3}$, consistent with chemical equilibrium
$\bar\mu_A=\bar\mu_B$ and $a_F=a_W$.
Disabling both reactions would instead conserve the prepared $B$
amount, allowing passive phase coexistence at that composition.
Thus, the observed homogenization involves chemical conversion,
not a change in interactions.

\section{Membrane--cytosol reaction--diffusion: recruitment and model reduction}
\label{sec:duality}

In this section, we examine what reduced state and channel descriptions retain about dynamics and chemical input power $\mathcal P_{\rm in}$.

\subsection{Physical free energy, recruitment kinetics, and chemical power} \label{sub:membrane_model}

We use a convex ideal free energy to distinguish recruitment-driven
redistribution from the interaction-driven phase separation of Sec.~\ref{sec:mixture}.
The cytosolic pool is retained because the membrane density alone does not determine attachment~\cite{Brauns2020,Weyer2026}.
We first construct the physical two-field model and examine whether
composition trajectories alone determine the chemical input power
$\mathcal P_{\rm in}$.
We then compare an invertible change of fields with a density-only
representation under a separate kinetic assumption.
The fuel-channel realization below is a constitutive example, not an identified protein-specific mechanism.
Coupled bulk--surface thermodynamics has also been used to study
wetting with surface binding and chemically driven changes of droplet
shape.\cite{Zhao2024,Liese2025Wetting}
Here, we isolate recruitment and internal redistribution on a fixed
geometry, rather than droplet wetting or contact-line motion.

\textbf{Reaction channels, fields, and constraints.}
Consider the passive and fueled conversions of Secs.~\ref{sec:catalyst} and \ref{sec:mixture}, now between cytosolic and membrane-bound protein states:
\begin{equation}\label{eq:CM-reaction}
C\rightleftharpoons M,\qquad C+F\rightleftharpoons M+W.
\end{equation}
Here, $C,M$ label the protein states and $F,W$ the fuel and waste;
$C$ is not the solvent of Sec.~\ref{sec:mixture}.
The subscripts $0,F$ distinguish passive and fueled conversion.

For a fixed flat membrane beneath a thin solution layer
[Fig.~\ref{fig:overview}(d)], define the amounts per unit area by
\begin{equation}\label{eq:height}
c(\boldsymbol x,t)=\int_0^h C_b(\boldsymbol x,z,t)\,dz,
\qquad m(\boldsymbol x,t)=\sigma_m(\boldsymbol x,t).
\end{equation}
Here, $C_b$ is the bulk molar concentration, $\sigma_m$ the membrane
molar surface density, $h$ the fixed layer height, and
$\boldsymbol x,z$ the lateral and perpendicular coordinates.
This height-averaged description requires perpendicular diffusion
to relax faster than binding and lateral redistribution;
a spatially resolved bulk requires a separate transport
problem.~\cite{Zhao2024}

The trial net attachment rates $r_0,r_F$ are positive for $C\to M$
and have units of moles per area per time.
On a periodic membrane domain $\Sigma$, the material balances are
\begin{subequations}\label{eq:memconstraints}
\begin{equation}\label{eq:memconstraintsa}
\partial_t c=-\nabla_s\cdot\boldsymbol J_c-r_0-r_F,
\end{equation}
\begin{equation}\label{eq:memconstraintsb}
\partial_t m=-\nabla_s\cdot\boldsymbol J_m+r_0+r_F.
\end{equation}
\end{subequations}
The lateral molar fluxes $\boldsymbol J_c,\boldsymbol J_m$ have units
of moles per length per time, and $\nabla_s$ differentiates laterally.
Total protein is conserved; membrane protein alone is not.

\textbf{Free energy and its rate.}
Choose ideal component thermodynamics:
\begin{equation}
\mathcal F=RT\int_\Sigma
\left[c\left(\ln\frac{c}{m_s}-1\right)
+m\left(\ln\frac{m}{k_{\rm b}m_s}-1\right)\right]dA.
\label{eq:memF}
\end{equation}
Here, $m_s>0$ is a reference areal density, also chosen as the
half-saturation density of the recruitment function $g(m)$ below.
The dimensionless $k_{\rm b}>0$ specifies the passive equilibrium
partition $m=k_{\rm b}c$; it incorporates the standard binding
free-energy difference and the fixed layer height.
The molar chemical potentials are
$\mu_c=\delta\mathcal F/\delta c=RT\ln(c/m_s)$ and
$\mu_m=\delta\mathcal F/\delta m=RT\ln[m/(k_{\rm b}m_s)]$.
The free energy is strictly convex for $c,m>0$.
Using the material balances gives
\begin{equation}\label{eq:memFdot}
\dot{\mathcal F}=\int_\Sigma\big[\sum_{i=c,m}\nabla_s\mu_i\cdot\boldsymbol J_i-(\mu_c-\mu_m)(r_0+r_F)\big]dA.
\end{equation} 

\textbf{Dissipation and rate selection.}
Combining the resistances for ideal dilute lateral transport with the reaction dissipation potential $\varphi$ of
Eq.~\eqref{eq:potential} gives the total dissipation potential $\Phi$: 
\begin{equation}\label{eq:memPhi}
\Phi=\int_\Sigma\left[
\frac{|\boldsymbol J_c|^2}{2\Lambda_c}
+\frac{|\boldsymbol J_m|^2}{2\Lambda_m}
+\sum_{\rho=0,F}\varphi(r_\rho;\Gamma_\rho)
\right]dA,
\end{equation}
where we take the mobility coefficients to be $\Lambda_c=D_c c/(RT)$ and $\Lambda_m=D_m m/(RT)$, with $D_c>D_m>0$ being the diffusion coefficients. Each mobility has units $\mathrm{mol}^2/(\mathrm J\,\mathrm s)$.
With areal reaction rates, $\varphi$ has power-per-area units. The positive kinetic scales $\Gamma_\rho$ may depend on the state variable fields and fuel controls; all coefficients are
held fixed during trial-rate variation.

For a spatially uniform maintained fuel--waste chemical-potential
difference $\Delta\mu_F=\mu_F^{\rm res}-\mu_W^{\rm res}$,
the chemical input power and the driving contribution to the Rayleighian are
\begin{equation}\label{eq:memR}
\mathcal P_{\rm in}=\Delta\mu_F\int_\Sigma r_F\,dA,
\end{equation} 
\begin{equation}\label{eq:memR2}
\dot{\mathcal F}-\mathcal P_{\rm in}
=\int_\Sigma\left[
\sum_{i=c,m}\nabla_s\mu_i\cdot\boldsymbol J_i
-\mathcal A_0r_0-\mathcal A_Fr_F
\right]dA.
\end{equation}
where $\mathcal A_0=\mu_c-\mu_m$ and
$\mathcal A_F=\mu_c-\mu_m+\Delta\mu_F$.
Minimizing the Rayleighian, $\mathcal R=\Phi+\dot{\mathcal F}-\mathcal P_{\rm in}$, at fixed fields and controls gives 
\begin{subequations}\label{eq:memAffinityRates}
\begin{equation}\label{eq:memAffinityRatesa}
\boldsymbol J_{c,*}=-D_c\nabla_s c,  \quad
\boldsymbol J_{m,*}=-D_m\nabla_s m,
\end{equation}
\begin{equation}\label{eq:memAffinityRatesb} 
r_{\rho,*}=\Gamma_\rho
\sinh\frac{\mathcal A_\rho}{2RT}, 
\end{equation}
\end{subequations}
for $\rho=0,F$. The reaction law specifies the affinity dependence at each fixed state, but still requires the kinetic scales.

\textbf{Kinetic specialization and recruitment.}
To obtain the cooperative-recruitment law at the prescribed fuel bias, choose
\begin{subequations}\label{eq:memGamma}
\begin{equation}\label{eq:memGammaa}
\Gamma_0=2k_{\rm d}[1-\theta(m)]\sqrt{k_{\rm b}cm}, 
\end{equation}
\begin{equation}\label{eq:memGammab} 
\Gamma_F=2k_{\rm d}\theta(m)e^{\Delta\mu_F/(2RT)}\sqrt{k_{\rm b}cm}.  
\end{equation}
\end{subequations}
Here, $k_{\rm d}>0$ is a detachment frequency, $\theta(m)={g(m)}/
\{k_{\rm b}[e^{\Delta\mu_F/(RT)}-1]\}$, and $g(m)={m^2}/{(m_s^2+m^2)}$.
The sufficient condition $\Delta\mu_F>RT\ln(1+1/k_{\rm b})$
ensures $0<\theta(m)<1$ for $m>0$. The square-root factors encode ideal reactant and product availability, while $g(m)$ represents saturating cooperative
recruitment and $\theta(m)$ allocates the kinetic prefactors between passive and fueled channels. This is a composition-dependent specialization of Eq.~\eqref{eq:activityGamma}, unlike the constant-prefactor mass-action model of Sec.~\ref{sec:catalyst}.
This allocation is a thermodynamically consistent realization of a prescribed constitutive recruitment law, not a prediction of that law from OVP or LDB.
Its functional form must not be readjusted when varying the fuel bias of the same mechanism.

Substituting Eq.~\eqref{eq:memGamma} into Eq.~\eqref{eq:memAffinityRatesb} gives
\begin{subequations}\label{eq:memselected}
\begin{equation}\label{eq:memselecteda}
r_{0,*}
=\underbrace{k_{\rm d}(1-\theta)k_{\rm b}c}_{r_0^+}
-\underbrace{k_{\rm d}(1-\theta)m}_{r_0^-},
\end{equation}
\begin{equation}\label{eq:memselectedb}
r_{F,*}
=\underbrace{k_{\rm d}\theta k_{\rm b}
e^{\Delta\mu_F/(RT)}c}_{r_F^+}
-\underbrace{k_{\rm d}\theta m}_{r_F^-}.
\end{equation}
\end{subequations}
The underbraced terms are the mean one-way rates of the specified
event model: $+$ denotes attachment and $-$ detachment.
Only the fueled channel changes the fuel and waste counts.
At the prescribed bias, the total net attachment rate is $r_*=r_{0,*}+r_{F,*}
=k_{\rm d}\{[k_{\rm b}+g(m)]c-m\}$, the dimensional cooperative-recruitment law~\cite{Brauns2020}.
Together with Eq.~\eqref{eq:memconstraints}, this yields 
\begin{subequations}
\begin{equation} 
\partial_t c=D_c\Delta_s c-r_*,  
\end{equation}  
\begin{equation}  
\partial_t m=D_m\Delta_s m+r_*,
\end{equation}  
\end{subequations}
where $\Delta_s=\nabla_s^2$.
The same membrane density with a different cytosolic pool therefore has a different attachment rate. The physical free-energy balance $\dot{\mathcal F}=\mathcal P_{\rm in}-\mathcal D$ of Eq.~\eqref{eq:Frate} holds with 
\begin{equation}\label{eq:memD}
\mathcal D=\int_\Sigma\left[\frac{|\boldsymbol J_{c,*}|^2}{\Lambda_c}
+\frac{|\boldsymbol J_{m,*}|^2}{\Lambda_m}+\mathcal A_0r_{0,*}+\mathcal A_Fr_{F,*}\right]dA\geq0.
\end{equation}
Each term is a conjugate force--rate product, although the reaction
contribution is generally not twice its dissipation potential. 
With the fuel bias removed, $\Delta\mu_F=0$, and the same kinetic allocation $\theta(m)$ held fixed, the convex free energy $\mathcal F$ provides no equilibrium demixing mechanism.
Recruitment-driven domains are therefore distinct from condensation caused by attractive interactions.

\textbf{Composition dynamics without channel labels.}
Passive and fueled events produce the same attachment or
detachment changes in $(c,m)$, so the observed protein-density dynamics do not distinguish their channel identities. The total one-way rates are
\begin{subequations}\label{eq:memaggregate}
\begin{equation}
r^+=r_0^++r_F^+=k_{\rm d}[k_{\rm b}+g(m)]c,
\end{equation}
\begin{equation}
r^-=r_0^-+r_F^-=k_{\rm d}m.
\end{equation}
\end{subequations}
Writing $u=\Delta\mu_F/(RT)$, the allocation
$\theta_u(m)=g(m)/[k_{\rm b}(e^u-1)]$ leaves both total
one-way rates independent of $u$, whereas the fueled net rate is
\begin{equation*}
r_{F,*}(u)=k_{\rm d}g(m)
\frac{e^u c-m/k_{\rm b}}{e^u-1}.
\end{equation*}
Consequently, the chemical input power
$\mathcal P_{\rm in}(u)=RTu\int_\Sigma r_{F,*}(u)\,dA$
generally differs between these realizations. 
They represent distinct channel-level kinetic models,
not changes of fuel bias within a fixed mechanism.
For identical initial distributions and transport statistics, the same total one-way rates give the same probability law for $(c,m)$ within the specified Markov jump model.
Thus, a model without channel labels can predict the retained protein dynamics under maintained conditions without identifying the underlying chemical input power.
Determining fuel consumption, or predicting dynamics when the fuel pool itself evolves, requires additional channel-resolved kinetic and reservoir information.

\subsection{Changing variables without changing physical forces}
\label{sub:fieldtransform}

An invertible change of variables preserves the state information.
Transform the densities and their fluxes together:
\begin{subequations}\label{eq:transformation}
\begin{equation}\label{eq:transformationa} 
\phi =c+m, \quad \eta =c+dm,  
\end{equation}
\begin{equation}\label{eq:transformationb}  
\boldsymbol J_\phi =\boldsymbol J_c+\boldsymbol J_m, \quad
\boldsymbol J_\eta =\boldsymbol J_c+d\boldsymbol J_m. 
\end{equation}
\end{subequations}
Here, $d =D_m/D_c$, $\phi,\eta$ are areal molar densities, not volume fractions,
and $d<1$ is dimensionless.
The inverse fields are $c=(\eta-d\phi)/(1-d)$ and
$m=(\phi-\eta)/(1-d)$, requiring $d\phi<\eta<\phi$ for positive
components. The transformed material balances are
\begin{subequations}\label{eq:transconstraints}
\begin{equation}\label{eq:transconstraintsa}
\partial_t\phi=-\nabla_s\cdot\boldsymbol J_\phi, 
\end{equation}
\begin{equation} \label{eq:transconstraintsb}
\partial_t\eta=-\nabla_s\cdot\boldsymbol J_\eta-(1-d)(r_0+r_F).
\end{equation}
\end{subequations}
The conjugate chemical potentials are $\mu_\phi=(\mu_m-d\mu_c)/(1-d)$ and
$\mu_\eta=(\mu_c-\mu_m)/(1-d)$; their transport forces are $\boldsymbol X_a=-\nabla_s\mu_a$, with $a\in\{\phi,\eta\}$. In particular, $\eta$ is not $\mu_\eta$.
The chain rule preserves the transport power: $\boldsymbol J_c\cdot\boldsymbol X_c+\boldsymbol J_m\cdot\boldsymbol X_m=\boldsymbol J_\phi\cdot\boldsymbol X_\phi+
\boldsymbol J_\eta\cdot\boldsymbol X_\eta$.

In the new variables, the same physical Rayleighian is
\begin{equation}\label{eq:Rtrans}
\begin{aligned}
\mathcal R&=\int_\Sigma\bigg[
\frac12\sum_{a,b}(\boldsymbol L_z^{-1})_{ab}
\boldsymbol J_a\cdot\boldsymbol J_b+\sum_a\nabla_s\mu_a\cdot\boldsymbol J_a\\
&+\sum_{\rho=0,F}\varphi(r_\rho;\Gamma_\rho)-(1-d)\mu_\eta(r_0+r_F)-\Delta\mu_F\,r_F\bigg]dA. 
\end{aligned}
\end{equation}
Here, $a,b\in\{\phi,\eta\}$ label flux coordinates, not chemical species;
$\Gamma_\rho$ is evaluated at the inverse component fields.
The mobility matrix is
\begin{equation}
\boldsymbol L_z=\begin{pmatrix}
\Lambda_c+\Lambda_m&\Lambda_c+d\Lambda_m\\
\Lambda_c+d\Lambda_m&\Lambda_c+d^2\Lambda_m
\end{pmatrix}.
\label{eq:transmob}
\end{equation}
It is positive definite, with
$\det\boldsymbol L_z=(1-d)^2\Lambda_c\Lambda_m>0$, and selects
$\boldsymbol J_{a,*}=\sum_b(L_z)_{ab}\boldsymbol X_b$.
The transport contribution to the physical dissipation in
Eq.~\eqref{eq:memD} is
\begin{equation*} 
\mathcal D_{\rm tr}
=\int_\Sigma\sum_{i=c,m}
\frac{|\boldsymbol J_{i,*}|^2}{\Lambda_i}\,dA=\int_\Sigma\sum_{a=\phi,\eta}
\boldsymbol J_{a,*}\cdot\boldsymbol X_a\,dA\geq0. 
\end{equation*}
The coordinate change introduces cross coefficients while preserving
the physical transport dissipation and chemical input power.
Variation and the material balances give
\begin{subequations}\label{eq:twofieldgeneral}
\begin{equation}\label{eq:twofieldgenerala} 
\partial_t\phi =D_c\Delta_s\eta,  
\end{equation}
\begin{equation}\label{eq:twofieldgeneralb}  
\partial_t\eta =(D_c+D_m)\Delta_s\eta-D_m\Delta_s\phi-(1-d)r_*(c,m). 
\end{equation}
\end{subequations}
Equation~\eqref{eq:twofieldgenerala} shows that the instantaneous total-density rate $\partial_t\phi$ also depends on $\Delta_s\eta$. 
Thus, $\phi$ alone does not determine its subsequent evolution.

% The first equation depends on the unobserved redistribution field even when the present total density is fixed. Identifying this dependence, while transforming resistance and power consistently, is the practical
% value of retaining and then changing the slow variables.

\subsection{What a density-only equation retains}
\label{sub:compression}

Equation~\eqref{eq:twofieldgenerala} still requires $\eta$ to determine the total-density rate. To examine what survives its elimination, consider the special Zhou--Frey kinetic law~\cite{ZhouFrey2026},
\begin{equation}
r_*(c,m)=\frac{\eta-\eta_*(\phi)}{\tau(1-d)},
\qquad \tau>0.
\label{eq:ZFspecial}
\end{equation}
Here, $\eta_*(\phi)$ is the value of $\eta$ at zero net conversion for a given total density $\phi$.
Without diffusion, $\phi$ is constant and
$\partial_t\eta=-[\eta-\eta_*(\phi)]/\tau$,
so $\tau$ is the internal relaxation time.
This is a separate kinetic assumption from
Eq.~\eqref{eq:memselected}.
Near a linearly stable zero-rate state, a smooth reaction rate
has this form to first order, with
$\tau^{-1}=\partial_c r_*-\partial_m r_*>0$
evaluated there.
Separate channel rates and their chemical work must be specified to give this benchmark a physical reaction model.

Using the special kinetic law in Eq.~\eqref{eq:ZFspecial},
differentiating Eq.~\eqref{eq:twofieldgenerala} and substituting Eq.~\eqref{eq:twofieldgeneralb} eliminates $\eta$ exactly: 
\begin{equation}
\begin{aligned}
\tau\partial_t^2\phi+
[1-\tau(D_c+D_m)\Delta_s]\partial_t\phi\\
=D_c\Delta_s\eta_*(\phi)-\tau D_cD_m\Delta_s^2\phi.
\end{aligned}
\label{eq:projected}
\end{equation}
No fast-relaxation approximation is used.
The eliminated field supplies the required initial rate
$\partial_t\phi(t_0)=D_c\Delta_s\eta(t_0)$;
the initial spatial mean of $\eta$ is additionally needed to
reconstruct $c,m$. 
Thus, writing a single equation does not remove the internal
information needed for density evolution.
The two additional terms relative to ordinary Cahn--Hilliard
dynamics describe internal relaxation and resistance to spatial variations of the density rate; both are fixed by $\tau,D_c,D_m$.

\textbf{Formal variational representation and physical information.}
For a constant reference mobility $M>0$ with units
$\mathrm{mol}^2/(\mathrm J\,\mathrm s)$, choose an effective functional
such that
\begin{equation*}
M\frac{\delta\mathcal F_{\rm eff}}{\delta\phi}
=D_c\eta_*(\phi)-\tau D_cD_m\Delta_s\phi.
\end{equation*}
Together with the auxiliary storage
$K_{\rm aux}=\tau\int_\Sigma|\boldsymbol J_\phi|^2dA/(2M)$ and
\begin{equation*}
\begin{aligned}
\Phi_s&=\int_\Sigma\left[
\frac{|\boldsymbol J_\phi|^2}{2M}
+\frac{\beta}{2}(\partial_t\phi)^2\right]dA,
\end{aligned}
\end{equation*}
with $\beta={\tau(D_c+D_m)}/{M}$, the scalar
$\mathcal R_0=\Phi_s+\dot{\mathcal F}_{\rm eff}+\dot K_{\rm aux}$ reproduces Eq.~\eqref{eq:projected}.
Variation is over $\partial_t\phi,\boldsymbol J_\phi$ subject to $\partial_t\phi+\nabla_s\cdot\boldsymbol J_\phi=0$, at fixed $\phi$ and flux acceleration $\partial_t\boldsymbol J_\phi$. 
This equation-level representation does not identify
$\mathcal F_{\rm eff}$ or $K_{\rm aux}$ with the physical energy, or the Rayleighian excess with a fluctuation cost. 
For example, $\mathcal R_q=q\mathcal R_0$ with constant $q>0$
preserves the deterministic minimizer while multiplying the excess
and Hessian by $q$; interpreting that excess with a fixed thermal
normalization would add a stochastic assumption.
The recruitment model in Sec.~\ref{sub:membrane_model}
specifies a physical free energy, chemical work, and reaction event
rates. A corresponding identification for
Eq.~\eqref{eq:ZFspecial} requires its own compatible channel model.

% \textbf{Physical consequences.}
% At a uniform state $\phi_0$, negative $\eta_*'(\phi_0)$ destabilizes sufficiently long wavelengths, if the domain supports them. Positive $\tau$ alone does not imply waves. Oscillatory linear relaxation requires a negative characteristic discriminant; for this scalar normal form, $\eta_*'(\phi_0)\leq1$ excludes it, including the locally attracting branch of the stated recruitment law. Moreover, $d(\mathcal F_{\rm eff}+K_{\rm aux})/dt=-2\Phi_s$ excludes nonconstant steadily translating periodic profiles. These restrictions do not rule out transient damped waves for other parameters, fronts on unbounded domains, or  multicomponent active waves. The conclusion is that an unobserved internal relaxation changes density evolution and its initial data, not that every effective inertial term generates a traveling pattern. These comparisons distinguish a representation of the density dynamics from identification of its physical Rayleighian.

\section{Discussion and conclusions}
\label{sec:discussion}

We have used OVP to construct coupled reaction--transport models and identify the distinct inputs required for deterministic dynamics, fluctuations, and chemical input power. The physical Rayleighian combines the retained free energy $\mathcal F$, input power $\mathcal P_{\rm in}$, and dissipation potential $\Phi$. 
The statistical interpretation requires a specified stochastic mechanism and its matching to these physical constitutive inputs.
For the constant-coefficient Gaussian model, matching is expressed by the Einstein relation $\boldsymbol D=k_{\mathrm B}T\boldsymbol L$.
For Poisson reactions, $\varphi(r_\rho;\Gamma_\rho)$ matches the rate-reversal-even part of the counting cost $\ell_\rho(r_\rho)$, while channelwise local detailed balance (LDB) identifies the statistical bias with the physical affinity by $\mathcal A_\rho=RT\vartheta_\rho$. 
For the latter, Eq.~\eqref{eq:reactionkernel} identifies
$(\mathcal R-\mathcal R_{\min})/(2k_{\mathrm B}T)$ as the local reaction-rate
fluctuation cost per unit time in the frozen-state, many-event regime.
This is a leading-exponential probability relation, not an exact
normalized finite-count transition density.

The construction builds on established nonlinear variational and reaction-fluctuation structures~\cite{Gyarmati1970,Lengyel1988,MPPR2017,LMPR2017}.
For the same convex dissipation potential and its conjugate, the Gyarmati-type functional and the Rayleighian excess differ only by a sign, as shown in Eq.~\eqref{eq:GyarmatiReactionCost}.  
Conditional local equilibrium (LE) permits retained thermodynamics; LDB identifies the statistical bias $\vartheta_\rho$ with the channel affinity $\mathcal A_\rho$; and the kinetic scales $\Gamma_\rho$ supply
the additional input needed to determine the mean one-way rates $r_\rho^\pm$. 
The quadratic Onsager form is recovered near equilibrium, while the law of mass action requires the corresponding concentration or activity dependence of these scales.

The examples separate complementary questions. In
Sec.~\ref{sec:catalyst}, a net-affinity response can be nonreciprocal under specified chemical-potential controls even though a convex dissipation potential $\Phi$ exists at each state. Changing composition changes both the affinities and kinetic scales, whereas OVP varies trial rates at a fixed state. Lengyel-type partial-affinity symmetries concern different derivatives and do not establish universal nonlinear reciprocity.
At a fixed state, the Poisson and logarithmic-mean Gaussian models select the same rate $r_*$ but have different fluctuation costs $\ell_P$ and $\ell_G$. The local Gaussian cost $\ell_{\rm loc}$ approximates $\ell_P$ near $r_*$. The inverse-log potential $\varphi_{\mathrm I}$ also selects $r_*$, but its Rayleighian excess $\Delta\mathcal R_{\mathrm I}$ is not identified with a fluctuation cost. 

Section~\ref{sec:mixture} couples nonideal thermodynamics to resolved
reaction and transport.
For the parameters studied, a chemically maintained dense domain
dissolves after external replenishment stops through conversion and
redistribution, without changing the interactions.
Retaining the finite fuel pool identifies $\mathcal P_{\rm in}$
as the free-energy transfer rate between pool and mixture, which can
briefly reverse while
$\dot{\mathcal F}+\dot{\mathcal F}_R=-\mathcal D\leq0$.
Thus, chemical input power and dissipation remain distinct even
during rapid domain dissolution.

Section~\ref{sec:duality} distinguishes reaction-channel resolution from state-variable reduction. Different passive/fueled channel realizations can have the same total attachment and detachment rates $r^\pm$ and, with identical transport statistics and initial distributions, the same trajectory probability law for $(c,m)$, but different chemical input power $\mathcal P_{\rm in}$. 
An invertible change of fields and fluxes preserves the physical Rayleighian $\mathcal R$. In contrast, the density-only equation for $\phi=c+m$, obtained using the special kinetic law in Eq.~\eqref{eq:ZFspecial}, requires the initial rate $\partial_t\phi(t_0)$; reconstructing $c,m$ also requires the initial spatial mean of $\eta$.
Thus, fewer explicit fields do not necessarily mean less dynamical information, and an equation-level variational representation does not by itself identify chemical input power or fluctuations.  

% Together, these results distinguish deterministic kinetics, fluctuation statistics, and resolved chemical work without treating them as successive consequences of one equation. A physical Rayleighian assigns storage, dissipation, and input power to specified mechanisms; statistical matching gives its excess a probability interpretation at the same resolution.
For the resolved isothermal macroscopic reaction model,
$\mathcal D=T\dot S_{\rm prod}^{\rm mac}$, but it is not generally a heat current. Heat flows into individual reservoirs require the corresponding internal-energy, entropy, and matter-exchange information~\cite{Seifert2011,Seifert2012,RaoEsposito2016}. 
Neither a mean rate nor a trajectory probability law without reaction-channel labels supplies all these physical inputs.

% The scope of the statistical identification is equally important.
Appendix~\ref{app:reciprocity} relates reaction-event fluctuations to equilibrium and higher-order response.
State-dependent event rates naturally produce multiplicative noise, but their chemical Langevin approximation need not preserve the full Poisson cost or its thermodynamic matching (Appendix~\ref{app:ensemble}).
Additional active fluctuations can leave an independently justified deterministic Rayleighian valid while changing the total fluctuation cost (Appendix~\ref{app:activecost}). Conversely, chemical driving does not itself violate LDB. These distinctions depend on the resolved mechanisms, not simply on whether increments are Gaussian or Poisson.

OVP provides a systematic framework for constructing and
approximating coupled reaction--transport dynamics from specified thermodynamic and kinetic inputs, while consistently incorporating material constraints and power balances. 
For resolved reaction channels with independent Poisson event drivers, the dissipation potentials are additive, although the channels remain coupled through shared states and material balances.
Identifying the Rayleighian excess with a fluctuation cost additionally requires statistical and thermodynamic matching, whereas chemical input power requires channel-resolved work information. 
Restricting admissible trial fields before variation provides a variational approximation method~\cite{Doi2015Approximation,Doi2019},
whereas eliminating state variables requires dynamical closure or retention of the relevant memory and initial data.
The common passive/fueled reaction topology of our examples therefore does not uniquely fix the kinetics: constant-coefficient mass action, composition-selective fuel coupling, and cooperative recruitment are distinct kinetic constitutive inputs that can affect driven steady states and chemical power, not merely relaxation times. 

These structures also suggest computational developments. Deep Ritz methods use neural-network trial functions for variational problems, including active elastic solids~\cite{EYu2018DeepRitz,Wang2022DeepRitz},
while the deep Onsager--Machlup method (DOMM) applies neural trial
trajectories to soft-matter dynamics~\cite{Li2025DOMM}.
Statistical and covariant Onsager--Machlup formulations and neural
optimization of active-particle paths provide complementary developments
~\cite{Yasuda2024Statistical,Yasuda2026Covariant,Zheng2026Paths}.
Extending DOMM to the nonquadratic reaction action of
Eq.~\eqref{eq:gap} would require admissible trial trajectories that respect material balances, resolved channels, and chemical work. Its development and validation remain future work. A further goal is to constrain physical models jointly by reaction counts, mechanical response, and chemical input power, and to determine which physical information survives reduction to effective active-field theories of phase separation ~\cite{Wittkowski2014,Tjhung2018,PisegnaGolestanian2026}.
In particular, such reductions should distinguish irreversibility observed at the coarse-grained field level from the physical energetic cost of the underlying reservoir coupling~\cite{Nardini2017,Markovich2021}.

\section*{Acknowledgements}
We dedicate this work to the memory of Rudolf Podgornik, in recognition of his contributions to statistical physics, soft matter physics, and biological physics. We acknowledge unpublished notes by Rudolf Podgornik, brought to our attention by B.~Zheng, concerning reaction dissipation.
X.X. thanks Tiezheng Qian and Ping Sheng for their longstanding support and mentorship, and Masao Doi and Christoph A. Weber for valuable and insightful discussions.
X.X. also thanks the organizers of the 2026 International Workshop on Soft Matter and Biophysics Theories at the Institute of Theoretical Physics, Chinese Academy of Sciences, Beijing, for their invitation and hospitality. Discussions with Erwin Frey and Daniel Zhou, together with the stimulating talks at the workshop, helped shape this work. For X.X., these occasions reaffirm the enduring value of honoring scientists' legacies and exchanging ideas in person, even in the age of artificial intelligence. X.X. acknowledges support from the National Natural Science Foundation of China (NSFC, No.~12374209 and No.
12131010). S.K. acknowledges support from the National Natural Science Foundation of China (Grant No.~12274098) and from the Zhejiang Key Laboratory of Soft Matter Biomedical Materials (2025ZY01036 and 2025E10072). This work was supported by the JSPS Core-to-Core Program ``Advanced core-to-core network for the physics of self-organizing active matter" (JPJSCCA20230002).

% \section*{Data availability}
% XXX
% The numerical data and scripts supporting the figures are included in the accompanying reproducibility package.
% % Before submission, add the author-approved permanent repository identifier.

\section*{Declaration of generative AI use}

X.X. used ChatGPT (GPT-6 Pro; OpenAI) for assistance with analytical derivations and consistency checks, numerical-code development, literature search and synthesis, and language editing. The tool supported exploration of mathematical formulations, implementation of the numerical model, and preparation of the paper. 
It also generated an initial draft of the schematic in
Fig.~\ref{fig:overview}, which the authors substantially revised. 

The authors reviewed and corrected all AI-assisted material included in this paper. Derivations were checked against the stated assumptions, numerical implementations and results against the governing equations, and references and literature-based statements against the cited sources. The reported quantitative results were obtained by explicit analytical evaluation or numerical computation of the stated
models. The authors made the final scientific judgments and take full responsibility for the scientific content, source attribution, and conclusions.
 
\section*{Author declarations}

\subsection*{Conflict of interest}
The authors have no conflicts to disclose.
 
\subsection*{Author contributions}
Xinpeng Xu: Conceptualization; Methodology; Formal analysis;
Investigation; Software; Validation; Visualization;
Writing -- original draft; Writing -- review \& editing.
Changhao Li, Zhongqiang Xiong, Bin Zheng, Kento Yasuda, and
Shigeyuki Komura: Conceptualization; Writing -- review \& editing.
The research ideas and physical interpretations were refined through
extensive discussions among all authors, who approved the final
manuscript.
% Xinpeng Xu initiated and led the study, developed the theoretical
% framework, performed the analytical and numerical work, and prepared
% the original manuscript. The research ideas, theoretical formulations,
% and physical interpretations were developed and refined through
% extensive discussions among all authors. All authors contributed
% to critically reviewing and revising the manuscript and approved
% the final version.

\section*{Data availability}

The numerical data and code supporting the findings of this study,
including the initial profiles, model parameters, and plotting
scripts, are available from the corresponding author upon
reasonable request.

\appendix

\section{Fluctuation constraints on nonlinear response}
\label{app:reciprocity}

We examine selected rates as functions of independent conjugate forces,
using the same force-control protocol for all derivatives.
For time-reversal-even chemical progress coordinates near equilibrium,
$\boldsymbol X=\boldsymbol0$, write
\begin{equation}
\dot\alpha_{i,*}
=\sum_jL_{ij}X_j
+\frac12\sum_{jk}M_{ijk}X_jX_k+O(|\boldsymbol X|^3).
\label{eq:higherresponse}
\end{equation}
Here, $L_{ij}=(\partial_{X_j}\dot\alpha_{i,*})_0$ and
$M_{ijk}=(\partial_{X_j}\partial_{X_k}\dot\alpha_{i,*})_0$,
with other independent controls fixed and $0$ denoting equilibrium.
Commuting force derivatives gives $M_{ijk}=M_{ikj}$, but does not
exchange the output index $i$ with a force index.

Use the frozen Poisson channels of Sec.~\ref{sub:Poisson} in fixed
volume $V$, with mean one-way molar rate densities $r_\rho^\pm$.
Let $\delta_\rho\boldsymbol\alpha$ be the fixed change in one forward event.
The retained coordinates and forces must resolve every channel affinity:
$\boldsymbol X\cdot\delta_\rho\boldsymbol\alpha
=\mathcal A_\rho/N_{\mathrm A}$.
Over $\Delta t$, the total increment is
$\Delta\boldsymbol\alpha
=\sum_\rho\delta_\rho\boldsymbol\alpha
(\Delta N_\rho^+-\Delta N_\rho^-)$,
where the independent Poisson counts have means
$N_{\mathrm A}Vr_\rho^\pm\Delta t$.
Exchanging every forward and reverse count and using LDB gives
\begin{equation}
\frac{P_{\boldsymbol X}(\Delta\boldsymbol\alpha)}
{P_{\boldsymbol X}(-\Delta\boldsymbol\alpha)}
=\exp\left(
\frac{\boldsymbol X\cdot\Delta\boldsymbol\alpha}{k_{\mathrm B}T}
\right).
\label{eq:incrementReversal}
\end{equation}
Here, $P_{\boldsymbol X}$ is the increment probability at the controlled
forces. All count allocations with the same increment have the same
ratio, so the relation survives summation over them.
It is exact for the frozen counting process; application to an evolving
composition uses the frozen-state approximation.

The selected mean rate is
$\dot\alpha_{i,*}=\langle\Delta\alpha_i\rangle_{\boldsymbol X}/\Delta t$.
Define the increment covariance per unit time by $C_{ij}(\boldsymbol X)
={\operatorname{Cov}_{\boldsymbol X}
(\Delta\alpha_i,\Delta\alpha_j)}/{\Delta t}$ and then
\begin{equation}\label{eq:incrementCovariance}
C_{ij}(\boldsymbol X) =N_{\mathrm A}V\sum_\rho
(\delta_\rho\alpha_i)(\delta_\rho\alpha_j)\sigma_\rho, 
\end{equation}
where $\sigma_\rho=r_\rho^++r_\rho^-$, and averages are over the frozen counting law. 
Equation~\eqref{eq:incrementCovariance} follows from independent Poisson variances and includes cross-correlations from shared events. 
Changing the summation variable
$\Delta\boldsymbol\alpha\to-\Delta\boldsymbol\alpha$ in
Eq.~\eqref{eq:incrementReversal} gives
\begin{equation}
2\langle\Delta\alpha_i\rangle_{\boldsymbol X}
=\left\langle
\Delta\alpha_i
\left[1-e^{-\boldsymbol X\cdot
\Delta\boldsymbol\alpha/(k_{\mathrm B}T)}\right]
\right\rangle_{\boldsymbol X}.
\label{eq:meanFromReversal}
\end{equation}
At zero force, odd increment moments vanish.
Expanding to second order in $\boldsymbol X$, the third moment therefore
contributes only at higher order. Products of mean increments are
$O(|\boldsymbol X|^2)$, so raw second moments and covariances have the
same value and first force derivative at equilibrium.
Comparison with Eq.~\eqref{eq:higherresponse} yields
\begin{equation}
\begin{aligned}
L_{ij}&=\frac{C_{ij}(0)}{2k_{\mathrm B}T},\\
M_{ijk}
&=\frac{[\partial_{X_k}C_{ij}+\partial_{X_j}C_{ik}]_0}{2k_{\mathrm B}T},\\
M_{ijk}-M_{jik}
&=\frac{[\partial_{X_j}C_{ik}-\partial_{X_i}C_{jk}]_0}{2k_{\mathrm B}T}.
\end{aligned}
\label{eq:secondResponseConstraint}
\end{equation}
Thus equilibrium fluctuations give $L_{ij}=L_{ji}$, whereas their force
dependence constrains, but does not generally symmetrize, second-order
response.~\cite{AndrieuxGaspard2007,Basu2015SecondOrder}
In particular,
\begin{equation}\label{eq:highercurl}
\frac{\partial\dot\alpha_{i,*}}{\partial X_j}
-\frac{\partial\dot\alpha_{j,*}}{\partial X_i}=\sum_k(M_{ijk}-M_{jik})X_k+O(|\boldsymbol X|^2). 
\end{equation}
These response derivatives compare controlled states; they are not
trial-rate derivatives within one fixed-state Rayleighian.

For a single molar reaction extent $\alpha$, each event changes $\alpha$ by $\pm1/N_{\mathrm A}$, so $C=V\sigma/N_{\mathrm A}$.
At equilibrium, $\sigma=\Gamma_{\rm eq}$ and
$L=V\Gamma_{\rm eq}/(2RT)$, giving the local increment diffusion coefficient $D_{\rm eq}\equiv C(0)/2=k_{\mathrm B}T L$.
This equilibrium Einstein relation requires neither Gaussian increments nor large counts; it is not a long-time diffusivity of bounded composition.

\section{State-dependent mobility and stochastic matching}\label{app:ensemble}

\textbf{Thermal diffusion.}
Using the retained coordinates and forces of Sec.~\ref{sub:OM}, consider the It\^o model
\begin{equation}\label{eq:generalMultiplicative}
\mathrm d\boldsymbol\alpha
=\dot{\boldsymbol\alpha}_{\rm I}(\boldsymbol\alpha,t)\,dt+\sqrt2\boldsymbol B(\boldsymbol\alpha,t)\,\mathrm d\boldsymbol W. 
\end{equation}
Here, $\boldsymbol W$ contains independent standard Wiener processes,
$\boldsymbol D=\boldsymbol B\boldsymbol B^{\mathsf T}$ is positive definite, and
$\dot{\boldsymbol\alpha}_{\rm I}$ is the conditional mean increment per
time. The frozen Gaussian cost is Eq.~\eqref{eq:GaussianCost} with
$\boldsymbol L\boldsymbol X$ replaced by
$\dot{\boldsymbol\alpha}_{\rm I}$.

Let $p(\boldsymbol\alpha,t)$ be the probability density relative to flat
coordinate volume and $j_i$ its probability flux, with
$\partial_t p=-\sum_i\partial_{\alpha_i}j_i$.
The It\^o model gives
\begin{equation*}
j_i=
\left(\dot\alpha_{{\rm I},i}
-\sum_j\partial_{\alpha_j}D_{ij}\right)p
-\sum_jD_{ij}\partial_{\alpha_j}p.
\end{equation*}
For symmetric positive-definite $\boldsymbol L$, a uniform bath at
temperature $T$, and no additional reversible current, the thermal flux
is $j_i=\sum_jL_{ij}(X_jp-k_{\mathrm B}T\partial_{\alpha_j}p)$.
Comparing coefficients gives~\cite{LauLubensky2007} 
\begin{equation}
\begin{aligned}
D_{ij}&=k_{\mathrm B}T L_{ij}, \\
\dot\alpha_{i,*}&=\sum_jL_{ij}X_j, \\
\dot\alpha_{{\rm I},i}
&=\dot\alpha_{i,*}+\sum_j\partial_{\alpha_j}D_{ij}.
\end{aligned}
\end{equation}  
Einstein matching fixes the quadratic curvature, but a nonzero noise-induced drift $\nabla_{\boldsymbol\alpha}\cdot\boldsymbol D$
separates the minimizer $\dot{\boldsymbol\alpha}_{\rm I}$ of the frozen Gaussian increment cost from the minimizing rate $\dot{\boldsymbol\alpha}_*$ of the physical Rayleighian.
The matching in Eq.~\eqref{eq:Einstein} for all trial rates additionally requires $\sum_j\partial_{\alpha_j}D_{ij}=0$ for every $i$ at that state.

\textbf{Chemical events and their diffusion approximation.} 
For a closed $A\rightleftharpoons B$ reactor of fixed volume $V$, write $c=c_B$ and $c_A=c_{\rm tot}-c$, where $c_{\rm tot}$ is the conserved total molar concentration.
Let $r^\pm(c)$ be the mean one-way molar rate densities.
During a frozen step $\Delta t$, independent Poisson counts $\Delta N^\pm$ have means $N_{\mathrm A}Vr^\pm\Delta t$, and
$\Delta c=(\Delta N^+-\Delta N^-)/(N_{\mathrm A}V)$.
Their moments are therefore
\begin{equation*}
\begin{aligned}
\langle\Delta c\rangle&=(r^+-r^-)\Delta t,\\
\operatorname{Var}(\Delta c)&=\frac{r^++r^-}{N_{\mathrm A}V}\Delta t.
\end{aligned}
\end{equation*}
When the rates vary little during the step and the relevant counts
are numerous, their Gaussian approximation gives the It\^o
chemical Langevin equation~\cite{Gillespie2000}
\begin{equation}\label{eq:chemicalCLE}
\mathrm dc=r_*(c)\,dt+\sqrt{2D_c(c)}\,dW_t. 
\end{equation}
Here, $r_*=r^+-r^-$, $D_c={\sigma}/{(2N_{\mathrm A}V)}$, $\sigma=r^++r^-$, $W_t$ is a standard Wiener process and $D_c$ has units of concentration squared per time. 
This diffusion occurs in composition space. Its drift and covariance follow from the difference and sum of the mean one-way rates $r^\pm$; no separate Brownian bath is introduced. 
The first jump moment fixes $\dot c_{\rm I}=r_*$, so adding
$\partial_cD_c$ would change this approximation.

Thermodynamic matching remains a separate step.
With $\mathcal A=RT\ln(r^+/r^-)$ and
$\Gamma=2\sqrt{r^+r^-}$, the physical Poisson Rayleighian
$\mathcal R_P=V[\varphi(r;\Gamma)-\mathcal A r]$ satisfies
\begin{equation*}
\left.\frac{\partial^2\mathcal R_P}{\partial r^2}\right|_{r_*}
=\frac{2RTV}{\sqrt{r_*^2+\Gamma^2}}
=\frac{2RTV}{\sigma}.
\end{equation*}
The derivatives hold the state and controls fixed. 
Consequently, its local quadratic excess gives
\begin{equation*}
\frac{\mathcal R_P-\mathcal R_{P,\min}}{2k_{\mathrm B}T}
\simeq
\frac{(\dot c-r_*)^2}{4D_c}
=N_{\mathrm A}V\ell_{\rm loc}(\dot c),
\end{equation*}
where the trial reaction rate is $r=\dot c$ and $\ell_{\rm loc}$ is defined in Eq.~\eqref{eq:localPoissonCost}.
Thus the event-based Rayleighian already reproduces the central
chemical fluctuations at the bath temperature.
The diffusion approximation does not retain the full Poisson tails
or automatically inherit all finite-affinity stochastic
thermodynamics~\cite{Horowitz2015}.
For multiple resolved channels, their contributions add to both
the drift and covariance; a fueled channel is not generally an
additional noise source with zero mean conversion.

% \textbf{Diffusion approximation of reaction counts.}
% For a closed $A\rightleftharpoons B$ reactor of fixed volume $V$, write
% $c=c_B$ and $c_A=c_{\rm tot}-c$, where $c_{\rm tot}$ is conserved.
% Let $r^\pm(c)$ be the mean one-way molar rate densities.
% At a frozen state, independent Poisson counts $\Delta N^\pm$ have means
% $N_{\mathrm A}Vr^\pm\Delta t$, and
% $\Delta c=(\Delta N^+-\Delta N^-)/(N_{\mathrm A}V)$.
% Hence, $\langle\Delta c\rangle=(r^+-r^-)\Delta t$ and $\operatorname{Var}(\Delta c)=\frac{r^++r^-}{N_{\mathrm A}V}\Delta t$. 
% When both counts are numerous, and the rates vary little during the step,
% their Gaussian approximation gives~\cite{Gillespie2000}
% \begin{equation} 
% \mathrm dc=r_*(c)\,dt+\sqrt{2D_c(c)}\,dW_t, 
% \label{eq:chemicalCLE}
% \end{equation}
% Here, $r_*=r^+-r^-$, $D_c=(r^++r^-)/{2N_{\mathrm A}V}$, $W_t$ is a standard Wiener process and $D_c$ has units of
% concentration squared per time.
% The first jump moment already fixes $\dot c_{\rm I}=r_*$;
% adding $\partial_cD_c$ would change this approximation.
% Its frozen rate cost is $N_{\mathrm A}V\ell_{\rm loc}$ from
% Eq.~\eqref{eq:localPoissonCost}; the distinction from the full Poisson cost
% and the matched frozen Gaussian model is established in
% Sec.~\ref{sub:ratefluct}.

\section{Additional noise and the limits of physical matching}
\label{app:activecost}

\textbf{Experimental motivation and stochastic model.}
Active fluctuations can be enhanced without a comparable change in
viscous response. In dilute suspensions of motile \emph{E. coli},
Chen et al. observed enhanced tracer fluctuations while independent
optical-tweezer measurements found approximately unchanged
viscosity~\cite{Chen2007Bacterial}.
This motivates a reduced description with additional fluctuations
but unchanged mobility, although such a separation is not universal.
Moreover, bacterial-bath experiments show that approximately
instantaneous friction can coexist with temporally correlated
active fluctuations~\cite{Maggi2017Memoryless}.
A white Gaussian approximation therefore requires a further
timescale separation
$\tau_a\ll\Delta t\ll\tau_{\rm slow}$,
where $\tau_a$ is the active-force correlation time,
$\Delta t$ the observation interval, and $\tau_{\rm slow}$ the
characteristic timescale of the retained dynamics.
The accumulated active forcing must also be approximately Gaussian.

Under these assumptions, consider the retained state
$\boldsymbol\alpha$, physical force $\boldsymbol X$, and constant
symmetric positive-definite mobility $\boldsymbol L$ of
Sec.~\ref{sub:OM}.
Suppose an additional mechanism produces independent fluctuations
without changing the prescribed conditional drift
$\boldsymbol L\boldsymbol X$ or mobility:
\begin{subequations}\label{eq:generalActiveNoise}
\begin{equation} 
\mathrm d\boldsymbol\alpha
=\boldsymbol L\boldsymbol X\,dt
+\sqrt2\boldsymbol B_{\rm th}\,\mathrm d\boldsymbol W_{\rm th}+\sqrt2\boldsymbol B_a\,\mathrm d\boldsymbol W_a,  
\end{equation}
\begin{equation}  
\boldsymbol D_{\rm tot}=k_{\mathrm B}T\boldsymbol L+\boldsymbol D_a,
\qquad
\boldsymbol D_a=\boldsymbol B_a\boldsymbol B_a^{\mathsf T}. 
\end{equation}
\end{subequations}
Here, $\boldsymbol W_{\rm th},\boldsymbol W_a$ are independent
standard Wiener processes, the noise amplitudes are constant,
and $\boldsymbol B_{\rm th}\boldsymbol B_{\rm th}^{\mathsf T}
=k_{\mathrm B}T\boldsymbol L$ describes the thermal contribution.
The additional noise is not assumed to originate from an
equilibrium bath. Externally imposed random forces provide one
realization~\cite{Martinez2013}; eliminating a coupled
nonequilibrium environment generally requires its response and
fluctuations to be determined together~\cite{MaesSteffenoni2015}.
Thus, unchanged drift and mobility are modeling assumptions,
not consequences of Gaussian statistics.

The additional noise represents unresolved forcing, not an
identified chemical reaction channel. As shown in
Appendix~\ref{app:ensemble}, resolved Poisson reaction events
generally contribute to both drift and covariance.
Their counting fluctuations must not be counted again as
independent active noise.

\textbf{Fluctuation cost and effective temperature.}
For a frozen time step, define
$\delta\dot{\boldsymbol\alpha}
=\dot{\boldsymbol\alpha}-\boldsymbol L\boldsymbol X$,
where $\dot{\boldsymbol\alpha}$ is a trial rate.
The Gaussian rate cost and the physical Rayleighian excess are
\begin{subequations}\label{eq:activeCurvatureComparison}
\begin{equation}
\mathscr L_{\rm tot}
=\frac14\delta\dot{\boldsymbol\alpha}^{\mathsf T}
\boldsymbol D_{\rm tot}^{-1}\delta\dot{\boldsymbol\alpha},
\end{equation}
\begin{equation}
\mathcal R-\mathcal R_{\min}
=\frac12\delta\dot{\boldsymbol\alpha}^{\mathsf T}
\boldsymbol L^{-1}\delta\dot{\boldsymbol\alpha}.
\end{equation}
\end{subequations}
Their minimizers coincide, but matching at the physical bath
temperature fails for nonzero $\boldsymbol D_a$.
Define $T_{\rm eff}>0$ only as a possible scalar rescaling of the Gaussian rate cost. Comparing the quadratic forms gives
\begin{equation}
\begin{gathered}
\mathscr L_{\rm tot}
=\frac{\mathcal R-\mathcal R_{\min}}{2k_{\mathrm B}T_{\rm eff}}
\quad\text{for all independent trial rates}\\
\Longleftrightarrow\quad
\boldsymbol D_{\rm tot}=k_{\mathrm B}T_{\rm eff}\boldsymbol L.
\end{gathered}
\label{eq:scalarTeffCondition}
\end{equation}
For one coordinate with constant coefficients, this ratio
always exists: $T_{\rm eff}=D_{\rm tot}/(k_{\mathrm B}L)$.
For multiple coordinates, however, all independent directions
must share the same ratio: proportional anisotropic tensors
qualify, whereas unequal directional ratios do not.
With state-dependent coefficients, a global constant
$T_{\rm eff}$ additionally requires the same ratio throughout
the considered state space and consistent drift centers,
as discussed in Appendix~\ref{app:ensemble}.
Scalar rescaling cannot correct a drift mismatch or reproduce
general colored or non-Gaussian fluctuations.

\textbf{Comparison with chemical reaction fluctuations.}
The distinction can be illustrated without postulating additional
reaction noise. For the concentration coordinate of
Eq.~\eqref{eq:chemicalCLE}, Poisson counting gives
$D_c=\sigma/(2N_{\mathrm A}V)$, where $\sigma=r^++r^-$ is the total
one-way rate density.
The logarithmic-mean quadratic model of
Sec.~\ref{sub:ratefluct} instead uses
$X=V\mathcal A$ and $L=L_{\log}/V$, with
$L_{\log}=r_*/\mathcal A$ understood by continuity at
$\mathcal A=0$.
Applying Eq.~\eqref{eq:scalarTeffCondition} to match the quadratic
Rayleighian $\mathcal R_{\log}$ to the local Poisson cost
$\ell_{\rm loc}$ gives  
\begin{equation}\label{eq:chemicalTeffComparison}
\frac{T_{\rm eff}}{T}
=\frac{\sigma}{2RT L_{\log}}
=\frac{\mathcal A}{2RT}
\coth\left(\frac{\mathcal A}{2RT}\right),
\end{equation}
with limiting value $1$ at $\mathcal A=0$.
This state-dependent rescaling matches the local quadratic
counting cost, but neither changes the bath temperature nor
recovers the non-Gaussian Poisson tails.
By contrast, the nonquadratic Poisson Rayleighian gives the
matched reaction-rate cost at the physical temperature $T$
under the assumptions of Sec.~\ref{sub:Poisson}.

Thus, even when an effective temperature reproduces a specified
Gaussian rate cost, it does not identify the work or heat exchanged by unresolved mechanisms. More generally, a deterministic Rayleighian determines neither fluctuation probabilities nor an Onsager--Machlup path weight without the corresponding stochastic model and physical thermodynamic matching.

\bibliographystyle{aipnum4-2}
\bibliography{references}

@article{Onsager1931a,
  author = {Lars Onsager},
  title = {Reciprocal Relations in Irreversible Processes. {I}.},
  journal = {Phys. Rev.},
  year = {1931},
  volume = {37},
  pages = {405--426},
  doi = {10.1103/PhysRev.37.405},
}

@article{Onsager1931b,
  author = {Lars Onsager},
  title = {Reciprocal Relations in Irreversible Processes. {II}.},
  journal = {Phys. Rev.},
  year = {1931},
  volume = {38},
  pages = {2265--2279},
  doi = {10.1103/PhysRev.38.2265},
}

@article{Casimir1945,
  author = {H. B. G. Casimir},
  title = {On {Onsager}'s Principle of Microscopic Reversibility},
  journal = {Rev. Mod. Phys.},
  year = {1945},
  volume = {17},
  pages = {343--350},
  doi = {10.1103/RevModPhys.17.343},
}

@article{OnsagerMachlup1953,
  author = {Lars Onsager and Stefan Machlup},
  title = {Fluctuations and Irreversible Processes},
  journal = {Phys. Rev.},
  year = {1953},
  volume = {91},
  pages = {1505--1512},
  doi = {10.1103/PhysRev.91.1505},
}

@article{MachlupOnsager1953,
  author = {Stefan Machlup and Lars Onsager},
  title = {Fluctuations and Irreversible Process. {II}. Systems with Kinetic Energy},
  journal = {Phys. Rev.},
  year = {1953},
  volume = {91},
  pages = {1512--1515},
  doi = {10.1103/PhysRev.91.1512},
}

@article{Doi2011,
  author = {Masao Doi},
  title = {{Onsager}'s variational principle in soft matter},
  journal = {J. Phys.: Condens. Matter},
  year = {2011},
  volume = {23},
  pages = {284118},
  doi = {10.1088/0953-8984/23/28/284118},
}

@article{Doi2021,
  author = {Masao Doi},
  title = {{Onsager} principle in polymer dynamics},
  journal = {Prog. Polym. Sci.},
  year = {2021},
  volume = {112},
  pages = {101339},
  doi = {10.1016/j.progpolymsci.2020.101339},
}

@article{Doi2019,
  author = {Masao Doi and Jiajia Zhou and Yana Di and Xianmin Xu},
  title = {Application of the {Onsager--Machlup} integral in solving dynamic equations in nonequilibrium systems},
  journal = {Phys. Rev. E},
  year = {2019},
  volume = {99},
  pages = {063303},
  doi = {10.1103/PhysRevE.99.063303},
}

@article{Wang2021,
  author = {Haiqin Wang and Tiezheng Qian and Xinpeng Xu},
  title = {{Onsager}'s variational principle in active soft matter},
  journal = {Soft Matter},
  year = {2021},
  volume = {17},
  pages = {3634--3653},
  doi = {10.1039/D0SM02076A},
}

@misc{PisegnaGolestanian2026,
  author        = {Pisegna, Giulia and Golestanian, Ramin},
  title         = {Active phase separation from
                   concentration-dependent diffusion
                   in chemical networks},
  year          = {2026},
  eprint        = {2610.30958},
  archivePrefix = {arXiv},
  primaryClass  = {cond-mat.stat-mech}
}

@article{Kruse2005,
  author = {Karsten Kruse and Jean-Fran\c{c}ois Joanny and Frank J\"{u}licher and Jacques Prost and Ken Sekimoto},
  title = {Generic theory of active polar gels: A paradigm for cytoskeletal dynamics},
  journal = {Eur. Phys. J. E},
  year = {2005},
  volume = {16},
  pages = {5--16},
  doi = {10.1140/epje/e2005-00002-5},
}

@article{Julicher2018,
  author = {Frank J\"{u}licher and Stephan W. Grill and Guillaume Salbreux},
  title = {Hydrodynamic theory of active matter},
  journal = {Rep. Prog. Phys.},
  year = {2018},
  volume = {81},
  pages = {076601},
  doi = {10.1088/1361-6633/aab6bb},
}

@article{Weber2019,
  author = {Christoph A. Weber and David Zwicker and Frank J\"{u}licher and Chiu Fan Lee},
  title = {Physics of active emulsions},
  journal = {Rep. Prog. Phys.},
  year = {2019},
  volume = {82},
  pages = {064601},
  doi = {10.1088/1361-6633/ab052b},
}

@article{Li1962,
  author = {James C. M. Li},
  title = {Stable Steady State and the Thermokinetic Potential},
  journal = {J. Chem. Phys.},
  year = {1962},
  volume = {37},
  pages = {1592--1595},
  doi = {10.1063/1.1733345},
}

@article{Bataille1978,
  author = {J. Bataille and D. G. B. Edelen and J. Kestin},
  title = {Nonequilibrium Thermodynamics of the Nonlinear Equations of Chemical Kinetics},
  journal = {J. Non-Equilib. Thermodyn.},
  year = {1978},
  volume = {3},
  pages = {153--168},
  doi = {10.1515/jnet.1978.3.3.153},
}

@article{LengyelGyarmati1981,
  author = {S. Lengyel and I. Gyarmati},
  title = {On the thermodynamics of elementary chemical reactions in homogeneous systems},
  journal = {J. Chem. Phys.},
  year = {1981},
  volume = {75},
  pages = {2384--2389},
  doi = {10.1063/1.442301},
}

@article{Lengyel1988,
  author = {S\'{a}ndor Lengyel},
  title = {Deduction of the {Guldberg--Waage} mass action law from {Gyarmati}'s governing principle of dissipative processes},
  journal = {J. Chem. Phys.},
  year = {1988},
  volume = {88},
  pages = {1617--1621},
  doi = {10.1063/1.454140},
}

@article{Goddard2015,
  author = {J. D. Goddard},
  title = {Dissipation Potentials for Reaction--Diffusion Systems},
  journal = {Ind. Eng. Chem. Res.},
  year = {2015},
  volume = {54},
  pages = {4078--4083},
  doi = {10.1021/ie503661b},
}

@article{Goddard2023,
  author = {J. D. Goddard},
  title = {On nonlinear {Onsager} symmetry and mass-action kinetics},
  journal = {Combust. Sci. Technol.},
  year = {2023},
  volume = {195},
  pages = {3627--3637},
  doi = {10.1080/00102202.2022.2041610},
}

@article{Mielke2011,
  author = {Alexander Mielke},
  title = {A gradient structure for reaction--diffusion systems and for energy--drift--diffusion systems},
  journal = {Nonlinearity},
  year = {2011},
  volume = {24},
  pages = {1329--1346},
  doi = {10.1088/0951-7715/24/4/016},
}

@article{MPR2014,
  author = {Alexander Mielke and Mark A. Peletier and D. R. Michiel Renger},
  title = {On the Relation between Gradient Flows and the Large-Deviation Principle, with Applications to {Markov} Chains and Diffusion},
  journal = {Potential Anal.},
  year = {2014},
  volume = {41},
  pages = {1293--1327},
  doi = {10.1007/s11118-014-9418-5},
}

@article{MPR2016,
  author = {Alexander Mielke and Mark A. Peletier and D. R. Michiel Renger},
  title = {A Generalization of {Onsager}'s Reciprocity Relations to Gradient Flows with Nonlinear Mobility},
  journal = {J. Non-Equilib. Thermodyn.},
  year = {2016},
  volume = {41},
  pages = {141--149},
  doi = {10.1515/jnet-2015-0073},
}

@article{MPPR2017,
  author = {Alexander Mielke and Robert I. A. Patterson and Mark A. Peletier and D. R. Michiel Renger},
  title = {Non-equilibrium Thermodynamical Principles for Chemical Reactions with Mass-Action Kinetics},
  journal = {SIAM J. Appl. Math.},
  year = {2017},
  volume = {77},
  pages = {1562--1585},
  doi = {10.1137/16M1102240},
}

@article{LMPR2017,
  author = {Matthias Liero and Alexander Mielke and Mark A. Peletier and D. R. Michiel Renger},
  title = {On microscopic origins of generalized gradient structures},
  journal = {Discrete Contin. Dyn. Syst. S},
  year = {2017},
  volume = {10},
  pages = {1--35},
  doi = {10.3934/dcdss.2017001},
}

@article{WangLiu2020,
  author = {Yiwei Wang and Chun Liu and Pei Liu and Bob Eisenberg},
  title = {Field theory of reaction-diffusion: Law of mass action with an energetic variational approach},
  journal = {Phys. Rev. E},
  year = {2020},
  volume = {102},
  pages = {062147},
  doi = {10.1103/PhysRevE.102.062147},
}

@article{PattersonRenger2019,
  author = {Robert I. A. Patterson and D. R. Michiel Renger},
  title = {Large Deviations of Jump Process Fluxes},
  journal = {Math. Phys. Anal. Geom.},
  year = {2019},
  volume = {22},
  pages = {21},
  doi = {10.1007/s11040-019-9318-4},
}

@article{Kobayashi2022Hessian,
  author  = {Kobayashi, Tetsuya J. and Loutchko, Dimitri
             and Kamimura, Atsushi and Sughiyama, Yuki},
  title   = {{Hessian} geometry of nonequilibrium chemical reaction
             networks and entropy production decompositions},
  journal = {Phys. Rev. Res.},
  volume  = {4},
  pages   = {033208},
  year    = {2022},
  doi     = {10.1103/PhysRevResearch.4.033208}
}

@article{Schnakenberg1976,
  author  = {Schnakenberg, J.},
  title   = {Network theory of microscopic and macroscopic behavior
             of master equation systems},
  journal = {Rev. Mod. Phys.},
  volume  = {48},
  pages   = {571--585},
  year    = {1976},
  doi     = {10.1103/RevModPhys.48.571}
}

@article{SchmiedlSeifert2007,
  author  = {Schmiedl, Tim and Seifert, Udo},
  title   = {Stochastic thermodynamics of chemical reaction networks},
  journal = {J. Chem. Phys.},
  volume  = {126},
  pages   = {044101},
  year    = {2007},
  doi     = {10.1063/1.2428297}
}

@article{PRS2024,
  author = {Robert I. A. Patterson and D. R. Michiel Renger and Upanshu Sharma},
  title = {Variational Structures Beyond Gradient Flows: A Macroscopic Fluctuation-Theory Perspective},
  journal = {J. Stat. Phys.},
  year = {2024},
  volume = {191},
  pages = {18},
  doi = {10.1007/s10955-024-03233-8},
}

@article{MaesNetocny2008,
  author = {Christian Maes and Karel Neto\v{c}n\'{y}},
  title = {Canonical structure of dynamical fluctuations in mesoscopic nonequilibrium steady states},
  journal = {EPL},
  year = {2008},
  volume = {82},
  pages = {30003},
  doi = {10.1209/0295-5075/82/30003},
}

@article{MaesNetocny2026,
  author = {Christian Maes and Karel Neto\v{c}n\'{y}},
  title = {Relaxation to nonequilibrium},
  journal = {J. Non-Equilib. Thermodyn.},
  year = {2026},
  volume = {51},
  number = {3},
  pages = {279--291},
  doi = {10.1515/jnet-2026-0037}
}

@article{Maes2021,
  author = {Christian Maes},
  title = {Local detailed balance},
  journal = {SciPost Phys. Lect. Notes},
  year = {2021},
  pages = {32},
  doi = {10.21468/SciPostPhysLectNotes.32},
}

@article{Polettini2014,
  author = {Matteo Polettini and Massimiliano Esposito},
  title = {Irreversible thermodynamics of open chemical networks. {I}. Emergent cycles and broken conservation laws},
  journal = {J. Chem. Phys.},
  year = {2014},
  volume = {141},
  pages = {024117},
  doi = {10.1063/1.4886396},
  note = {See also erratum, J. Chem. Phys. \textbf{142}, 229901 (2015)},
}

@article{RaoEsposito2016,
  author = {Riccardo Rao and Massimiliano Esposito},
  title = {Nonequilibrium Thermodynamics of Chemical Reaction Networks: Wisdom from Stochastic Thermodynamics},
  journal = {Phys. Rev. X},
  year = {2016},
  volume = {6},
  pages = {041064},
  doi = {10.1103/PhysRevX.6.041064},
}

@article{Gillespie2000,
  author = {Daniel T. Gillespie},
  title = {The chemical {Langevin} equation},
  journal = {J. Chem. Phys.},
  year = {2000},
  volume = {113},
  pages = {297--306},
  doi = {10.1063/1.481811},
}

@article{Horowitz2015,
  author = {Jordan M. Horowitz},
  title = {Diffusion approximations to the chemical master equation only have a consistent stochastic thermodynamics at chemical equilibrium},
  journal = {J. Chem. Phys.},
  year = {2015},
  volume = {143},
  pages = {044111},
  doi = {10.1063/1.4927395},
}

@article{Cates2022,
  author = {Michael E. Cates and Fodor, {\'E}tienne and Tomer Markovich and Cesare Nardini and Elsen Tjhung},
  title = {Stochastic Hydrodynamics of Complex Fluids: Discretisation and Entropy Production},
  journal = {Entropy},
  year = {2022},
  volume = {24},
  pages = {254},
  doi = {10.3390/e24020254},
}

@article{Avanzini2021,
  author = {Francesco Avanzini and Emanuele Penocchio and Gianmaria Falasco and Massimiliano Esposito},
  title = {Nonequilibrium thermodynamics of non-ideal chemical reaction networks},
  journal = {J. Chem. Phys.},
  year = {2021},
  volume = {154},
  pages = {094114},
  doi = {10.1063/5.0041225},
}

@article{Avanzini2024,
  author = {Francesco Avanzini and Timur Aslyamov and Fodor, {\'E}tienne and Massimiliano Esposito},
  title = {Nonequilibrium thermodynamics of non-ideal reaction--diffusion systems: Implications for active self-organization},
  journal = {J. Chem. Phys.},
  year = {2024},
  volume = {161},
  pages = {174108},
  doi = {10.1063/5.0231520},
}

@article{Bauermann2022,
  author = {Jonathan Bauermann and Sudarshana Laha and Patrick M. McCall and Frank J\"{u}licher and Christoph A. Weber},
  title = {Chemical Kinetics and Mass Action in Coexisting Phases},
  journal = {J. Am. Chem. Soc.},
  year = {2022},
  volume = {144},
  pages = {19294--19304},
  doi = {10.1021/jacs.2c06265},
}

@article{Haugerud2026,
  author = {Ivar S. Haugerud and Hidde D. Vuijk and Job Boekhoven and Christoph A. Weber},
  title = {Excitability and Oscillations of Active Droplets},
  journal = {Phys. Rev. Lett.},
  year = {2026},
  volume = {136},
  pages = {218401},
  doi = {10.1103/g9dh-ymx9},
}

@misc{ZhouFrey2026,
  author = {Daniel Zhou and Erwin Frey},
  title = {Duality Between Chemical Potential Dynamics and Reaction--Diffusion Systems},
  year = {2026},
  howpublished = {arXiv:2605.15158},
  doi = {10.48550/arXiv.2605.15158},
  note = {Preprint, version 1},
}

@article{Weyer2026,
  author = {Henrik Weyer and Tobias A. Roth and Erwin Frey},
  title = {Protein pattern morphology and dynamics emerging from effective interfacial tension},
  journal = {Nat. Phys.},
  year = {2026},
  volume = {22},
  pages = {94--102},
  doi = {10.1038/s41567-025-03101-6},
}

@article{LauLubensky2007,
  author = {A. W. C. Lau and T. C. Lubensky},
  title = {State-dependent diffusion: Thermodynamic consistency and its path integral formulation},
  journal = {Phys. Rev. E},
  volume = {76},
  pages = {011123},
  year = {2007},
  doi = {10.1103/PhysRevE.76.011123},
}

@article{Mori1965,
  author = {Hazime Mori},
  title = {Transport, Collective Motion, and {Brownian} Motion},
  journal = {Prog. Theor. Phys.},
  year = {1965},
  volume = {33},
  pages = {423--455},
  doi = {10.1143/PTP.33.423},
}

@article{Brauns2020,
  author = {Fridtjof Brauns and Jacob Halatek and Erwin Frey},
  title = {Phase-Space Geometry of Mass-Conserving Reaction-Diffusion Dynamics},
  journal = {Phys. Rev. X},
  volume = {10},
  pages = {041036},
  year = {2020},
  doi = {10.1103/PhysRevX.10.041036}
}

@article{Zhao2024,
  author = {Xueping Zhao and Susanne Liese and Alf Honigmann and Frank J\"{u}licher and Christoph A. Weber},
  title = {Theory of wetting dynamics with surface binding},
  journal = {New J. Phys.},
  volume = {26},
  pages = {103025},
  year = {2024},
  doi = {10.1088/1367-2630/ad80bb}
}

@article{Nardini2017,
  author = {Nardini, Cesare and Fodor, {\'E}tienne and Tjhung, Elsen and
            van Wijland, Fr{\'e}d{\'e}ric and Tailleur, Julien and Cates, Michael E.},
  title = {Entropy Production in Field Theories without Time-Reversal Symmetry:
           Quantifying the Non-Equilibrium Character of Active Matter},
  journal = {Phys. Rev. X},
  volume = {7},
  pages = {021007},
  year = {2017},
  doi = {10.1103/PhysRevX.7.021007}
}

@book{Gyarmati1970,
  author = {Istv{\'a}n Gyarmati},
  title = {Non-equilibrium Thermodynamics: Field Theory and Variational Principles},
  publisher = {Springer},
  address = {Berlin, Heidelberg},
  year = {1970},
  doi = {10.1007/978-3-642-51067-0}
}

@article{HartichGodec2023,
  author = {David Hartich and Alja{\v z} Godec},
  title = {Violation of local detailed balance upon lumping despite a clear timescale separation},
  journal = {Phys. Rev. Research},
  volume = {5},
  pages = {L032017},
  year = {2023},
  doi = {10.1103/PhysRevResearch.5.L032017}
}

@article{AndrieuxGaspard2007,
  author = {David Andrieux and Pierre Gaspard},
  title = {A fluctuation theorem for currents and non-linear response coefficients},
  journal = {J. Stat. Mech.},
  volume = {2007},
  number = {02},
  pages = {P02006},
  year = {2007},
  doi = {10.1088/1742-5468/2007/02/P02006}
}

@article{GrabertGreen1979,
 author = {Hermann Grabert and Melville S. Green},
 title = {Fluctuations and nonlinear irreversible processes},
 journal = {Phys. Rev. A},
 volume = {19}, pages = {1747}, year = {1979},
 doi = {10.1103/PhysRevA.19.1747}
}

@article{Bazant2013,
 author = {Martin Z. Bazant},
 title = {Theory of Chemical Kinetics and Charge Transfer based on Nonequilibrium Thermodynamics},
 journal = {Acc. Chem. Res.},
 volume = {46}, number = {5}, pages = {1144--1160}, year = {2013},
 doi = {10.1021/ar300145c}
}

@article{Yasuda2026Covariant,
  author        = {Yasuda, Kento and Zheng, Bin and Xiong, Zhongqiang and
                   Hou, Zhanglin and Ishimoto, Kenta and Xu, Xinpeng and
                   Andelman, David and Komura, Shigeyuki},
  title         = {Covariant {Onsager} and {Onsager--Machlup} principles
                   for active and inertial dynamics},
  journal       = {Chin. Phys. B},
  volume  = {35},
  pages   = {260750},
  year    = {2026},
  doi     = {10.1088/1674-1056/ae8da8} 
}

@article{WangLiu2022Review,
  author  = {Yiwei Wang and Chun Liu},
  title   = {Some Recent Advances in Energetic Variational Approaches},
  journal = {Entropy},
  volume  = {24},
  number  = {5},
  pages   = {721},
  year    = {2022},
  doi     = {10.3390/e24050721}
}

@article{Wittkowski2014,
  author={Wittkowski, Raphael and Tiribocchi, Adriano and Stenhammar, Joakim and Allen, Rosalind J. and Marenduzzo, Davide and Cates, Michael E.},
  title={Scalar {$\phi^4$} field theory for active-particle phase separation},
  journal={Nature Communications}, volume={5}, pages={4351}, year={2014},
  doi={10.1038/ncomms5351}
}

@article{Tjhung2018,
  author={Tjhung, Elsen and Nardini, Cesare and Cates, Michael E.},
  title={Cluster Phases and Bubbly Phase Separation in Active Fluids: Reversal of the {Ostwald} Process},
  journal={Physical Review X}, volume={8}, pages={031080}, year={2018},
  doi={10.1103/PhysRevX.8.031080}
}

@article{Martinez2013,
  author  = {Mart{\'i}nez, Ignacio A. and Rold{\'a}n, {\'E}dgar
             and Parrondo, Juan M. R. and Petrov, Dmitri},
  title   = {Effective heating to several thousand kelvins of an
             optically trapped sphere in a liquid},
  journal = {Phys. Rev. E},
  volume  = {87},
  pages   = {032159},
  year    = {2013},
  doi     = {10.1103/PhysRevE.87.032159}
}

@article{Chen2007Bacterial,
  author  = {Chen, D. T. N. and Lau, A. W. C. and Hough, L. A.
             and Islam, M. F. and Goulian, M. and Lubensky, T. C.
             and Yodh, A. G.},
  title   = {Fluctuations and Rheology in Active Bacterial Suspensions},
  journal = {Phys. Rev. Lett.},
  volume  = {99},
  pages   = {148302},
  year    = {2007},
  doi     = {10.1103/PhysRevLett.99.148302}
}

@article{Maggi2017Memoryless,
  author  = {Maggi, Claudio and Paoluzzi, Matteo and Angelani, Luca
             and Di Leonardo, Roberto},
  title   = {Memory-less response and violation of the
             fluctuation-dissipation theorem in colloids
             suspended in an active bath},
  journal = {Scientific Reports},
  volume  = {7},
  pages   = {17588},
  year    = {2017},
  doi     = {10.1038/s41598-017-17900-2}
}

@article{MaesSteffenoni2015,
  author  = {Maes, Christian and Steffenoni, Stefano},
  title   = {Friction and noise for a probe in a nonequilibrium fluid},
  journal = {Phys. Rev. E},
  volume  = {91},
  pages   = {022128},
  year    = {2015},
  doi     = {10.1103/PhysRevE.91.022128}
}

@article{Bazant2009,
  author  = {Bazant, Martin Z. and Kilic, Mustafa Sabri
             and Storey, Brian D. and Ajdari, Armand},
  title   = {Towards an understanding of induced-charge electrokinetics
             at large applied voltages in concentrated solutions},
  journal = {Advances in Colloid and Interface Science},
  volume  = {152},
  number  = {1--2},
  pages   = {48--88},
  year    = {2009},
  doi     = {10.1016/j.cis.2009.10.001}
}

@incollection{AndersonKurtz2011,
  author    = {Anderson, David F. and Kurtz, Thomas G.},
  title     = {Continuous Time {Markov} Chain Models for Chemical Reaction Networks},
  booktitle = {Design and Analysis of Biomolecular Circuits},
  editor    = {Koeppl, Heinz and Densmore, Douglas
               and Setti, Gianluca and di Bernardo, Mario},
  publisher = {Springer},
  address   = {New York},
  pages     = {3--42},
  year      = {2011},
  doi       = {10.1007/978-1-4419-6766-4_1}
}

@article{Gingrich2016,
  author  = {Gingrich, Todd R. and Horowitz, Jordan M.
             and Perunov, Nikolay and England, Jeremy L.},
  title   = {Dissipation Bounds All Steady-State Current Fluctuations},
  journal = {Physical Review Letters},
  volume  = {116},
  pages   = {120601},
  year    = {2016},
  doi     = {10.1103/PhysRevLett.116.120601}
}

@book{Shiraishi2023,
  author    = {Shiraishi, Naoto},
  title     = {An Introduction to Stochastic Thermodynamics: From Basic to Advanced},
  series    = {Fundamental Theories of Physics},
  volume    = {212},
  publisher = {Springer},
  address   = {Singapore},
  year      = {2023},
  doi       = {10.1007/978-981-19-8186-9}
}

@article{vanRoosbroeck1950,
  author  = {van Roosbroeck, W.},
  title   = {Theory of the Flow of Electrons and Holes in Germanium
             and Other Semiconductors},
  journal = {The Bell System Technical Journal},
  volume  = {29},
  number  = {4},
  pages   = {560--607},
  year    = {1950},
  doi     = {10.1002/j.1538-7305.1950.tb03653.x}
}

@article{BazantThorntonAjdari2004,
  author  = {Bazant, Martin Z. and Thornton, Katsuyo and Ajdari, Armand},
  title   = {Diffuse-charge dynamics in electrochemical systems},
  journal = {Physical Review E},
  volume  = {70},
  pages   = {021506},
  year    = {2004},
  doi     = {10.1103/PhysRevE.70.021506}
}

@article{Markovich2021,
  author={Markovich, Tomer and Fodor, {\'E}tienne and Tjhung, Elsen and Cates, Michael E.},
  title={Thermodynamics of Active Field Theories: Energetic Cost of Coupling to Reservoirs},
  journal={Physical Review X}, volume={11}, pages={021057}, year={2021},
  doi={10.1103/PhysRevX.11.021057}
}

@article{WangDobnikarFrenkel2022,
  author = {Wang, Xipeng and Dobnikar, Jure and Frenkel, Daan},
  title = {Numerical Test of the {Onsager} Relations in a Driven System},
  journal = {Physical Review Letters},
  volume = {129},
  pages = {238002},
  year = {2022},
  doi = {10.1103/PhysRevLett.129.238002}
}

@article{Gabrielli1996,
  author = {Gabrielli, Davide and Jona-Lasinio, Giovanni and Landim, Claudio},
  title = {{Onsager} Reciprocity Relations without Microscopic Reversibility},
  journal = {Physical Review Letters},
  volume = {77},
  pages = {1202--1205},
  year = {1996},
  doi = {10.1103/PhysRevLett.77.1202}
}

@article{DeCorato2022,
  author = {{De Corato}, Marco and Pagonabarraga, Ignacio},
  title = {{Onsager} reciprocal relations and chemo-mechanical coupling for chemically active colloids},
  journal = {The Journal of Chemical Physics},
  volume = {157},
  pages = {084901},
  year = {2022},
  doi = {10.1063/5.0098425}
}

@article{KaiserJackZimmer2018,
  author = {Kaiser, Marcus and Jack, Robert L. and Zimmer, Johannes},
  title = {Canonical Structure and Orthogonality of Forces and Currents in Irreversible {Markov} Chains},
  journal = {Journal of Statistical Physics},
  volume = {170},
  pages = {1019--1050},
  year = {2018},
  doi = {10.1007/s10955-018-1986-0}
}

@article{Prinz2011,
  author = {Prinz, Jan-Hendrik and Wu, Hao and Sarich, Marco and Keller, Bettina and Senne, Martin and Held, Martin and Chodera, John D. and Sch{\"u}tte, Christof and No{\'e}, Frank},
  title = {{Markov} models of molecular kinetics: Generation and validation},
  journal = {The Journal of Chemical Physics},
  volume = {134},
  pages = {174105},
  year = {2011},
  doi = {10.1063/1.3565032}
}

@article{RogersPitman1981,
  author = {Rogers, L. C. G. and Pitman, J. W.},
  title = {{Markov} Functions},
  journal = {The Annals of Probability},
  volume = {9},
  number = {4},
  pages = {573--582},
  year = {1981},
  doi = {10.1214/aop/1176994363}
}

@article{Basu2015SecondOrder,
  author = {Basu, Urna and Kr{\"u}ger, Matthias and Lazarescu, Alexandre and Maes, Christian},
  title = {Frenetic aspects of second order response},
  journal = {Physical Chemistry Chemical Physics},
  volume = {17},
  pages = {6653--6666},
  year = {2015},
  doi = {10.1039/C4CP04977B}
}

@article{Doi2015Approximation,
  author  = {Doi, Masao},
  title   = {{Onsager} principle as a tool for approximation},
  journal = {Chin. Phys. B},
  volume  = {24},
  number  = {2},
  pages   = {020505},
  year    = {2015},
  doi     = {10.1088/1674-1056/24/2/020505}
}

@misc{Lyu2026Glassy,
  author        = {Lyu, Peihan and Ding, Zhaoyu and Doi, Masao and Man, Xingkun},
  title         = {Shear Unfreezing Explains Yielding, Plasticity and
                   Neck Initiation of Glassy Polymers},
  year          = {2026},
  eprint        = {2607.09160},
  archivePrefix = {arXiv},
  primaryClass  = {cond-mat.soft},
  doi           = {10.48550/arXiv.2607.09160},
  note          = {Preprint}
}

@article{Yasuda2024Statistical,
  author  = {Yasuda, Kento and Ishimoto, Kenta and Komura, Shigeyuki},
  title   = {Statistical formulation of the {Onsager--Machlup}
             variational principle},
  journal = {Phys. Rev. E},
  volume  = {110},
  pages   = {044104},
  year    = {2024},
  doi     = {10.1103/PhysRevE.110.044104}
}

@article{Zheng2026Paths,
  author  = {Zheng, Bin and Xiong, Zhongqiang and Li, Changhao and
             Hou, Zhanglin and Zhang, Ziluo and Xu, Xinpeng and
             Lin, Li-Shing and Ishimoto, Kenta and Yasuda, Kento and
             Komura, Shigeyuki},
  title   = {Neural optimization of the most probable paths of
             three-dimensional active {Brownian} particles},
  journal = {Phys. Rev. E},
  volume  = {114},
  pages   = {L012103},
  year    = {2026},
  doi     = {10.1103/ys1b-1lrd}
}

@article{Hao2025GrayScott,
  author  = {Hao, Wenrui and Liu, Chun and Wang, Yiwei and Yang, Yahong},
  title   = {On pattern formation in the thermodynamically-consistent
             variational {Gray--Scott} model},
  journal = {Math. Biosci.},
  volume  = {385},
  pages   = {109453},
  year    = {2025},
  doi     = {10.1016/j.mbs.2025.109453}
}

@article{Liese2025Wetting,
  author  = {Liese, Susanne and Zhao, Xueping and Weber, Christoph A. and
             J{\"u}licher, Frank},
  title   = {Chemically active wetting},
  journal = {Proc. Natl. Acad. Sci. U.S.A.},
  volume  = {122},
  number  = {15},
  pages   = {e2403083122},
  year    = {2025},
  doi     = {10.1073/pnas.2403083122}
}

@article{EYu2018DeepRitz,
  author  = {E, Weinan and Yu, Bing},
  title   = {The Deep Ritz Method: A Deep Learning-Based
             Numerical Algorithm for Solving Variational Problems},
  journal = {Communications in Mathematics and Statistics},
  volume  = {6},
  pages   = {1--12},
  year    = {2018},
  doi     = {10.1007/s40304-018-0127-z}
}

@article{Wang2022DeepRitz,
  author  = {Wang, Haiqin and Zou, Boyi and Su, Jian
             and Wang, Dong and Xu, Xinpeng},
  title   = {Variational methods and deep Ritz method
             for active elastic solids},
  journal = {Soft Matter},
  volume  = {18},
  pages   = {6015--6031},
  year    = {2022},
  doi     = {10.1039/D2SM00404F}
}

@article{Li2025DOMM,
  author  = {Li, Zhihao and Zou, Boyi and Wang, Haiqin
             and Su, Jian and Wang, Dong and Xu, Xinpeng},
  title   = {Deep Learning-Based Computational Method for
             Soft Matter Dynamics: Deep Onsager-Machlup Method},
  journal = {Communications in Computational Physics},
  volume  = {37},
  number  = {2},
  pages   = {353--382},
  year    = {2025},
  doi     = {10.4208/cicp.OA-2023-0294}
}

@article{Seifert2011,
  author = {Udo Seifert},
  title = {Stochastic thermodynamics of single enzymes and
           molecular motors},
  journal = {Eur. Phys. J. E},
  volume = {34},
  pages = {26},
  year = {2011},
  doi = {10.1140/epje/i2011-11026-7}
}

@article{Seifert2012,
  author = {Udo Seifert},
  title = {Stochastic thermodynamics, fluctuation theorems,
           and molecular machines},
  journal = {Rep. Prog. Phys.},
  volume = {75},
  pages = {126001},
  year = {2012},
  doi = {10.1088/0034-4885/75/12/126001}
}

@article{TenaSolsona2017,
  author = {Marta Tena-Solsona and Benedikt Rie{\ss} and Raphael K. Gr{\"o}tsch
            and Franziska C. L{\"o}hrer and Caren Wanzke and Benjamin K{\"a}sdorf
            and Andreas R. Bausch and Peter M{\"u}ller-Buschbaum
            and Oliver Lieleg and Job Boekhoven},
  title = {Non-equilibrium dissipative supramolecular materials with a tunable lifetime},
  journal = {Nature Communications},
  volume = {8},
  pages = {15895},
  year = {2017},
  doi = {10.1038/ncomms15895}
}

@incollection{Arroyo2018,
  author = {Marino Arroyo and Nikhil Walani and Alejandro Torres-S{\'a}nchez
            and Dimitri Kaurin},
  title = {{Onsager}'s Variational Principle in Soft Matter: Introduction and
           Application to the Dynamics of Adsorption of Proteins onto Fluid Membranes},
  booktitle = {The Role of Mechanics in the Study of Lipid Bilayers},
  editor = {David J. Steigmann},
  series = {CISM International Centre for Mechanical Sciences},
  volume = {577},
  pages = {287--332},
  publisher = {Springer},
  address = {Cham},
  year = {2018},
  doi = {10.1007/978-3-319-56348-0_6}
}
\end{document}